\documentclass[twocolumn,english,superscriptaddress,amsmath,floatfix,longbibliography,floats,prx]{revtex4-1}
\usepackage[colorlinks=true,urlcolor=blue,citecolor=blue,linkcolor=blue]{hyperref} 
\usepackage[T1]{fontenc}
\usepackage[utf8]{inputenc}
\usepackage{amssymb}
\usepackage{graphicx}
\usepackage{amsmath,color}
\usepackage{mathrsfs}
\usepackage{float}
\usepackage{indentfirst}
\usepackage{txfonts}
\usepackage[normalem]{ulem}
\usepackage[table]{xcolor}
\usepackage{array,ragged2e}
\usepackage{multirow}
\makeatletter

\newcommand{\ave}[1]{\langle #1 \rangle}

\makeatother

\usepackage{babel}

\begin{document}

\hyphenpenalty=5000

\tolerance=1000

\title{The emergence of gapless quantum spin liquid from deconfined quantum critical point}

\author{Wen-Yuan Liu}
\affiliation{Department of Physics, The Chinese University of Hong Kong, Shatin, New Territories, Hong Kong, China}
\author{Juraj Hasik}
\affiliation{Laboratoire de Physique Th\'eorique, C.N.R.S. and Universit\'e de Toulouse, 31062 Toulouse, France}
\author{Shou-Shu Gong}
\affiliation{Department of Physics,  Beihang University, Beijing 100191, China}
\author{Didier Poilblanc}
\affiliation{Laboratoire de Physique Th\'eorique, C.N.R.S. and Universit\'e de Toulouse, 31062 Toulouse, France}
\author{Wei-Qiang Chen}
\affiliation{Shenzhen Key Laboratory of Advanced Quantum Functional Materials and Devices, Southern University of Science and Technology, Shenzhen 518055, China}
\affiliation{Institute for Quantum Science and Engineering and Department of Physics, Southern University of Science and Technology, Shenzhen 518055, China}
\author{Zheng-Cheng Gu}
\affiliation{Department of Physics, The Chinese University of Hong Kong, Shatin, New Territories, Hong Kong, China}

\date{\today }

\begin{abstract}
A quantum spin liquid (QSL) is a novel phase of matter with long-range entanglement where localized spins are highly correlated with the vanishing of magnetic order. Such exotic quantum states provide the opportunities to develop new theoretical frameworks for many-body physics and have the potential application in realizing robust quantum computations.
Here we show that a gapless QSL can naturally emerge from a deconfined quantum critical point (DQCP), which is originally proposed to describe Landau forbidden continuous phase transition between antiferromagnetic (AFM) and valence-bond solid (VBS) phases. Via large-scale tensor network simulations of a square-lattice spin-1/2 frustrated Heisenberg model, both QSL state and DQCP-type AFM-VBS transition are observed. With tuning coupling constants, the AFM-VBS transition vanishes and instead, a gapless QSL phase gradually develops in between.
Remarkably, along the phase boundaries of AFM-QSL and QSL-VBS transitions, we always observe the same correlation length exponents $\nu\approx 1.0$, which is intrinsically different from the one of the 
DQCP-type transition, indicating new types of universality classes. Our results explicitly demonstrate a new scenario for understanding the emergence of gapless QSL from an underlying DQCP.  The discovered QSL phase survives in a large region of tuning parameters and we expect its experimental realizations in solid state materials or quantum simulators.

\end{abstract}

\date{\today}
\maketitle

\noindent\textbf{Introduction}

Quantum fluctuations can melt antiferromagnetic (AFM) orders in frustrated spin systems at low temperature, leading to the emergence of new phases of quantum matter and unconventional quantum phase transitions. A prominent example is the quantum spin liquid (QSL) that preserves both spin rotation symmetry and lattice symmetry. QSL states can exhibit collective phenomena such as emergent gauge fields and fractional excitations, which are qualitatively different from an ordinary paramagnet~\cite{balents2010}. In the past several decades, QSL has attracted numerous attention since it was proposed as the parent state of the high-temperature cuprate superconductors~\cite{Anderson1987}. Nowadays ongoing efforts on searching for QSL are still being made also due to its exotic topological properties.

Another well known example is the zero-temperature continuous phase transition between an AFM state and  a valence-bond solid (VBS) state, which cannot be understood by the standard  Landau-Ginzburg-Wilson (LGW) paradigm.  To describe such a kind of continuous  phase transition between two ordered phases, a deconfined quantum critical point (DQCP) scenario was proposed, where fractional excitations and emergent gauge field also naturally arise~\cite{DQCP1,DQCP2}. In the DQCP theory, the fundamental degrees of freedom are fractionalized spinon excitations with spin-$1/2$. The condensation of spinons leads to the AFM phase while the confinement of spinons leads to the VBS phase. Right at the critical point, the deconfined spinons couple to the emergent $U(1)$ gauge field and enhanced symmetries have been observed numerically~\cite{JQ2007,loopmodel2,sreejith2019}.

Previously, the DQCP-related physics has been extensively studied in sign-problem-free models, including a  family of $J-Q$ models~\cite{JQ2007,JQ2008,JQ2008_2,JQ2009,JQ2010,JQ2011,JQ2013,JQ2013_2,JQ2013_3,JQ2015,JQ2020}, loop~\cite{loopmodel1,loopmodel2} and dimer~\cite{sreejith2019,charrier2010} models, as well as  fermionic models~\cite{liuyuhai2019}.
In most cases, the obtained physical quantities exhibit unusual scaling violations~\cite{JQ2008,JQ2010,JQ2011,JQ2015,JQ2020,loopmodel1,sreejith2019}, and it is unclear whether the observed phase transition between AFM and VBS states is continuous or weakly first-order. Different scenarios have been proposed for explaining these perplexing phenomena~\cite{JQ2016,wang2017,gorbenko2018_1,gorbenko2018_2,lizhijin2018,Ashvin2019,Wang2020,Nahum2020,sandvik2020,He2021,Fakher2021,wangling2021}. According to P. W.  Anderson, a QSL state can be regarded as a resonating valence bond (RVB) state~\cite{Anderson1973}, or the superposition of different kinds of VBS patterns, and the translational symmetry is restored by quantum fluctuations. 
In fact, DQCP can also be regarded as a special kind of unstable RVB state from which both VBS order and AFM order are developed. Such a physical picture raises a very important issue: Is there an intrinsic relation between the DQCP and a QSL phase? Recently conjectured quantum field theory suggests that a DQCP might expand into a stable gapless QSL phase~\cite{liuQSL}. Later, a numerical study of Shastry-Sutherland model combining analyses of results from other models also argues that the DQCP is a multi-critical point at the end of the QSL phase, connecting it to a first-order transition line~\cite{wangling2021}. However, it is  far from clear whether these speculations are correct or not.

 \begin{figure}[htbp]
 \centering
 \includegraphics[width=3.4in]{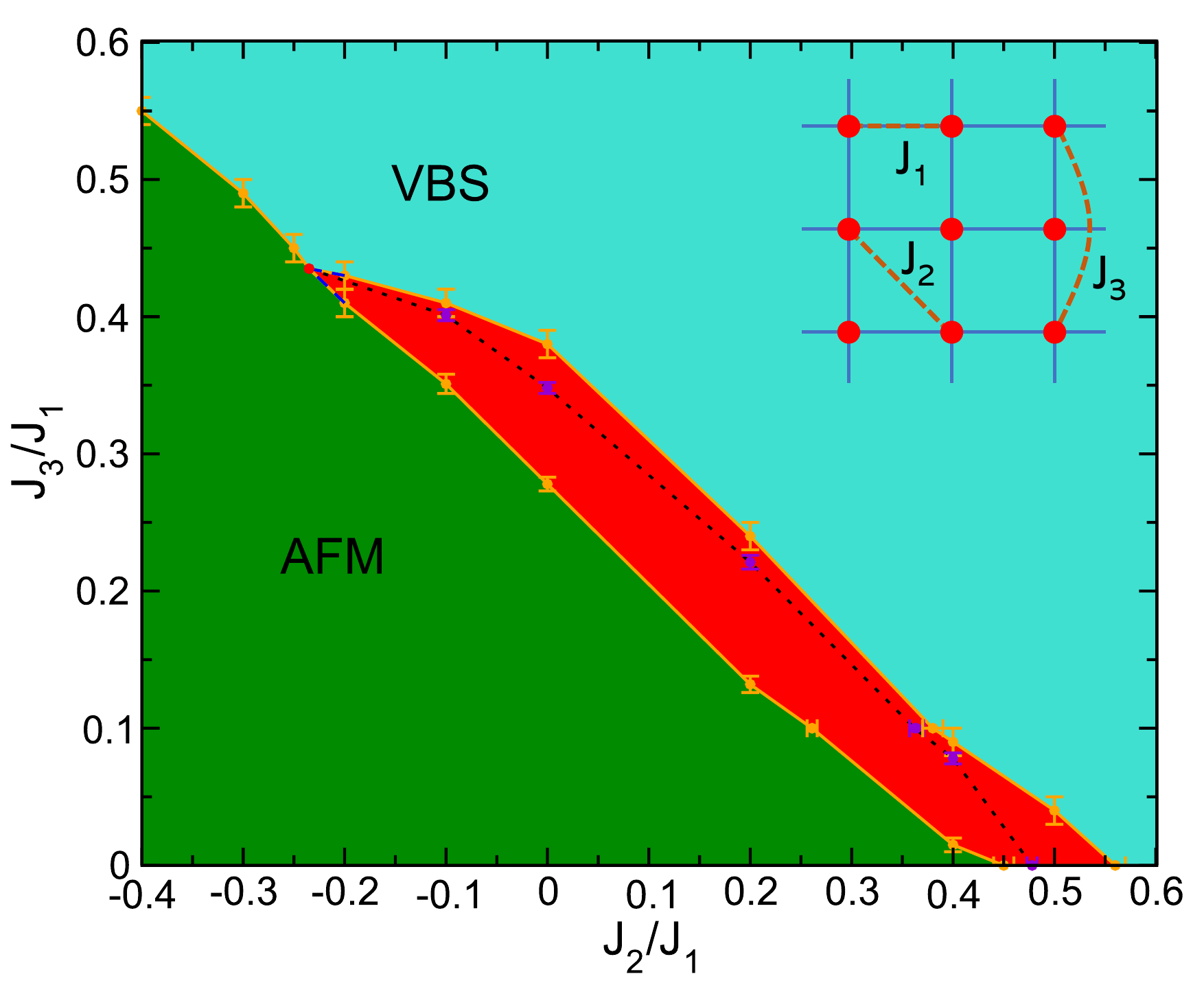}
 \caption{Ground state phase diagram of the $J_1$-$J_2$-$J_3$ model, including three phases: the AFM, VBS and gapless QSL (red region) phases. The blue dashed line denotes the unknown shape of the QSL phase close to the tricritical point. Violet points on the black dotted line in the QSL phase have the same decay power  for spin and dimer correlations, namely $\eta_s=\eta_d$.
 Error bars denote one standard deviation.}
 \label{fig:J1J2J3phaseDiagram}
 \end{figure}

In this work we present a specific example where both gapless QSL and DQCP can be observed in a single frustrated model. 
The gapless QSL gradually develops from the DQCP by tuning parameters.  To the best of our knowledge, this is the first concrete example to explicitly show the possible intrinsic relation between DQCP and QSL. Such a relation can provide crucial insights into the understanding of the underlying physics of DQCP and the gapless QSL phase.  In particular, we investigate a spin-$1/2$ square-lattice  model, which contains first, second and third-nearest neighbour Heisenberg exchange interaction couplings $J_1$, $J_2$ and $J_3$, respectively, described by the following Hamiltonian:
\begin{equation}
H=J_1\sum_{\langle i,j \rangle}\mathbf{S_i}\cdot\mathbf{S_j}+J_2\sum_{\langle\langle
i,j\rangle\rangle}\mathbf{S_i}\cdot\mathbf{S_j}+J_3\sum_{\langle\langle\langle
i,j\rangle\rangle\rangle}\mathbf{S_i}\cdot\mathbf{S_j}.\label{model}
\end{equation}
We set $J_1=1$, and the couplings $J_2, J_3$ are tuning parameters. This model drew much attention in the early days after high temperature superconductivity was discovered. Although there are analytic and small-size numerical results~\cite{rastelli1986,spinwave1988,RG1988,ed1990,largeN1991,nonlineareffect1992,ferrer1993,ed1996,DMRG2004,classical2004,PVB4,ed2010}, its phase diagram is still far from clear, especially for the region close to the classical critical line $J_1-2J_2=4J_3$. By using the state-of-the-art tensor network method, both DQCP and gapless QSL are observed near the classical critical line. 
Remarkably, along the phase boundaries of AFM-QSL and QSL-VBS transitions, we always observe the same correlation length exponents $\nu\approx 1.0$. In contrast, along the phase boundary of AFM-VBS transition, we observe the correlation length exponents are intrinsically close to the one from  other DQCP studies based on similar sizes which is $\nu\approx 0.8$~\cite{JQ2007}.  
These findings  reveal the deep relations between DQCP and QSL, and provide us with invaluable guidance for  understanding the gapless QSL developed from an underlying DQCP as well as the experimental realization of gapless QSL in square-lattice based materials or quantum simulators.

\begin{figure}[htbp]
 \centering
 \includegraphics[width=3.4in]{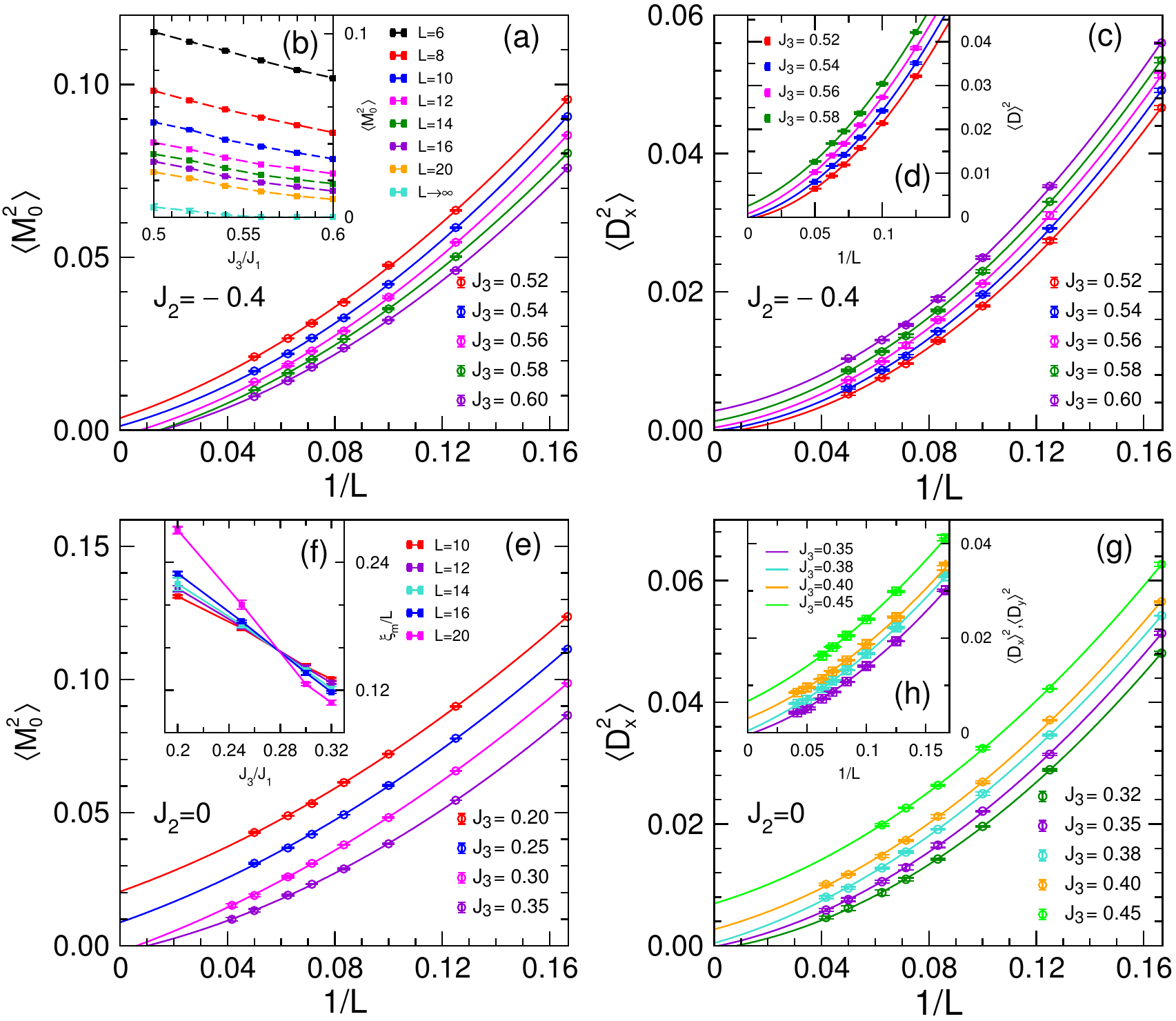}
 \caption{Scaling analysis of (squared) order parameters of  $J_1$-$J_2$-$J_3$ model at $J_2=-0.4$ (a-d) and at $J_2=0$ (e-h). (a,e) Finite size scaling of AFM order parameters, (b) $J_3$ dependence of AFM orders on different sizes; (c,g) horizontal dimer order parameters $\langle D_x^2 \rangle$ based on bond-bond correlations; (d) boundary-induced  dimer order parameters $\langle D \rangle^2$; (f) dimensionless quantity $\xi_m/L$; (h) boundary-induced  dimer order parameters along both $x-$ direction ($\langle D_x \rangle^2$, square symbols) and $y-$ direction ($\langle D_y\rangle^2$, diamond symbols).  All extrapolations with respect to $1/L$ are preformed through  second order polynomial fits. Error bars denote one standard deviation of the sampled mean values.}
 \label{fig:OrderParameterJ2_0_J2_m04}
 \end{figure}

\bigskip
\noindent\textbf{Continuous AFM to VBS transition}  \\
We first consider the phase diagram of the $J_1$-$J_2$-$J_3$ model in the region with a fixed significant ferromagnetic $J_2$ coupling, e.g. $J_2=-0.4$, $J_2=-0.3$ or $J_2=-0.25$. In this situation  $J_2$ coupling will enhance the AFM order, coordinating with the $J_1$ coupling. With increasing AFM coupling $J_3$, we observe a direct transition from the AFM to the VBS phase.  
 The (squared)  AFM order parameter  $\langle {\bf M}_0^2\rangle$ is defined as the value of the structure factor $S({\bf k})=\frac{1}{L^2}\sum_{\bf{ij}}\langle{\bf S}_{{\bf i}}\cdot {\bf S}_{{\bf j}}\rangle {e}^{i {\bf k}\cdot({\bf i}-{\bf j})}$ at the wave vector  $\bf k_0=(\pi,\pi)$, i.e., $\langle {\bf M}_0^2\rangle=\frac{1}{L^2} S({\bf k_0})$.
In Fig.~\ref{fig:OrderParameterJ2_0_J2_m04}(a), we present the AFM  order (squared) on different $L\times L$ systems up to $20\times 20$. Finite size scaling reveals the disappearance of the AFM order at $J_3 = J_{c1} \simeq 0.55$, for $J_2=-0.4$.

Then, we measure the dimer order parameter to detect the possible spontaneous appearance of a VBS order. The dimer order parameter (DOP) on open boundary conditions is defined as~\cite{zhao2020,liuQSL}
\begin{equation}
D_{\alpha}=\frac{1}{N_b}\sum_{{\bf i}}(-1)^{i_{\alpha}}B^{\alpha}_{\bf i},
\end{equation} 
where $B^{\alpha}_{\bf i}={\bf S}_{{\bf i}} \cdot {\bf S}_{{\bf i}+{\rm e_\alpha}}$ is the bond operator between site ${\bf i}$ and site ${\bf i}+{\rm e_\alpha}$ along $\alpha$ direction  with $\alpha=x$ or $y$, and $N_b=L(L-1)$ is the corresponding total number of counted bonds along the $\alpha$ direction. The horizontal DOP $\langle D^2_x \rangle$ based on the bond-bond correlations is presented in Fig.~\ref{fig:OrderParameterJ2_0_J2_m04}(c) with the largest size up to $20\times 20$. 
One can see that the DOP vanishes in the 2D limit at $J_3= 0.54$, but acquires a nonzero extrapolated value at $J_3=0.56$ indicating a VBS order. As a double check, boundary induced dimerizations $\langle D \rangle^2=\langle D_x \rangle^2+\langle D_y \rangle^2$ are also shown in Fig.~\ref{fig:OrderParameterJ2_0_J2_m04}(d), and it also suggests that VBS order sets in just above $J_3=J_{c2}\simeq 0.55$.
Our analysis then shows that $J_{c2}\simeq J_{c1}$, giving strong evidence for a direct AFM-VBS transition point, located at $J_3\simeq 0.55$ for $J_2=-0.4$. As shown in Fig.~\ref{fig:OrderParameterJ2_0_J2_m04}(b), the AFM order parameter on each size shows a smooth variation with respect to $J_3$. Therefore, based on our results, the AFM-VBS transition is very likely to be continuous, though the possibility of a weak first-order transition cannot be completely ruled out.

\bigskip
\noindent\textbf{Emergent QSL phase}  \\
Next, we set $J_2=0$, i.e. we investigate the phase diagram of the $J_1$-$J_3$ model. 
Through a finite-size scaling analysis shown in Fig.~\ref{fig:OrderParameterJ2_0_J2_m04}(e), we can see that the AFM order still survives at $J_3=0.25$, but vanishes at  $J_3= 0.3$ in the thermodynamic limit. To determine the phase transition point precisely and conveniently, a  dimensionless quantity $\xi_m/L$ is computed, where $\xi_m$ is a correlation length defined as $\xi_m=\frac{L}{2\pi}\sqrt{\frac{S(\pi,\pi)}{S(\pi,\pi+2\pi/L)}-1}$~\cite{FSS1},  which clearly shows that the AFM phase transition point is located at $J_3=J_{c1}\simeq 0.28$ in Fig.~\ref{fig:OrderParameterJ2_0_J2_m04}(f). 
 
We then measure dimer order parameters  to detect possible VBS order. The horizontal DOP $\langle D^2_x \rangle$ is presented in Fig.~\ref{fig:OrderParameterJ2_0_J2_m04}(c) with the largest size up to $24\times 24$. One can see that  the DOP in 2D limit is close to zero at $J_3=0.38$, but appears at $J_3=0.40$ with an obvious nonzero extrapolated value characterizing a VBS order. In addition, the boundary induced dimerizations $\langle D_x \rangle^2$ and $\langle D_y \rangle^2$ shown in Fig.~\ref{fig:OrderParameterJ2_0_J2_m04}(d) also suggest the absence of VBS order for $J_3\lesssim 0.38$. Note that the extrapolated values of  $\langle D_x^2 \rangle$ and  $\langle D_x \rangle^2$ are 0.0004(4) and 0.0004(3) at $J_2=0.38$; 0.0027(5) and 0.0031(4) at $J_3=0.40$; 0.0069(6) and 0.0067(4) at $J_3=0.45$, respectively, well consistent with each other in all three cases 
(for all $L\times L$ sizes presented here, the $x-$ and $y-$ directions are isotropic with $\langle D_x\rangle^2=\langle D_y\rangle^2$). 
These results demonstrate the reliability of our calculations. Therefore, by excluding spin and dimer orders, a QSL phase is suggested for $0.28\lesssim J_3 \lesssim 0.38$.

Now we turn to the $J_2 > 0$ case, in which $J_1$, $J_2$ and $J_3$ couplings compete with each other. To complete the full phase diagram of the $J_1$-$J_2$-$J_3$ model, we compute the relevant order parameters along two vertical lines, $J_2=0.2$ and $J_2=0.4$, varying $J_3$ (see Appendix~\ref{app:gapless} for details), as well as along a horizontal line $J_3=0.1$ varying $J_2$. Through finite-size scaling analysis of the order parameters, we locate the QSL phase in the regions $0.13\lesssim J_3\lesssim0.24$ along the line $J_2=0.2$, $0.015\lesssim J_3 \lesssim 0.09$ along the line $J_2=0.4$, and $0.26\lesssim J_2\lesssim 0.38$ along the line $J_3=0.1$, sandwiched by the AFM and VBS phases.
 
To confirm the existence of a potential QSL phase, we further compare the finite PEPS results with infinite PEPS (iPEPS) results at two typical points $(J_2, J_3)=(0, 0.35)$ and (0.2, 0.2), which we find to be inside the QSL phase from our finite PEPS calculations. At these two points, the thermodynamic limit ground state energies from finite size scaling  are $-0.56995(9)$ and $-0.53982(9)$, in very good agreement with the corresponding iPEPS ground state energies  $-0.56956(2)$ and $-0.53966(2)$. Furthermore, by measuring order parameters, the iPEPS results also support that the two points are in the QSL phase. More details can be seen in Appendix~\ref{app:iPEPS}. Thus, our results strongly indicate a QSL phase in the $J_1$-$J_2$-$J_3$ model in an extended region of the two-dimensional tuning parameter space $(J_2,J_3)$. Finite-size effects have been effectively reduced by a detailed comparison of systems of increasing size up to $24\times 24$, backed-up by supplementary iPEPS computations directly in the thermodynamic limit.

Finally, we focus on the strip between the vertical line $J_2=-0.3$, hosting a direct AFM-VBS transition at $J_3\simeq 0.49$ and another vertical line $J_2=0$ with a wide QSL phase for $0.28\lesssim J_3 \lesssim 0.38$. By analysing order parameters, at fixed $J_2=-0.25$ a direct AFM-VBS transition occurs at $J_3\simeq 0.45$, and  at  fixed $J_2=-0.2$ the QSL potentially appears in a small region $0.41\lesssim J_3 \lesssim 0.43$ and evidently expands to a relatively large region $0.35\lesssim J_3 \lesssim 0.41$ at  fixed $J_2=-0.1$  (see Appendix~\ref{app:gapless} for details). This reveals how the QSL can gradually emerge by increasing the $J_2$ coupling from the DQCP which describes the continuous transition line between AFM and VBS phases. The phase diagram summarizing these results is shown in Fig.~\ref{fig:J1J2J3phaseDiagram}.

\bigskip
\noindent\textbf{Critical exponents}\\
To extract critical exponents for the quantum phase transitions between QSL and AFM/VBS phases, we  analyse the scaling of physical quantities according to the standard scaling formula with a possible subleading correction~\cite{JQ2007,FSS1}:
\begin{equation}
A(J_3, L)=L^{\kappa}(1+aL^{-\omega} )F[L^{1/\nu}(J_3-J_c)/J_c]. 
\end{equation}
where $A= \xi_m$, $\ave{M^2_0}$, or  $\ave{D^2_x}$, and $\kappa=1$ for $\xi_m$, $-(z+\eta_s^*)$ for $\ave{M^2_0}$, and  $-(z+\eta_d^*)$ for $\ave{D^2_x}$, in which   $\eta_s^*$ and $\eta_d^*$ are corresponding spin and dimer correlation function exponents and $z$ is the dynamic exponent at the transition. Factors $a$ and $\omega$ are tuning parameters of the subleading term. Here $J_3$ is the tuning parameter with a fixed $J_2$. 

\begin{figure*}[htbp]
 \centering
 \includegraphics[width=6.4in]{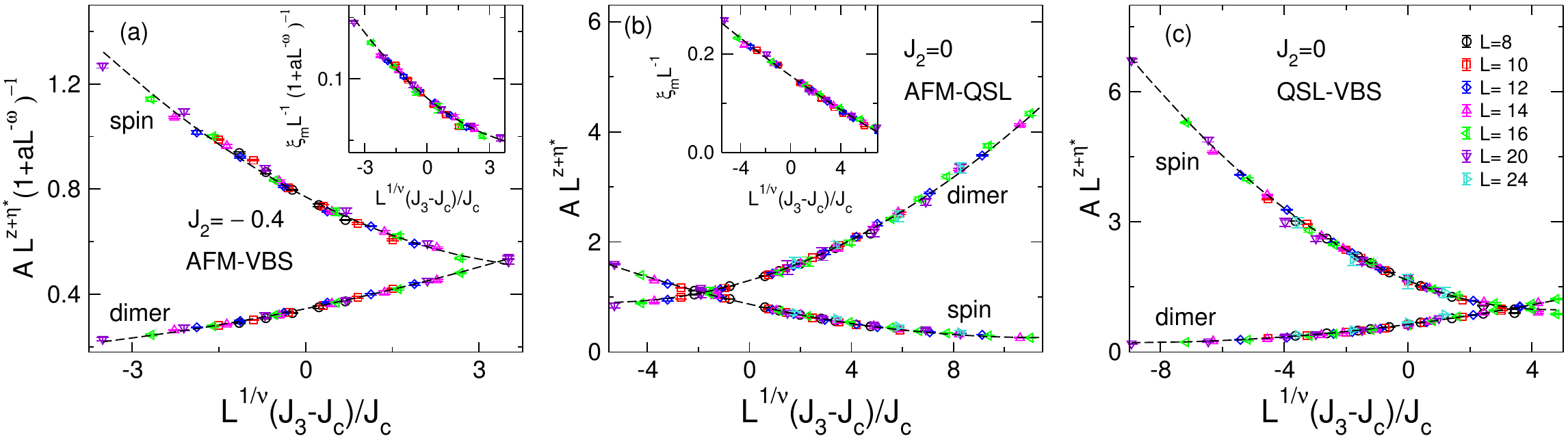}
 \caption{Scaling of physical quantities including order parameters and correlation lengths. (a) At the AFM-VBS transition point with a fixed $J_2=-0.4$, using $\nu=0.82$, $J_c=0.55$, $z+\eta_s^*=1.33$, $z+\eta_d^*=1.36$ and $\omega=1.5$. Prefactors $a$ are $11.5$, $3.8$, $11.3$ for the quantity $\xi_m$, $\ave{{\bf M}^2_0}$ and $\ave{D^2_x}$, correspondingly.  
 (b)  At the AFM-QSL transition point with a fixed $J_2=0$, using $\nu=1.02$, $J_c=0.278$, $z+\eta_s^*=1.21$, $z+\eta_d^*=1.89$. (c) At the QSL-VBS transition point with a fixed $J_2=0$, using $\nu=1.02$, $J_c=0.38$, $z+\eta_s^*=1.69$, $z+\eta_d^*=1.40$. For AMF-QSL and QSL-VBS transitions, subleading corrections are not used. Black dashed lines are quadratic curves using corresponding critical exponents.}
 \label{fig:DataCollapse}
 \end{figure*}

We first consider the quantities scaling for the AFM-VBS transition with fixed $J_2=-0.4$. The transition point is estimated at $J_3\simeq 0.55$ from finite-size scaling analysis of order parameters as mentioned previously, and it is actually also supported by the crossing of the dimensionless quantity  $\xi_m/L$.  To achieve a good data collapse for these quantities, we find a subleading correction is necessary. As seen in the inset of Fig.~\ref{fig:DataCollapse}(a),  the spin correlation length  $\xi_m$ at different sizes and couplings can be scaled using  $\nu=0.82(5)$  and $J_c=0.55(1)$. Next, we keep $\nu$ and $J_c$ fixed to extract spin and dimer correlation function exponents which leads to $z+\eta_s^*=1.33(6)$ and $z+\eta_d^*= 1.36(5)$. In these cases, a subleading term has been used with fixed $\omega= 1.5$ and different $a$  for  $\xi_m$, $\ave{M^2_0}$ and $\ave{D^2_x}$, respectively. Note the obtained critical exponents including $\nu$, $z+\eta_s^*$ and $z+\eta_d^*$ are close to those from $J$-$Q$ model based on the similar system sizes. 

Now we consider the critical exponents for AFM-QSL transition at $J_{c1}$ and QSL-VBS transition at $J_{c2}$. In this case, we find that a single correlation length exponent $\nu$ can scale the physical quantities very well at the two transition points, as  shown in the  Figs.~\ref{fig:DataCollapse}(b) and (c) for the data collapse of the $J_1-J_3$ model, i.e. at fixed $J_2=0$. We also choose other fixed values including $J_2=-0.1$, $0.2$ and $0.4$ with the tuning parameter $J_3$, as well as fixed $J_3=0.1$ with the tuning parameter $J_2$, to extract  critical exponents at their transition points,  and they have the same behaviour.  In these cases,  a good data collapse can be obtained without subleading correction terms. The critical exponents are  listed in Table.~\ref{tab:criticalexponents}.

From Table.~\ref{tab:criticalexponents}, we can see that, for all cases of the AFM-QSL and QSL-VBS transitions at a fixed $J_2$ or $J_3$,  the corresponding spin and dimer correlation exponents $z+\eta_s^*$ and $z+\eta_d^*$ are consistent for each kind of transitions. Roughly, $z+\eta_{s1}^*\sim 1.2$ and $z+\eta_{d1}^*\sim 1.9$ for the AFM-QSL transition, and $z+\eta_{s2}^*\sim 1.6$ and $z+\eta_{d2}^*\sim 1.5$ for the QSL-VBS transition. The correlation exponents for the $J_1-J_2$ model (i.e. for fixed $J_3=0$) show slight differences, probably caused by a very large correlation length. In this case the DMRG results have not yet converged well even with as many as $M=14000$ SU(2) kept states (equivalent to about 56000 U(1) states)) on $12\times 28$ strips~\cite{liuQSL}, unlike the $J_1-J_2-J_3$ model for which $M=10000$ works very well for two typical points $(J_2,J_3)=(0, 0.35)$ and (0.2, 0.2) that are also in the QSL phase (see Appendix.~\ref{app:dmrg}). Most importantly, all of these cases support the same correlation length exponent, i.e., $\nu\approx 1.0$. In particular, $\nu\approx 1$ is apparently different from that of the AFM-VBS transition obtained in the $J-Q$ model or in the $J_1-J_2-J_3$ model  with similar system sizes.  These features strongly suggest new universality classes for the AFM-QSL and QSL-VBS transitions. 
 
   \begin{table}[htbp]
   \centering
 \caption { Critical exponents of $J_1$-$J_2$-$J_3$ model  at the AFM-QSL and QSL-VBS transition point using fixed $J_2$ or $J_3$, or at the AFM-VBS transition using  fixed $J_2=-0.25$, $-0.3$ and $-0.4$.  $J_3=0$ results are taken from Ref.~\cite{liuQSL}. Results of $J$-$Q$ model up to $32\times 32$~\cite{JQ2007} are listed for comparison. Numbers in brackets are one standard deviation. Errors of critical point $J_c$ are estimated from finite-size scaling analysis of order parameters or fitting correlation lengths. Values of $\nu$ for QSL-related transitions are an average of several values from fitting order parameters at the transition points, see Appendix.~\ref{app:extract}, and spin and dimer exponents $z+\eta^*$ are obtained by data collapse using  the listed values of $\nu$ and $J_c$. }
	\begin{tabular*}{\hsize}{@{}@{\extracolsep{\fill}}lccccl@{}}
		\hline\hline
	   model &type  &   $z+\eta_s^*$ & $z+\eta_d^*$    & $\nu$ & $J_c$  \\ \hline
 		 $J-Q$ &AFM-VBS & 1.26(3) & 1.26(3) & 0.78(3) & ~\\
         $J_2=-0.4$ &AFM-VBS & 1.33(6) & 1.36(5) & 0.82(5) & 0.55(1)\\
         $J_2=-0.3$ &AFM-VBS & 1.35(3) & 1.32(4) & 0.86(6) & 0.49(1)\\
         $J_2=-0.25$ &AFM-VBS & 1.31(3) & 1.34(3) & 0.89(5) & 0.45(1)\\ \hline\hline
          $J_2=-0.1$ & AFM-QSL  & 1.31(1)   & 1.83(1)  & 1.03(6) & 0.351(7) \\
	   	$J_2=-0.1$ &QSL-VBS &1.60(1)   & 1.53(1)  & 1.03(6) & 0.41(1) \\ \hline
	     $J_2=0$ & AFM-QSL  & 1.21(1)   & 1.89(2)  & 1.02(5) & 0.278(5) \\
	   	$J_2=0$ &QSL-VBS &1.69(2)   & 1.40(2)  & 1.02(5) & 0.38(1) \\ \hline
      $J_2=0.2$ & AFM-QSL  & 1.18(1)   & 1.95(1)  & 1.01(4) & 0.132(6)   \\
	   	$J_2=0.2$ &QSL-VBS  &1.63(3)   & 1.45(3)  & 1.01(4) &0.24(1)  \\ \hline
        $J_2=0.4$ & AFM-QSL  & 1.31(1)   & 1.88(1)  & 1.04(3) &0.015(5)  \\
	   	$J_2=0.4$ &QSL-VBS  &1.63(1)   & 1.51(2)  & 1.04(3) & 0.09(1) \\ \hline\hline
	   	 $J_3=0.1$ & AFM-QSL  & 1.17(2)   & 1.93(1)  & 1.00(7) & 0.261(5) \\
 		$J_3=0.1$ &QSL-VBS  &1.60(1)   & 1.54(1)  & 1.00(7) & 0.38(1) \\  \hline
 		$J_3=0$ & AFM-QSL  & 1.38(3)   & 1.72(4)  & 0.99(6) & 0.45(1) \\
 		$J_3=0$ &QSL-VBS  &1.96(4)   & 1.26(3)  & 0.99(6) & 0.56(1) \\  
 		\hline\hline
	\end{tabular*}
\label{tab:criticalexponents}	
\end{table}

\begin{figure}[htbp]
 \centering
 \includegraphics[width=3.4in]{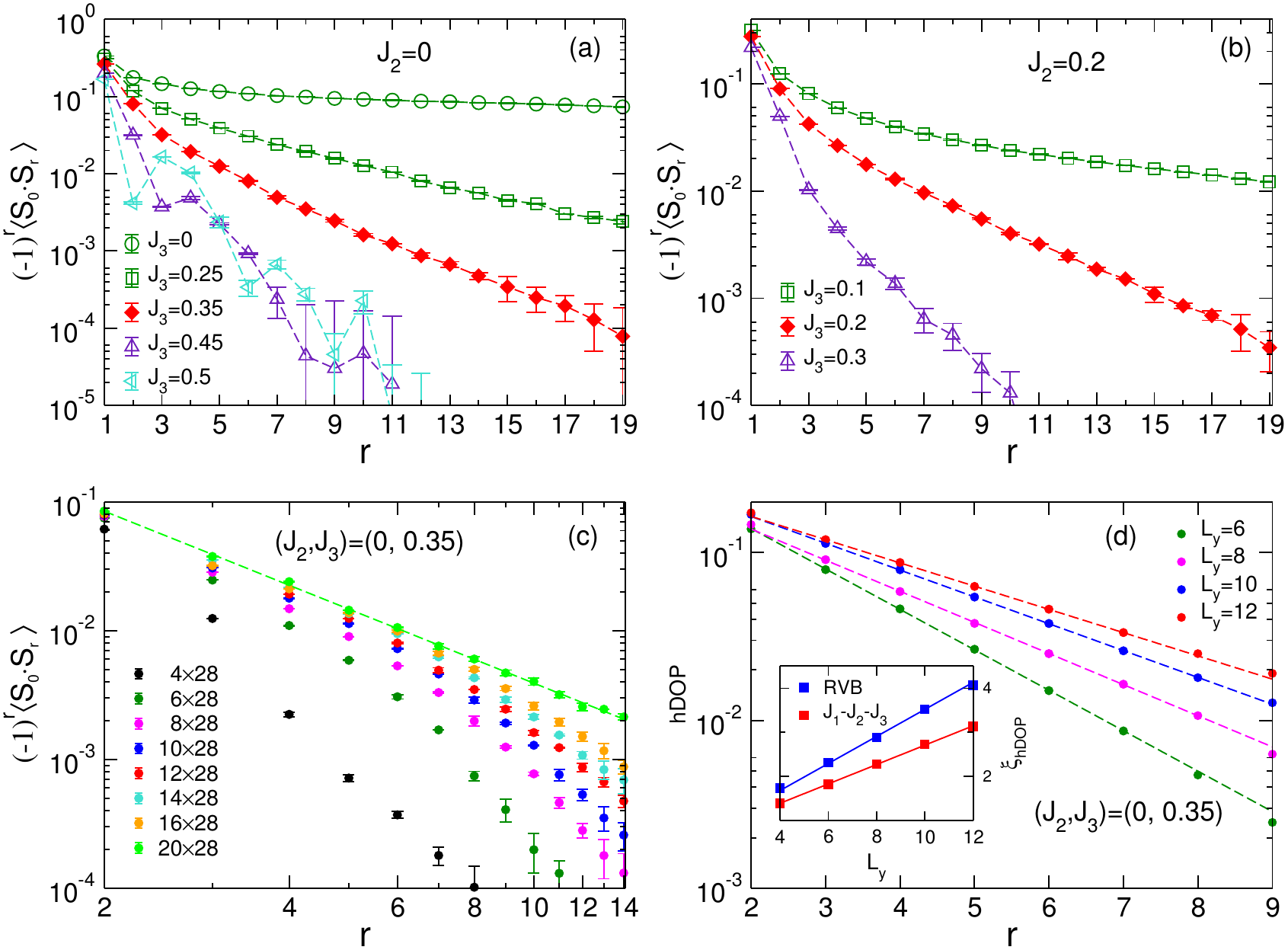}
 \caption{Decay behaviour of correlations. With a fixed $J_2=0$ (a) and a fixed $J_2=0.2$ (b), spin correlation functions of $J_1$-$J_2$-$J_3$ model on a $12\times 28$ strip along the central row at different $J_3$. (c) Fxing $L_x=28$, the variation of spin correlations on strip $L_y\times L_x$ with $L_y$ increasing  from 4 to 20 at a typical point $(J_2,J_3)=(0, 0.35)$ in the QSL phase. Green dashed line denotes the power law fit $y=cr^{-\alpha_s}$ for $L_y=20$. (d) At $(J_2,J_3)=(0, 0.35)$, the log-linear plot of hDOP obtained on different $L_y\times 28$ with respect to the distance from the left edge along the central line. Corresponding decay lengths for different $L_y$ are extracted from exponential fits (shown as straight dashed lines). The inset shows a clear linear growth of the hDOP decay length for the gapless RVB spin liquid state (blue) and the QSL state at $(J_2,J_3)=(0, 0.35)$ (red)  with  $L_y$ increasing with a slope 0.301(4) and 0.219(2), respectively. Error bars denote one standard deviation of the sampled mean values.}
 \label{fig:longstrip}
 \end{figure}

\bigskip
\noindent\textbf{Correlation functions in the QSL phase}\\
To understand the physical nature of the QSL phase, we measure spin-spin and dimer-dimer correlation functions  along the central row on a $12\times 28$ strip where both $x-$ and $y-$ directions are open. Specifically, we first look at spin correlations at different $J_3$ with the fixed $J_2=0$, as shown in Fig.~\ref{fig:longstrip}(a). In the AFM phase, spin correlations at $J_3=0$ and 0.25 decay very slow and tend to saturate at long distance. In the VBS phase, spin correlations at $J_3=0.45$ and 0.5 have a clear exponential decay behaviour, though some oscillations appear due to the mixture of short-range spiral orders, which will be discussed elsewhere. Compared with these two cases, the spin correlations at $J_3=0.35$ which is in the QSL phase, exhibit a long tail indicating a likely power-law decay behaviour. Similarly, for the given $J_2=0.2$ which has a QSL phase in the region $0.13\lesssim J_3\lesssim 0.24$, spin correlations at $J_3=0.1$, $0.2$ and $0.3$, show three different kinds of decay behaviour, corresponding to the AFM, QSL and VBS phases, shown in Fig.~\ref{fig:longstrip}(b). Focusing at the two typical points in the QSL phase, $(J_2, J_3)=(0, 0.35)$ and $(0.2, 0.2)$, we make detailed comparisons with the results from density matrix renormalization group (DMRG) method based on $12\times 28$ strip. The PEPS energy, spin and dimer correlation functions all agree excellently with those of converged DMRG results, see Appendix~\ref{app:dmrg}.

To provide more evidence to show the decay behaviour of correlations in the QSL phase, taking $(J_2, J_3)=(0, 0.35)$ as an example, we consider how they change on different system sizes. In Fig.~\ref{fig:longstrip}(c), we present the spin correlations on different $L_y$ from $4$ to $20$. By fixing $L_x=28$, we expect their behaviour would approach to the real 2D one when increasing $L_y$. Increasing $L_y$ from 10 to 20, we can see the long-distance correlations increase significantly, tending to a power-law decay behaviour, and the power exponent is  $\alpha_s \simeq 1.91(2)$ from  $L_y=20$ correlations.

Then we detect the dimer behaviour by using the characteristic decay length of the local horizontal dimer order parameter (hDOP) for a given $L_y$. The hDOP is defined as the difference $\Delta B$ between nearest strong and weak horizontal bond energies
\begin{equation}
\Delta B(r)=|B^{x}_r-B^{x}_{r+e_x}|.
 \end{equation}
The hDOP decays exponentially from the left system boundary and the corresponding decay length $\xi_{\rm hDOP}$ can be extracted, shown in Fig.~\ref{fig:longstrip}(d) at $(J_2, J_3)=(0, 0.35)$. We find  the decay length $\xi_{\rm hDOP}$ grows linearly with increasing system size $L_y$, consistent with  a power law decay behaviour of the dimer-dimer correlation functions in the QSL phase. Actually, the same behaviour of  $\xi_{\rm hDOP}$ has already been observed in a short-range RVB state, whose dimer correlations decay in a  power law, as well as in the gapless QSL phase in the $J_1-J_2$ model ($J_3=0$)~\cite{liuQSL}. Note here $L_x=28$ is long enough to extract  correct $\xi_{\rm hDOP}$ for $L_y$ ranging from $4$ to $12$, while for larger $L_y$ a relatively large $L_x$  is necessary  to  minimize finite-size effects on $\xi_{\rm hDOP}$, which has also been observed in the calculations of RVB state~\cite{liuQSL}. In summary, these results suggest the discovered QSL is gapless with power-law decay behaviours of both spin and dimer correlations.

\bigskip
\noindent \textbf{Discussion}\\
In summary, by applying the state-of-the-art tensor network method, we study the phase diagram of spin-$1/2$ $J_1$-$J_2$-$J_3$ square-lattice AFM model. For negative and large $J_2$, we find strong evidence for a direct continuous AFM-VBS Landau-forbidden transition line. Along this critical line, exponents are close to those obtained in the $J-Q$ model~\cite{JQ2007,JQ2010} or in classical cubic-lattice dimer models~\cite{charrier2010,sreejith2019}, suggesting the same universality class described by DQCP. 
In particular, spin and dimer correlations decay with similar exponents, indicating an emergent SO(5) symmetry that rotates the AFM and the VBS order parameters into each other~\cite{loopmodel2,sreejith2019}.
Whether this symmetry is exact or approximate needs further inquiries. Surprisingly, we also found that the AFM-VBS transition line ends at -- what could be -- a tricritical point, from which a gapless QSL arises and forms an extended critical phase, separating the AFM and VBS phases. Remarkably, both AFM-QSL and QSL-VBS phase transitions have the same correlation length exponents $\nu\approx 1.0$, indicating new types of universality classes. 

We stress that the gapless QSL found here is very different from the gapless U(1) deconfined phase obtained by the compact quantum electrodynamics with fermionic matter on square lattices, including correlation behaviours and critical exponents~\cite{u1squarelattice2019,u1squarelattice2020}. A recent SU(2) gauge theory~\cite{shackleton2021} based on a fermionic parton construction~\cite{wen2002}, proposed a gapless $\mathbb{Z}_2$ spin liquid as a candidate for such an intermediate phase. However, variational Monte Carlo(VMC) simulations of the corresponding Gutwiller-projected ansatz~\cite{ferrari2020} found constant correlation-function exponents $z+\eta_s\sim z+\eta_d\sim 2$, in contrast to our findings in Fig.~\ref{fig:QSLexponent} in Appendix \ref{app:extract} showing smaller, varying exponents. Moreover, the SU(2) gauge theory further predicts a weak breaking of SO(5) symmetry for the AFM-QSL phase transition, which is very different from our results in Fig.~\ref{fig:J1J2J3phaseDiagram} where the line with $z+\eta_s=z+\eta_d\sim 1.55$ (consistent with the potential SO(5) symmetry) is rather far away from the AFM-QSL phase boundary. In addition, we also note that a tricritical point does not come out naturally from such a gauge theory which has to resort to a first-order transition to connect the AFM-VBS critical line to the QSL phase, while we have detected no sign of first order behavior in our simulations. 

Usually both QSL and DQCP are associated with deconfined gauge fluctuations and fractionalized spinon excitations. The unified phase diagram of QSL and DQCP revealed in the $J_1-J_2-J_3$ model suggests  the underlying field theory for QSL  may have close relation to the DQCP theory~\cite{DQCP1,DQCP2}, though it could be different from the SU(2) gauge theory description~\cite{shackleton2021}. Perhaps the possibility of emergent topological theta term near the DQCP is a very promising future direction~\cite{liuQSL}. The intrinsic relation between QSL and DQCP might also be helpful for solving the mystery of DQCP, by attacking from the QSL phase. Our results would intensify the interest of P. W. Anderson's famous proposal that doping a QSL might lead to superconductivity~\cite{Anderson1987}. Particularly, in a conjectured scenario of high-temperature superconductivity~\cite{SO5theory}, the SO(5) symmetry formed by AFM and superconductor order parameters, might be a reflection of the potential SO(5) symmetry in the QSL phase formed by AFM and VBS order parameters upon doping. Experimentally, the large region of QSL can be sought for based on square-lattice material, and quantum simulators are also a promising platform to realize the novel phases and phase transitions discovered here~\cite{lukin2021,browaeys2021,topologicalExp1}.
\bigskip

\noindent\textbf{Methods}\\
Tensor network states provide a powerful and efficient representation to encode low-energy physics based on their local entanglement structure, whose representation accuracy is systematically controlled by the bond dimension $D$ of the tensors~\cite{white1992,verstraete2008}. As a numerical approach, tensor network states are very suitable to simulate frustrated magnets, where Quantum Monte Carlo methods fail.  The tensor network state methods we used include finite size projected entangled pair state (PEPS), infinite PEPS (iPEPS) and the density matrix renormalization group (DMRG) methods. The DMRG is a well-established method to simulate 1D and quasi-1D systems, and here SU(2) spin rotation symmetries are incorporated to improve the accuracy~\cite{su2}. 

For the finite PEPS algorithm, we use open boundary conditions and each tensor is independent. The finite PEPS ansatz allows us to simulate uniform and non-uniform phases with incommensurate short-range or long-range spiral orders. In our calculations, the finite PEPS works in the scheme of variational Monte Carlo sampling, and the summation of physical freedoms is replaced by the Monte Carlo sampling~\cite{sandvik2007,schuch2008,liu2017,obcPEPS}. The physical quantities are evaluated by importance sampling according to the weights of  given spin configurations, which can be effectively obtained by contracting  single-layer tensor networks. When optimizing the PEPS, we first perform the simple update imaginary time evolution method for initializations~\cite{jiang2008}, and then use stochastic gradient method for accurate optimization to obtain the ground states~\cite{sandvik2007,liu2017,obcPEPS}. With the  obtained ground states, physical quantities including energies, correlations functions, and order parameters are computed via Monte Carlo sampling. More details can be found in Ref.~\cite{obcPEPS}. Without otherwise specified, we use $D=8$ for finite PEPS calculations, which is good enough to obtain convergent results, see Appendix~\ref{app:convergence}. When computing spin and dimer correlations on the central line of the lattice, we use about $6 000 000$ Monte Carlo sweeps, which can have a standard deviation of the mean about $1\times 10^{-4}$ or smaller for each value at different distances. When computing order parameters which contain all kinds of correlations, usually, we use $100 000$ Monte Carlo sweeps which produce one standard deviation of the mean about $2\times10^{-4}$ on a $20\times20$ square lattice, and it takes about 3 days using 500 Intel(R) Xeon(R) E5-2690 v3 CPU cores for such a calculation after the optimization process. This work spent about 10 million CPU hours. 

The iPEPS method is widely used to directly simulate infinite two-dimensional systems, which have translation invariance. The iPEPS used here has only a single unique tensor, which can describe antiferromagnetic phase and uniform paramagnetic phase. The largest bond dimension we used is $D=8$, and the thermodynamic properties can be evaluated with appropriate extrapolations with respect to the bond dimension or the corresponding correlation length.
\\

\noindent \textbf{Data availability.} The data that support the results of this study are available from the corresponding authors upon reasonable request.

\bigskip

\bigskip
\noindent \textbf{Acknowledgments}\\

We thank Anders Sandvik for useful comments. This work was supported by the NSFC/RGC Joint Research Scheme No. N-CUHK427/18 from the Hong Kong’s Research Grants Council and No. 11861161001 from the National Natural Science Foundation of China, the ANR/RGC Joint Research Scheme No. A-CUHK402/18 from the Hong Kong’s Research Grants Council and the TNSTRONG ANR-16-CE30-0025, TNTOP ANR-18-CE30-0026-01 grants awarded from the French Research Council. S.S.G. was supported by National Natural Science Foundation of China grants 11874078, 11834014, and the Fundamental Research Funds for the Central Universities. W.Q.C. was supported by the Science, Technology and Innovation Commission of Shenzhen Municipality (No. ZDSYS20190902092905285).  This work was also granted access to the HPC resources of CALMIP supercomputing center under the allocation 2017-P1231 and center for computational science and engineering at southern university of science and technology.

\bigskip
\noindent \textbf{Author contributions}\\
W.Y.L., W.Q.C. and Z.C.G. conceived the project. W.Y.L. developed the finite PEPS code and carried out the simulations. J.H. developed the iPEPS code. J.H. and D.P. carried out iPEPS simulations and S.S.G. carried out DMRG simulations. W.Y.L., D.P. and Z.C.G. wrote the manuscript with input from other authors. All the authors participated in the discussion.

\bigskip
\noindent \textbf{Competing interests}\\
The authors declare no competing interests.

\clearpage

\appendix

\section{gapless QSL region }

\label{app:gapless}

\begin{figure}[htbp]
 \centering
 \includegraphics[width=3.4in]{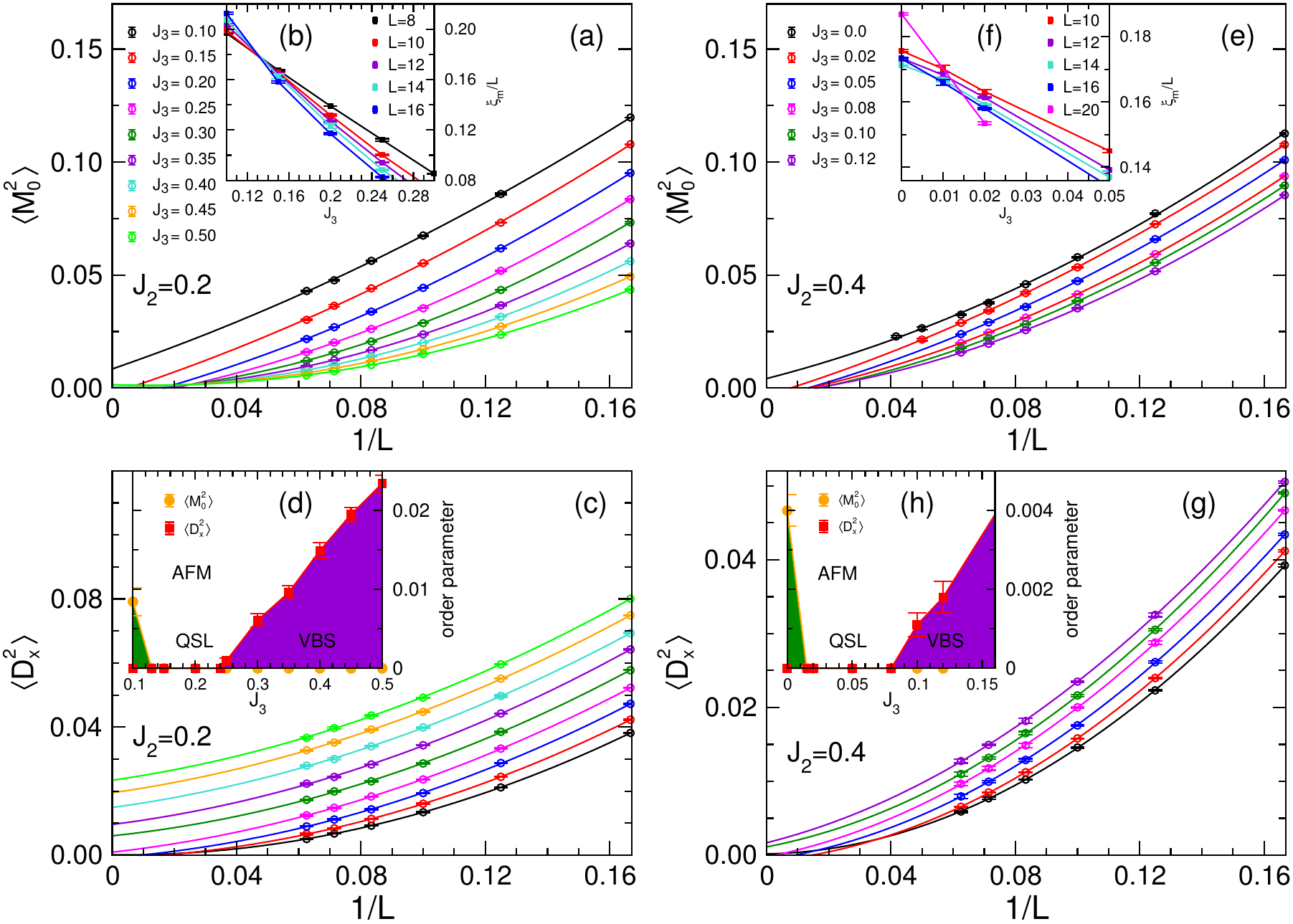}
 \caption{Order parameters and phase diagrams for $J_2=0.2$ shown as (a)-(d), and $J_2=0.4$ shown as (e)-(h). (a) and (c) denote AFM and VBS order parameters at $J_2=0.2$, respectively, and they share the same legend symbols. (b) denotes the dimensionless quantity $\xi_m/L$ at $J_2=0.2$ for different system sizes, and (d) is the phase diagram at $J_2=0.2$ including AFM, (gapless) QSL and VBS phases. (e)-(h) are corresponding ones at $J_2=0.4$. All extrapolations are quadratic fits.}
 \label{fig:OrderParameterJ2_02_J2_04}
 \end{figure}

We consider the $J_1$-$J_2$-$J_3$ model with a fixed $J_2>0$. Here we take $J_2=0.2$ and $J_2=0.4$ as examples. We first focus on the $J_2=0.2$ case. AFM and VBS order parameters, as well as correlation length $\xi_m$ of spin structure factor are shown in Fig.~\ref{fig:OrderParameterJ2_02_J2_04}(a)-(d). From Fig.~\ref{fig:OrderParameterJ2_02_J2_04}(a), one can find the AFM order vanishes between $J_3=0.1$ and $J_3=0.15$, and actually the behaviour of  $\xi_m/L$ shows that the AFM-QSL transition point is estimated at $J_{3}\simeq 0.13$. In the meanwhile, from finite size scaling anlysis, the VBS order starts to appear at $J_3\simeq 0.24$, as presented in Fig.~\ref{fig:OrderParameterJ2_02_J2_04}(c). That means, given $J_2=0.2$,  in the region $0.13\lesssim J_3 \lesssim 0.24$, it is a QSL phase.  Similar analysis is applied to $J_2=0.4$ and a QSL phase  in the region $0.015\lesssim J_3 \lesssim 0.09$ is also discovered, as shown in Fig.~\ref{fig:OrderParameterJ2_02_J2_04}(e)-(h). Note the dimer structure factor on open boundary systems is not well defined~\cite{zhao2020,liuQSL}, so in this context we can not use the dimer correlation lengths based on dimer structure factors to locate the onset of VBS orders. 

\begin{figure}[htbp]
 \centering
 \includegraphics[width=3.2in]{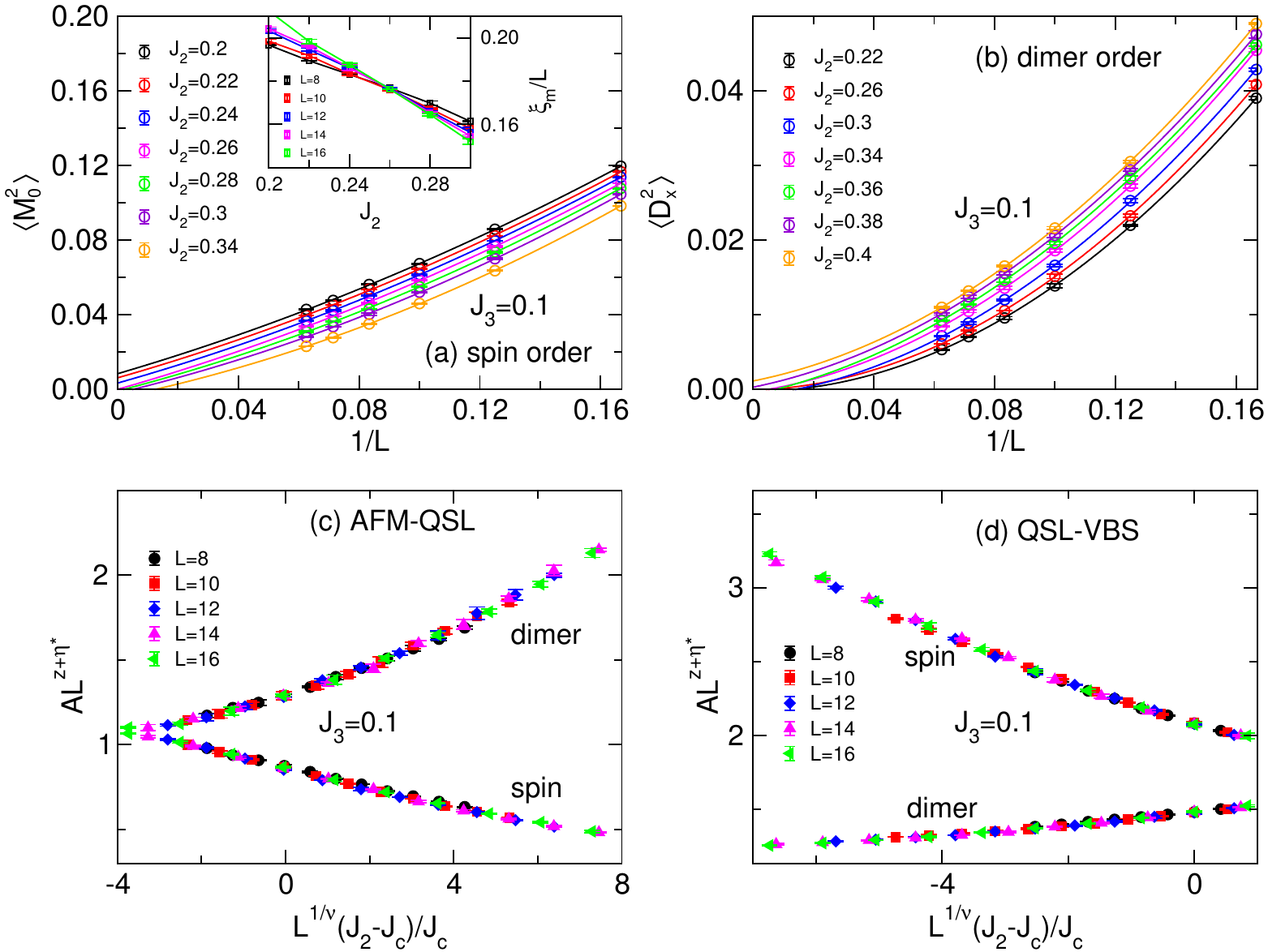}
 \caption{Given $J_3=0.1$, order parameters at different $J_2$. Finite size scaling  of spin (a) and dimer (b) order parameters through second order extrapolations. Inset of (a) shows the crossing of the dimensionless quantity $\xi_m/L$.   Data collpase of spin and dimer orders at AFM-QSL at transition point $J_{c1}=0.261$ (c), and QSL-VBS transition point $J_{c2}=0.38$ (d), with $\nu=1.00$. }
\label{fig:OrderParameterJ3_01}
 \end{figure}

Similarly, we can also explore the phase diagram by sweeping $J_2$ with a fixed $J_3$. We use $J_3=0.1$ as an example. Shown in Fig.~\ref{fig:OrderParameterJ3_01}, we compute the spin and dimer order parameters at different $J_2$ ranging from $J_2=0.2$ to $0.4$. In Fig.~\ref{fig:OrderParameterJ3_01}(a) and (b),  one can clearly see the AFM order vanishes at $J_2\simeq0.26$, and the VBS order begins to appear for $J_2>0.38$. That means for the region $0.26\lesssim J_2 \lesssim 0.38$ it is a QSL. Here we note that $(J_2,J_3)=(0.38, 0.1)$ is located at the QSL-VBS phase boundary, compatible with previous calculations along the vertical line $J_2=0.4$ where $(J_2,J_3)=(0.4, 0.1)$ is in the VBS phase. 

\begin{figure}[htbp]
 \centering
 \includegraphics[width=3.2in]{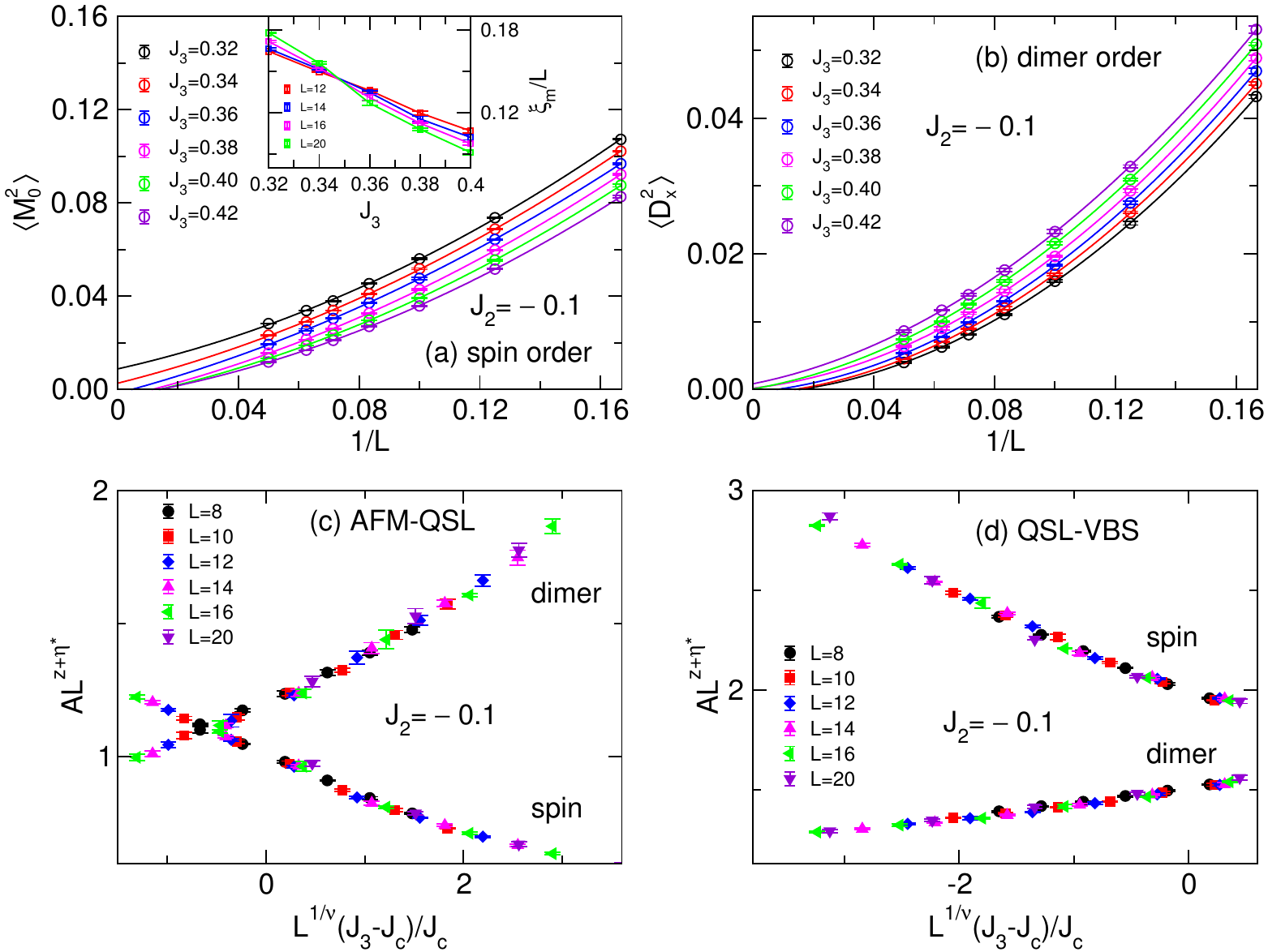}
 \caption{Given $J_2=-0.1$, order parameters at different $J_3$. Finite size scaling  of spin (a) and dimer (b) order parameters through second order extrapolations. Inset of (a) shows the crossing of the dimensionless quantity $\xi_m/L$.   Data collpase of spin and dimer orders at AFM-QSL at transition point $J_{c1}=0.351$ (c), and QSL-VBS transition point $J_{c2}=0.41$ (d), with $\nu=1.03$. }
\label{fig:OrderParameterJ3_m01}
 \end{figure}

We also consider ferromagnetic $J_2$ couplings. Using $J_2=-0.1$, a QSL in the region $0.35 \lesssim J_3/J_1\lesssim 0.41$ is suggested, sandwiched by the AFM phase and VBS phase, shown in Fig.~\ref{fig:OrderParameterJ3_m01}. With a stronger $J_2=-0.2$, the QSL phase further shrinks to a very narrow region $0.41 \lesssim J_3/J_1\lesssim 0.43$. Further enhancing $J_2=-0.25$ ($J_2=-0.3$), a continuous AFM-VBS transition is suggested at  $J_3\simeq0.45$ ($J_3\simeq0.49$) with the QSL phase disappearing. 
Actually, for the three cases with fixed  $J_2=-0.2$, $-0.25$ and $-0.3$ and for regions of $J_3$ close to the tricritical point, we compute the VBS order parameters with two definitions $\langle D \rangle^2=\langle D_x \rangle^2+\langle D_y\rangle^2$ and $\langle D_x\rangle^2$, and carefully check the finite-size scaling versus $1/L$ using different fitting functions and different sizes. The onset of VBS order is estimated at $J_3=0.43(1)$, $J_3=0.45(1)$ and $J_3=0.49(1)$, respectively. 

\begin{figure}[htbp]
 \centering
 \includegraphics[width=3.2in]{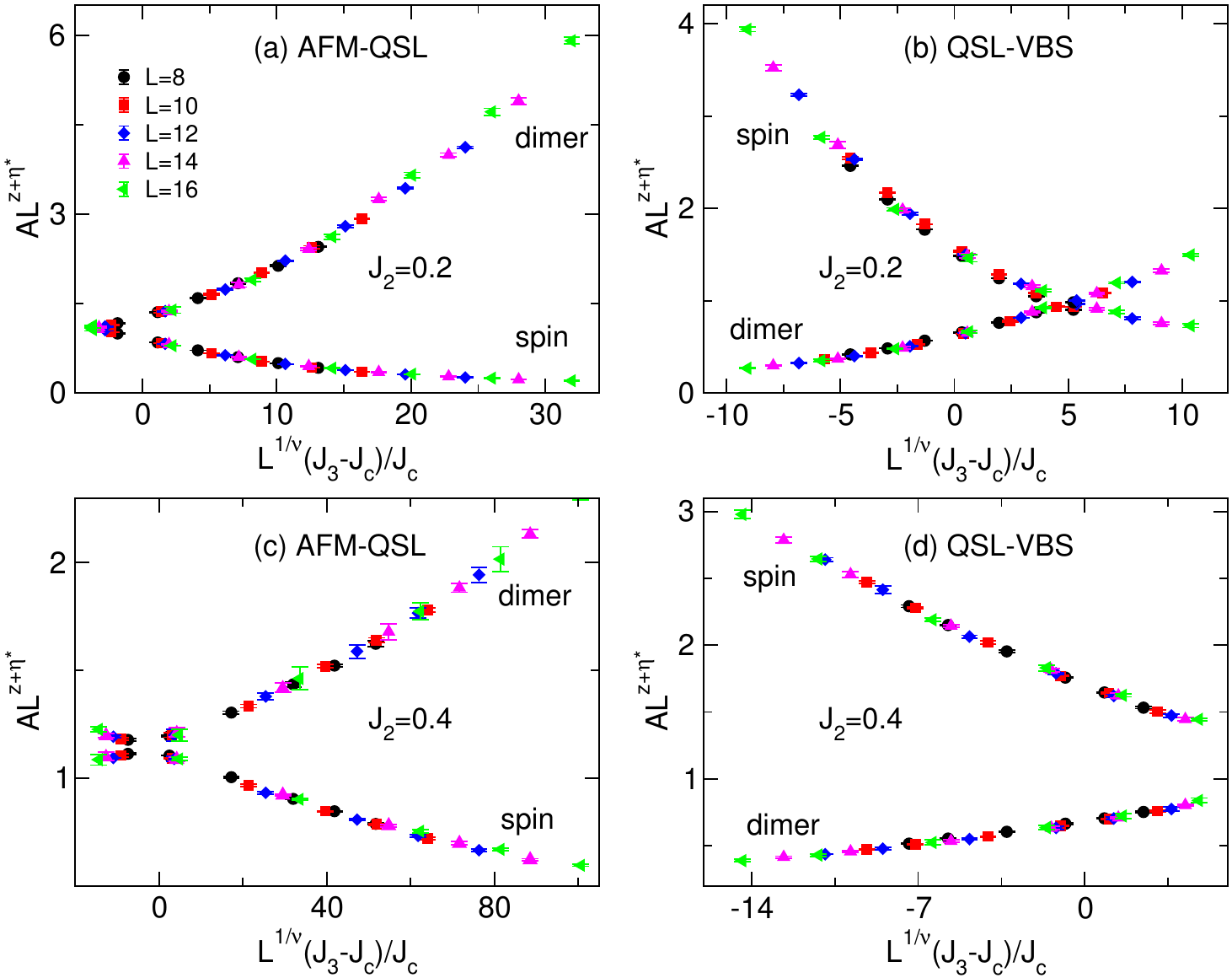}
 \caption{Data collpase of spin and dimer orders for AFM-QSL and QSL-VBS phase transition points at fixed $J_2=0.2$ (a-b) and $J_2=0.4$ (c-d). The scaling at AFM-QSL transition point $J_{c1}=0.132$ (a), and QSL-VBS transition point $J_{c2}=0.24$ (b) along $J_2=0.2$ with $\nu=1.01$. The scaling at AFM-QSL transition point $J'_{c1}=0.015$ (c), and QSL-VBS transition point $J'_{c2}=0.09$ (d) along $J_2=0.4$ with $\nu=1.04$. }
\label{fig:AFM_VBSdataCollapseJ2ne0}
 \end{figure}
   
\begin{figure}[htbp]
 \centering
 \includegraphics[width=3.2in]{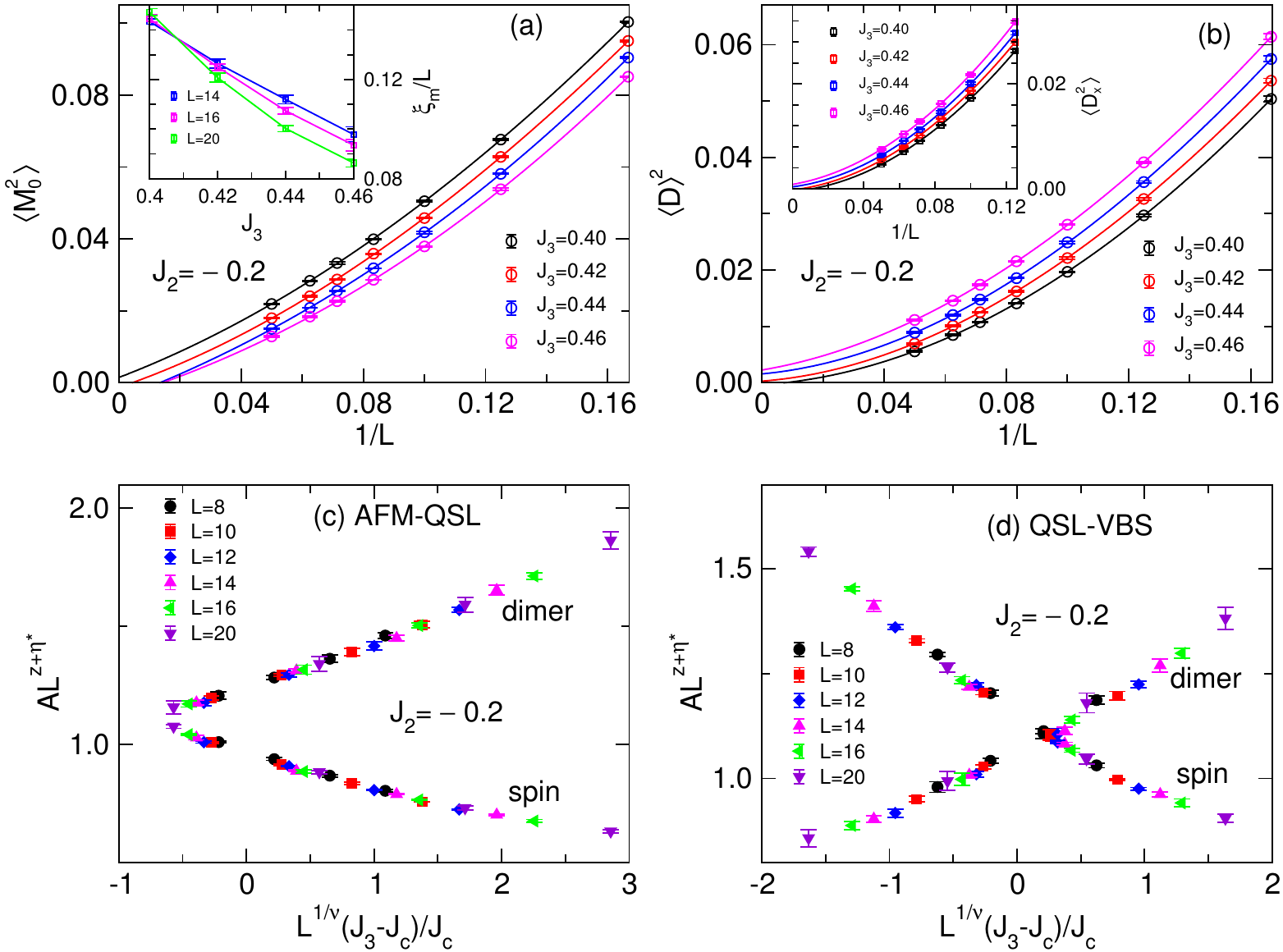}
 \caption{Scaling of quantities at  given $J_2=-0.2$.  (a) and (b) present AFM  and boundary-induced dimer order parameters $\langle D\rangle^2=\langle D_x\rangle^2+\langle D_y\rangle^2$ with cubic fits for $L=6-20$. Inset of (a) shows the crossing of the dimensionless quantity $\xi_m/L$, and the inset of (b) shows the dimer order parameter defined based on bond-bond correlations $\langle D_x^2\rangle$ with quadratic fits for $L=8-20$. (c) and (d) show the data collapse at the AFM-QSL and QSL-VBS transition points $J_{c1}=0.41$ and $J_{c2}=0.43$, respectively. At AFM-QSL transitions $J_{c1}=0.41$, the quantities can be collapsed using $z+\eta_{s1}^*=1.31(4)$, $z+\eta_{d1}^*=1.84(4)$ and $\nu=0.95(8)$.  At QSL-VBS transitions $J_{c2}=0.43$, the quantities can be collpased using $z+\eta_{s2}^*=1.42(3)$, $z+\eta_{d2}^*=1.74(5)$ and $\nu=0.95(8)$.}
\label{fig:OrderParameterJ3_m02}
 \end{figure}

\begin{figure}[htbp]
 \centering
 \includegraphics[width=3.2in]{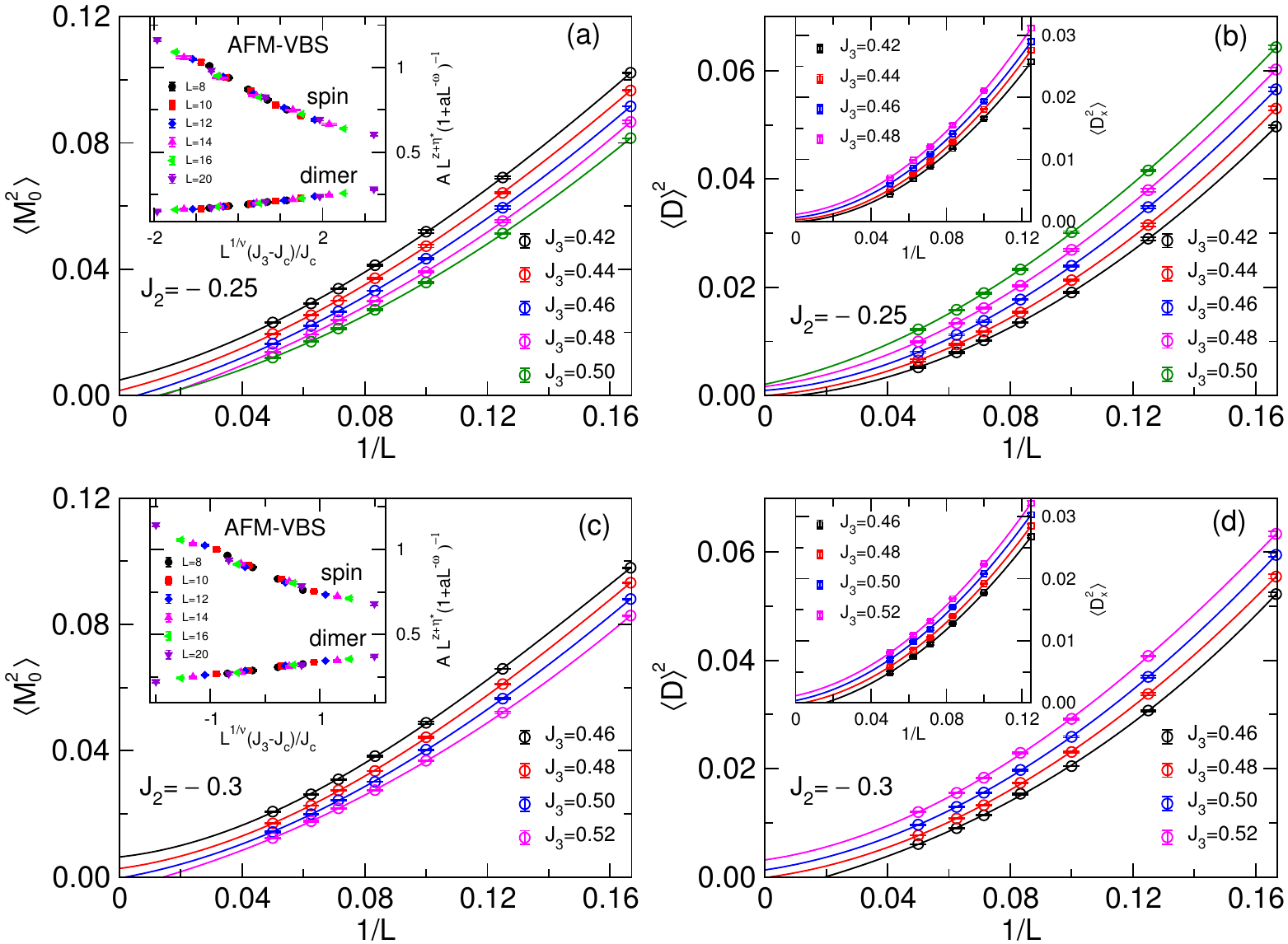}
 \caption{Scaling of quantities at given $J_2=-0.25$ (a-b) and $J_2=-0.3$ (c-d).  (a) and (b) present AFM and dimer order parameters  $\langle D\rangle^2$ at $J_2=-0.25$ with cubic fits for $L=6-20$.  The inset of (b) shows $\langle D_x^2\rangle$ with quadratic fits for $L=8-20$. The inset of (a)  shows the data collapse at the AFM-VBS transition point using $J_{c}=0.45$, $\nu=0.89$, $z+\eta_s^*=1.31$ with subleading factors $a=2.2$ and $\omega=1.9$, and $z+\eta_d^*=1.34$ with subleading factors $a=9.5$ and $\omega=0.9$.  (c) and (d) present AFM and dimer order parameters  $\langle D\rangle^2$ at $J_2=-0.3$ with cubic fits for $L=6-20$.  The inset of (b) shows $\langle D_x^2\rangle$ with quadratic fits for $L=8-20$. The inset of (a)  shows the data collapse at the AFM-VBS transition point using $J_{c}=0.49$, $\nu=0.86$, $z+\eta_s^*=1.35$ with subleading factors $a=3$ and $\omega=1.5$, and $z+\eta_d^*=1.32$ with subleading factors $a=12$ and $\omega=1.5$. }
\label{fig:OrderParameterJ3_m025_m03}
 \end{figure}

  \begin{figure*}[htbp]
 \centering
 \includegraphics[width=6.4in]{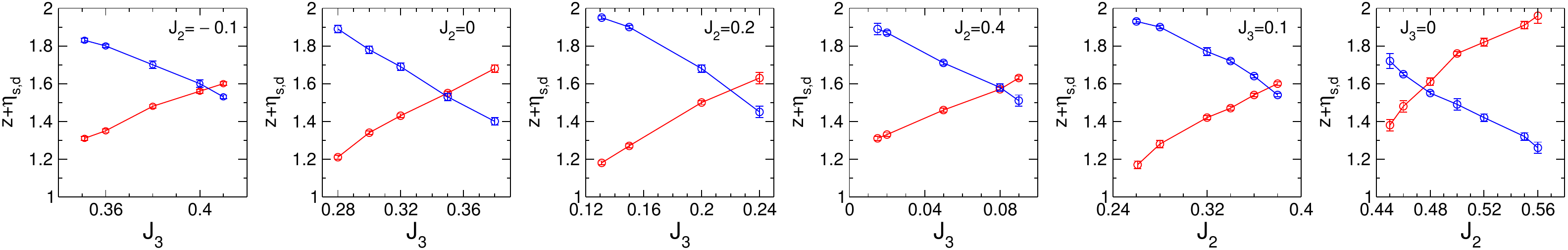}
 \caption{The exponent $z+\eta_{s,d}$ for spin (red) and dimer (blue) at different couplings at fixed $J_2$ or $J_3$, in the QSL phase, computed from order parameter $A$ according to the scaling $A\propto{L^{-(z+\eta)}}$.}
 \label{fig:QSLexponent}
 \end{figure*}
     
In the gapless QSL region, we can extract the spin and dimer decay powers $z+\eta_{s,d}$ according to a scaling $A\propto L^{-(z+\eta)}$. As seen in Fig.~\ref{fig:QSLexponent}, for a fixed $J_2$ or $J_3$, the extracted spin (dimer) power decreases (increases) with increasing $J_3$ or $J_2$. Such a character is not supportable for a gapless $\mathbb{Z}_2$ QSL, which has a constant decay power in the whole QSL region, according to the variational Monte Carlo(VMC) simulations of the corresponding Gutwiller-projected ansatz~\cite{ferrari2020}. An interesting feature for the cases in Fig.~\ref{fig:QSLexponent} is that $z+\eta_{s}$ and  $z+\eta_{d}$ always have a crossing at the value around 1.55. A reasonable speculation is that  an SO(5) symmetry emerges at these crossing points. Thus we guess there exists a line for different $(J_2,J_3)$ in the whole QSL region, on which it has SO(5) symmetry. 
Note in the calculations at $J_2=-0.2$, where the QSL region is very narrow $0.41\lesssim J_3 \lesssim 0.43$, very close to the tricritical point, the cross of spin and dimer decay power in the QSL phase does not occur (see the values in the caption of Fig.~\ref{fig:OrderParameterJ3_m02}), which might be caused by finite-size effects.

 \section{Extracting critical exponents}
 \label{app:extract}
 
 Accurately determining critical exponents in a numerical way is very challenging for unconventional 2D phase transitions, which often can only be realized in unbiased simulations like Quantum Monte Carlo computations~\cite{JQ2008,loopmodel2,JQ2016,JQ2020} and the accuracy depends on system sizes and sampling errors. In our tensor network results, the precision of physical quantities may also have some influences, but we still try to evaluate the reasonability of critical exponents, especially focusing on the correlation length exponent $\nu_1$ at the AFM-QSL and $\nu_2$ at the QSL-VBS transitions which we claim to be the same. The physical quantities from different sizes and different couplings are collectively fitted for collapse, according to the formula~\cite{JQ2007,FSS1}:
 \begin{equation}
A(J_3, L)=L^{\kappa}(1+aL^{-\omega} )F[L^{1/\nu}(J_3-J_c)/J_c],
\end{equation}
 where $A= \xi_m$, $\ave{M^2_0}$, or  $\ave{D^2_x}$, and $\kappa=1$ for $\xi_m$, $-(z+\eta_s^*)$ for $\ave{M^2_0}$, and  $-(z+\eta_d^*)$ for $\ave{D^2_x}$. Factors $a$ and $\omega$ are tuning parameters of the subleading term.  $F[]$ is a polynomial function, and here a second order expansion is used, considering our largest system size is  $24\times 24$ (a third order fitting actually is also tested and has a negligible third-order coefficient). Usually the subleading term is not included for the fitting of AFM-QSL and QSL-VBS transitions.  The transition point $J_c$ can be estimated from the scaling analysis of order parameters or the spin correlation length, mostly used as a fixed value for fitting critical exponents. 
 
We take the $J_1-J_3$ model, i.e. $J_2=0$,  as an example. The AFM-QSL transition occurs at $J_{c1}=J_3\simeq 0.28$. It can also be given from a collective fit of $J_{c1}$ and $\nu_1$, which gives the critical point $J_{c1}=0.278(5)$ and the correlation length exponent $1.01(9)$. Using a fixed $J_{c1}=0.278$, we can  evaluate  the correlation length exponent $\nu_1$ independently from AFM and VBS order parameters, respectively, by a collective fit of $\nu$ and $z+\eta^*$. Fitting AFM order gives  $\nu_{1,s}=1.10(3)$ and $z+\eta_{s1}^*=1.17(2)$, and fitting VBS order gives $\nu_{1,d}=0.98(6)$ and $z+\eta_{d1}^*=1.87(5)$. The three fits produce consistent $\nu_1$. 
At the QSL-VBS transition, the critical point is located at $J_{c2}=0.38(1)$ according to the finite-size scaling of VBS order parameters. In this case, we can not use a similar correlation length defined based on dimer structure factor to locate the  transition point, since the dimer structure factor on open boundary systems is not well defined~\cite{zhao2020,liuQSL}.  However, we can still check $\nu_2$ from the fitting of AFM and VBS order parameters by using a fixed $J_{c2}=0.38$.  The AFM order fit gives $\nu_{2,s}=1.05(4)$ and $z+\eta_{s2}^*=1.67(3)$, and VBS order fit gives $\nu_{2,d}=0.96(5)$ and $z+\eta_{d2}^*=1.44(2)$. The two fits also give consistent $\nu_2$ (assuming other $J_{c2}$ like $J_{c2}=0.37$ for fitting  gives almost the same $\nu_2$). Note the obtained $\nu_1$ at AFM-QSL transition point and $\nu_2$ at QSL-VBS transition point  indicate  $\nu_1\approx\nu_2\approx 1.0$. 

Similar analyses are applied to other cases for a fixed $J_2$ or a fixed $J_3$. By scaling AFM and VBS order parameters with fixed critical points, collective fits of two parameters $\nu$ and $z+\eta^*$  can give $\nu_{1,s}$, $\nu_{1,d}$,  $\nu_{2,s}$ and $\nu_{2,d}$, respectively, associated with their corresponding $z+\eta_{s1}^*$, $z+\eta_{d1}^*$, $z+\eta_{s2}^*$ and $z+\eta_{d2}^*$, as listed in Table.~\ref{tab:freenu}. Remarkably, the obtained $\nu$ all agree well, close to $1.0$, indicating the same correlation length exponent $\nu$ at the AFM-QSL and QSL-VBS transitions.  Meanwhile, the fitted spin and dimer critical exponents using different  fixed $J_2$ or $J_3$, are consistent, respectively, at the AFM-QSL and QSL-VBS transition points. Their rough estimated values are $z+\eta_{s1}^*~\sim 1.2$, $z+\eta_{d1}^*~\sim 1.9$, $z+\eta_{s2}^*~\sim 1.6$ and $z+\eta_{d2}^*~\sim 1.5$, as  shown in the last row of Table.~\ref{tab:freenu}. 
In order to clearly show a single correlation length exponent $\nu$ can scale all the physical quantities well for each case, we use the averaged value $\bar{\nu}$ over  $\nu_{1,s}$, $\nu_{1,d}$,  $\nu_{2,s}$ and $\nu_{2,d}$ as a fixed parameter, then to fit $z+\eta^*$. The scaled quantities using $\bar{\nu}$ for data collapse are shown in the figures in Appendix.~\ref{app:gapless}, and  the fitted values $z+\eta^*$ are listed in Table.~\ref{tab:criticalexponents}. They are also listed in Table.~\ref{tab:fixnu} for a convenient comparison with Table.~\ref{tab:freenu}, which shows  small differences in the two tables. This indicates a single $\nu$ close to $1.0$ indeed works well at the AFM-QSL and QSL-VBS transitions. 

 Finally, let us discuss the $\chi^2$ (per degree of freedom) of the fittings, which quantifies the goodness of a fit. Usually $\chi^2 \sim 1.0$ or smaller means a satisfactory fit.  In our optimal fittings, for some cases we indeed have $\chi^2\sim 1.0$ including the fits at the AFM-VBS transition using $J_2=-0.4$. But there are some fits with $\chi^2 \sim 3.0$, which is relatively too large for the number of degrees of freedom of the fit. As we know, $\chi^2$ depends on the sampling errors and the data window to fit~\cite{FSS1}. For our tensor network results, the obtained physical quantities have unavoidable slight deviations from their exact values due to imperfect optimizations, which also leads to extra influences on the values of $\chi^2$.  This is different from unbiased Quantum Monte Carlo simulations where there are no wavefunction optimization problems. However, our fits can still give  reasonable and correct information to understand the unconventional quantum phase transition, evidenced by a series of rather smooth curves formed by the scaled quantities  with  similar critical exponents, at different fixed $J_2$ or $J_3$, as is shown in the previous figures in Appendix.~\ref{app:gapless},  which is just the requirement of the well-behaved universal scaling functions.

 \begin{table}
    \centering
 \caption { Critical exponents obtained by a collective fitting of $\nu$ and $z+\eta^*$ using a fixed critical point. In each fit, $\nu$ and $z+\eta^*$ are free parameters. Fitting AFM order parameters gives $\nu_{1,s}$ and  $z+\eta_{s1}^*$ at the AFM-QSL critical point, and gives $\nu_{2,s}$ and  $z+\eta_{s2}^*$ at the QSL-VBS critical point. Fitting VBS order parameters gives $\nu_{1,d}$ and  $z+\eta_{d2}^*$ at the AFM-QSL critical point, and gives $\nu_{2,s}$ and  $z+\eta_{d2}^*$ at the QSL-VBS critical point. Errors are from fittings. The last column $\bar{\nu}$ is an average over $\nu_{1,s}$, $\nu_{1,d}$,  $\nu_{2,s}$ and $\nu_{2,d}$. The last row represents the averaged values of $z+\eta$ over different $J_2$ or $J_3$.}
  	\begin{tabular*}{\hsize}{@{}@{\extracolsep{\fill}}lccccc@{}}
		\hline\hline
		        &  $\nu_{1,s}$ & $\nu_{1,d}$   & $\nu_{2,s}$& $\nu_{2,d}$& $\bar{\nu}$   \\ \hline
	          $J_2=-0.1$& 0.96(6) & 1.01(7)   &  1.09(5) & 1.04(1)& 1.03(6)  \\ \hline 
	     $J_2=0$& 1.10(3) & 0.98(6)   &  1.05(4) & 0.96(5)& 1.02(5) \\ \hline
	     $J_2=0.2$& 1.04(3) & 0.94(4)   &  1.14(6) & 0.93(3)&1.01(4)   \\ \hline
  	    $J_2=0.4$& 1.09(4) & 0.96(3)   &  1.11(2) & 0.99(4)&1.04(3)  \\ \hline
	   $J_3=0.1$& 1.06(9) & 1.03(4)   &  0.96(6) & 0.95(8)& 1.00(7)  \\ \hline\hline
	        &  $z+\eta_{s1}^*$ & $z+\eta_{d1}^*$   & $z+\eta_{s2}^*$& $z+\eta_{d2}^*$   \\ \hline
	          $J_2=-0.1$& 1.28(2) & 1.82(1)   &  1.61(1) & 1.53(2) \\ \hline 
	     $J_2=0$& 1.17(2) & 1.87(5)   &  1.67(3) & 1.44(2)  \\ \hline
	     $J_2=0.2$& 1.17(2) & 1.99(2)   &  1.65(2) & 1.49(2)   \\ \hline
  	    $J_2=0.4$& 1.28(1) & 1.87(1)   &  1.62(1) & 1.49(1)  \\ \hline
	   $J_3=0.1$& 1.18(2) & 1.93(2)   &  1.62(2) & 1.52(2)  \\  \hline
	   average  & 1.22(2)& 1.90(2)& 1.63(2)& 1.49(2) \\
	   		\hline\hline
	\end{tabular*}
\label{tab:freenu}	
\end{table}

 \begin{table}
    \centering
 \caption { Critical exponents obtained by using a single correlation length exponent $\bar{\nu}$ at the AFM-QSL and QSL-VBS transition points.  In each fit, $z+\eta^*$ is a free parameter, and  $\bar{\nu}$ is fixed. Errors are from fittings. %The last row represents the averaged values of $z+\eta$ for different $J_2$ or $J_3$.
 }
  	\begin{tabular*}{\hsize}{@{}@{\extracolsep{\fill}}lccccc@{}}
		\hline\hline
	        &  $z+\eta_{s1}^*$ & $z+\eta_{d1}^*$   & $z+\eta_{s2}^*$& $z+\eta_{d2}^*$ & $\bar{\nu}$  \\ \hline
	          $J_2=-0.1$& 1.31(2) & 1.83(1)   &  1.60(1) & 1.53(1) & 1.03 \\ \hline 
	     $J_2=0$& 1.21(1) & 1.89(2)   &  1.69(2) & 1.40(2) & 1.02  \\ \hline
	     $J_2=0.2$& 1.18(1) & 1.95(1)   &  1.63(3) & 1.45(3)&1.01  \\ \hline
  	    $J_2=0.4$& 1.31(1) & 1.88(1)   &  1.63(1) & 1.51(2)&1.04  \\ \hline
	   $J_3=0.1$& 1.17(2) & 1.93(1)   &  1.60(1) & 1.54(1) & 1.00  \\  
	   %\hline average  & 1.24(1)& 1.90(1)& 1.63(2)& 1.49(2) \\
	   		\hline\hline
	\end{tabular*}
\label{tab:fixnu}	
\end{table}

\section{Comparison with DMRG}
\label{app:dmrg}
  
\begin{figure}[htbp]
 \centering
 \includegraphics[width=3.4in]{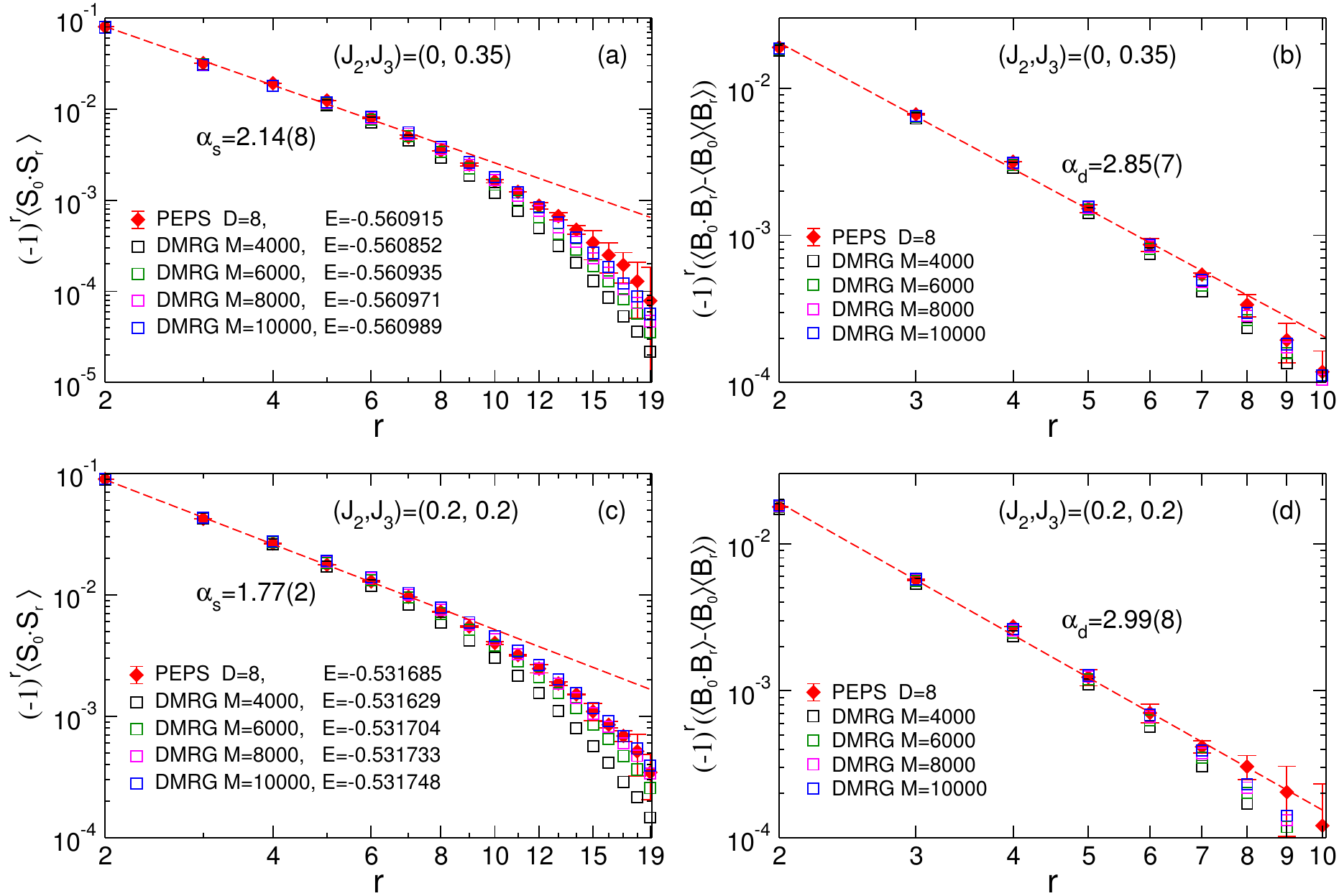}
 \caption{Correlation functions of $J_1$-$J_2$-$J_3$ model on a $12\times 28$ strip along the central line.  (a,b) show spin and dimer correlations at $(J_2, J_3)=(0, 0.35)$, and (c,d) correspond to those at $(J_2, J_3)=(0.2, 0.2)$. Red dashed lines denote  power law fits $y=cr^{-\alpha}$ for the correlation values with $r\le 7$, and corresponding exponents $\alpha_s$ ($\alpha_d$) for spin (dimer) are presented. }
 \label{fig:12x28corrJ1J2J3}
 \end{figure}

  We compare finite PEPS results with those from the density matrix renormaliztion group (DMRG) method with SU(2) spin rotation symmetry. Based on $J_1-J_2$ model, it has been shown that finite PEPS results agree very well with the convergent DMRG results in our previous work~\cite{obcPEPS,liuQSL}. For the $J_1-J_3$ model discussed here, the squared AFM  order  $\langle {\bf M}_0^2\rangle$  at $J_3/J_1=0.3$ calculated by PEPS, are  $0.09865$ on $6\times 6$ and $0.06568$ on $8\times 8$ lattice on open boundary conditions, also in excellent agreement with those calculated by DMRG on the same systems, which are $0.09886$ and $0.06588$, correspondingly. Now we consider larger sizes on $J_1-J_2-J_3$ model. Fig.~\ref{fig:12x28corrJ1J2J3}(a) and (c) depict the spin correlations at the two points $(J_2,J_3)=(0,0.35)$ and (0.2,0.2) in the QSL phase. DMRG results with different numbers of SU(2) kept state are presented up to $M=10000$ (equivalent to about 40000 U(1) states). The corresponding ground state energies for different $M$ are listed with each legend showing that DMRG and PEPS energies are highly consistent. When increasing $M$, the DMRG spin correlations gradually increase until convergence, which is also in excellent agreement with PEPS results. The connected dimer-dimer correlations along $x-$ direction in the QSL phase, defined as 
\begin{equation}
C_d(r)=\langle B^x_0B^x_r \rangle-\langle B^x_0\rangle\langle B^x_r \rangle \, ,
\label{eq:dimer_cor}
\end{equation}
are also computed,  shown in Fig.~\ref{fig:12x28corrJ1J2J3}(d) and (d). We can see the PEPS and DMRG dimer-dimer correlations also agree very well. We remark that such agreements are consistent with our previous results of the pure $J_1$-$J_2$ model, i.e., $J_3=0$. In the pure $J_1$-$J_2$ model,  DMRG needs a  very large bond dimension $M$ to converge spin correlations, at least up to $M=12000$ at $J_2/J_1=0.5$ and more than $M=14000$ (equivalent to about 56000 U(1) states) at $J_2/J_1=0.55$, but a $D=8$ PEPS can produce convergent results quite well compared to the $D=10$ PEPS results~\cite{liuQSL}. However, for the two points  $(J_2, J_3)=(0, 0.35)$ and (0.2, 0.2) we discuss here, DMRG with $M=10000$ already obtains well converged correlations, so it indicates the $D=8$ PEPS also already converges the results.

\section{Convergence of finite PEPS with bond dimension}   
  \label{app:convergence} 
\begin{figure}[htbp]
 \centering
 \includegraphics[width=3.4in]{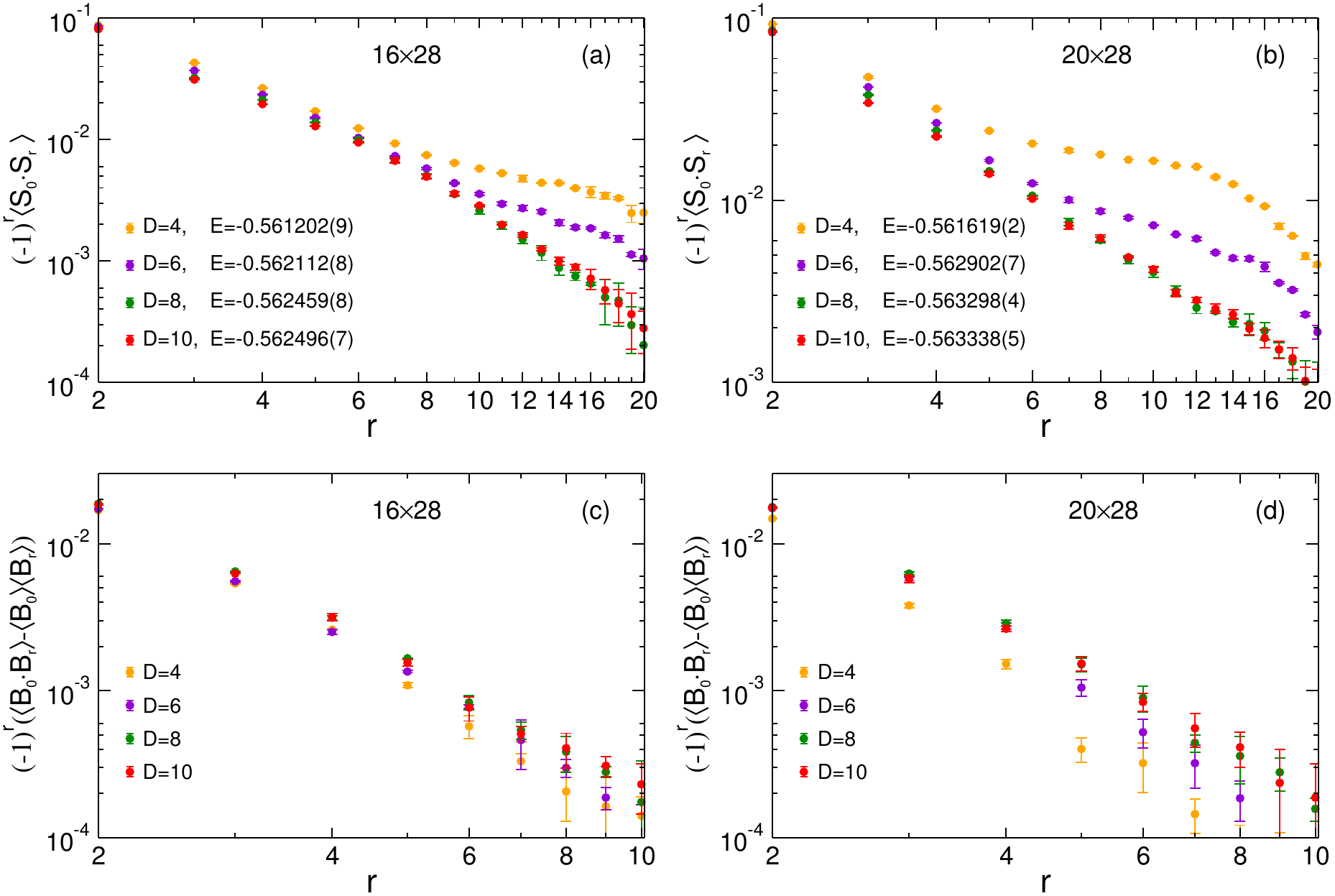}
 \caption{Convergence of finite PEPS calculations with respect to bond dimension $D$ at $(J_2,J_3)=(0,0.35)$. (a) and (b) shows spin correlation functions of $J_1$-$J_2$-$J_3$ model on  $16 \times 28$ and $20\times 28$ strips along the central line. (c) and (d) are corresponding dimer-dimer correlations. Energies using different $D$ are listed in the legend. }
 \label{fig:converge}
 \end{figure}

 To  check the $D-$convergence behaviour, we consider the spin and dimer correlations based on  large sizes including $16\times 28$ and $20\times 28$ sites using $D=4,6,8$ and 10. We choose the point $(J_2, J_3)=(0,0.35)$, which is critical and belongs to the most challenging for accurate simulations. For each case, we use simple update imaginary time evolution method for initialization, and then use stochastic gradient method for further optimization. From Fig.~\ref{fig:converge} we can see after increasing $D=4$ to $D=8$, the energy, spin and dimer correlations, all have significant improvement. Whereas, after increasing $D=8$ to $D=10$, the improvement is very small, indicating $D=8$ already converges the results. Since the size $20\times 28$ is among the largest ones in the finite size simulations and $(J_2, J_3)=(0,0.35)$ is critical, the above results suggest $D=8$ can also converge the presented results for  other finite sizes at different values of $J_2$ and $J_3$.  Actually, in our previous studies, we have demonstrated $D=8$ can well converge the results for Heisenberg model up to $32\times 32$ sites and frustrated $J_1-J_2$ model up to $24\times 24$ sites~\cite{obcPEPS,liuQSL}.  
   
  \section{IPEPS results}   
  \label{app:iPEPS}
  \subsection{Scaling analysis}
   
 For iPEPS, resorting to translation symmetry,  a single tensor with $U(1)$ and $C_{4v}$ symmetries  can be used to  describe the  AFM and QSL phases~\cite{hasik2021}. With different bond dimension $D$ of iPEPS, one can extract the corresponding correlation length, then use the finite-$D$ scaling or the so-called finite correlation length scaling (FCLS) to obtain the extrapolated physical quantities. Such an approach has been demonstrated to work well on Heisenberg model (shortly reviewed here). Next, we consider two typical points $(J_2, J_3)=(0, 0.35)$ and $(0.2, 0.2)$ for comparison. In that case, we show that the simple finite correlation length scaling has to be extended, including simultaneous $1/D$ corrections.

The wavefunction is completely parametrized 
by a single real rank-5 tensor $a$, with physical index $s$ for spin-$1/2$ DoF  
and four auxiliary indices $u,l,d,r$ of bond dimension $D$ corresponding to the four directions \textit{up}, \textit{left}, \textit{right}, and \textit{down} 
of the square lattice. The tensor $a$ is given by a linear combination $a = \sum_i \lambda_i t_i$ of elementary tensors $t$
which are distinct representatives of the fully symmetric ($A_1$) irrep of the $C_{4v}$ point group. The $t$ tensors also obey $U(1)$ charge conservation i.e., certain assignment of charges $\vec{u}$ to physical index $s$ and ${\vec{v}}$
to each of the four auxiliary indices, for some fixed $N$. 
For each considered bond dimension we choose charges $[\vec{u},\vec{v}]=[u_\uparrow, u_\downarrow, v_0, \ldots, v_{D-1}]$, 
listed in Tab.~\ref{tab:table1}, based on the analysis of optimal
states from the unrestricted simulations of the Néel phase of $J_1-J_2$ model~\cite{hasik2021}. The only variational parameters 
of this ansatz are the coefficients $\vec{\lambda}$ associated to the family of elementary tensors $t$.
\begin{table}[b]
\caption{\label{tab:table1}
$U(1)$ charges for the Néel phase taking $N=1$. Charges for $D=8$ are a prediction. The last column 
shows the number of elementary tensors $t_i$. }
\begin{ruledtabular}
\begin{tabular}{lll}
$D$ & $[ u_\uparrow, u_\downarrow, v_0,v_1,\cdots, v_{D-1}] $  & number of tensors\\
\colrule
2 & $[1,-1,0,2]$ & 2 \\
3 & $[1,-1,0,2,0]$& 12 \\
4 & $[1,-1,0,2,-2,0]$& 25 \\
5 & $[1,-1,0,2,-2,0,2]$& 52 \\
6 & $[1,-1,0,2,-2,0,2,-2]$&  93 \\
7 & $[1,-1,0,2,-2,0,2,-2,2]$&  165 \\
8 & $[1,-1,0,2,-2,0,2,-2,0,2]$& 294 \\
\end{tabular}
\end{ruledtabular}
\end{table}

Observables are evaluated by Corner transfer matrix (CTM) technique.
The optimization of energy per site $e(\vec{\lambda})$ is carried out using L-BFGS optimizer, 
which is a gradient-based quasi-Newton method. The gradients of the energy with respect to the parameters $\vec{\lambda}$
are evaluated by reverse-mode automatic differentiation (AD) which differentiates the entire CTM procedure, the construction 
of the reduced density matrices and finally the evaluation of spin-spin interactions $\mathbf{S}\cdot\mathbf{S}$ for NN, NNN, and NNNN terms
\footnote{The antiferromagnetic order is incorporated into this translationally invariant wavefunction by rotation of 
the physical space on each sublattice-B site: $\mathbf{S}\cdot\mathbf{S}\ \rightarrow\ \mathbf{S}\cdot\mathbf{\tilde{S}}$ with 
$\tilde{S}^\alpha = -\sigma_y S^\alpha (\sigma_y)^T$}. 
We typically perform gradient optimization until difference in the energy between two consecutive gradient steps becomes
smaller than $10^{-8}$.
The entire implementation of the ansatz and its optimization are available as a part of the open-source library \textit{peps-torch}~\cite{Hasik2021pepstorch}.

\paragraph{AFM Heisenberg point $(J_2,J_3) = (0, 0)$:}
AFM Heisenberg model, realized at point $(J_2,J_3) = (0, 0)$, provides a solid benchmark for extrapolation
techniques of finite-$(D,\chi)$ iPEPS data (note here $\chi$ is the cutoff bond dimension of contracted tensor network). The recently developed finite correlation-length scaling~\cite{Corboz2018, rader2018}
coupled with gradient optimization considerably improved upon initial thermodynamic estimates of order parameter $m^2$  
based on the plain $1/D$ scaling. The finite-$\chi$ estimate of the correlation length $\xi$ 
can be readily extracted from the leading part of the spectrum of transfer matrix 
as $\xi = -1/log |\Lambda_1/\Lambda_0|$ where $\Lambda_0, \Lambda_1$ are leading and sub-leading eigenvalues respectively.
In most recent variation of FCLS~\cite{vanhecke2021scaling}, one treats each $(D,\chi)$ optimization
as an individual data point. For sufficiently large correlation lengths, the data for the order parameter are expected 
to obey simple scaling hypothesis ansatz 
\begin{equation}
\label{eq:simple_FCLS_m2}
m^2 = m^2_0 + a/\xi + O(1/\xi^2)  
\end{equation}
inspired by the established finite-size corrections of nonlinear $O(3)$ sigma model.

\begin{figure}[tb]
\includegraphics[width=\columnwidth]{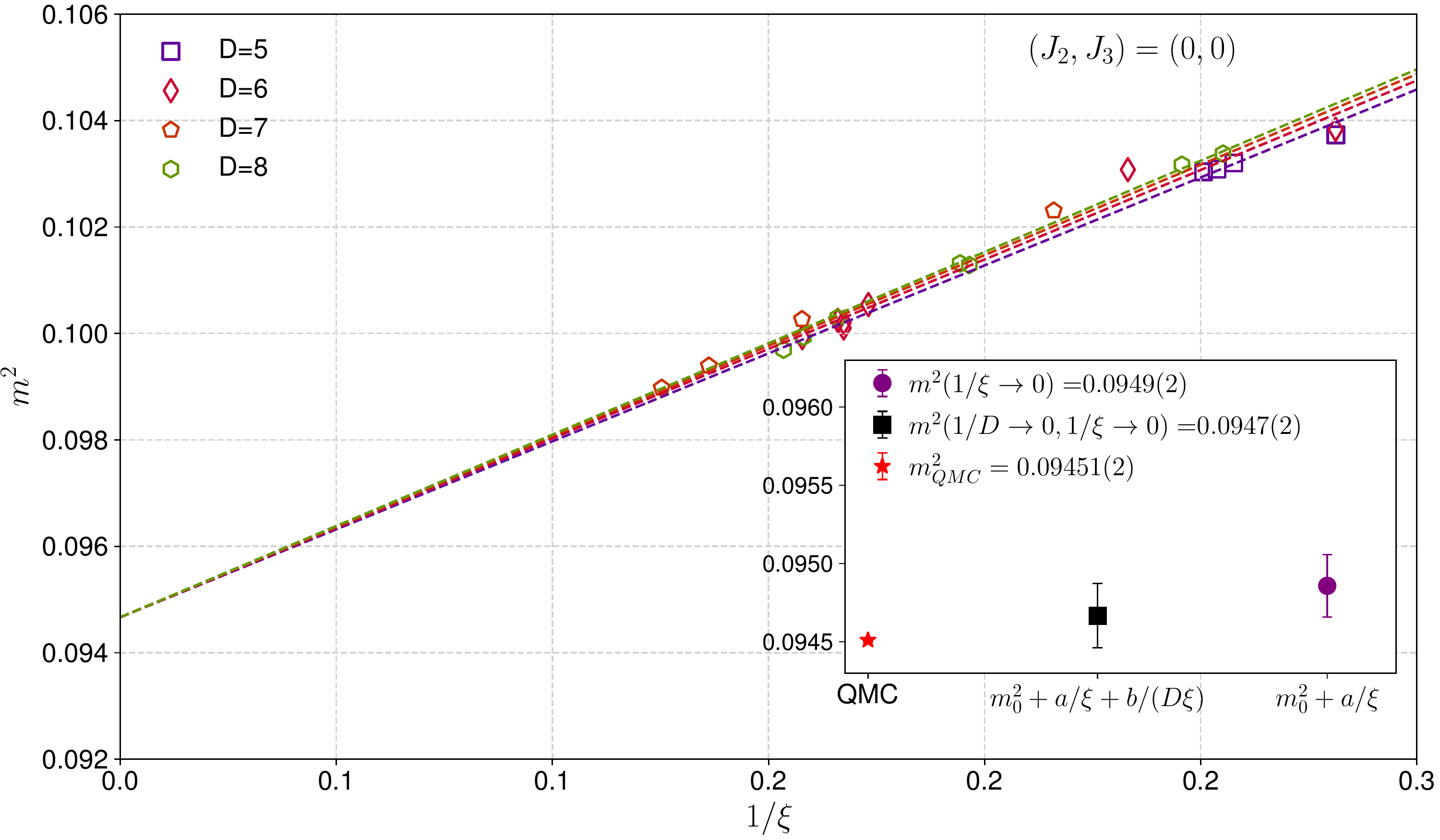}
\caption{\label{fig:surface_m2_vs_invxi_vs_invD} FCLS of magnetization for AFM Heisenberg model. 
Dashed lines represent fixed $D$ cuts of the surface $m^2=m^2_0 + a/\xi + b/(D\xi)$. The inset shows comparison of QMC estimate with 
estimates from original scaling hypothesis and its extended form.
Symbols are individual iPEPS simulations with $D=(5,6,7,8)$ and selected $\chi$ from 17 up to 200 for $D=(5,6,7)$ and up to 147 for $D=8$.}
\end{figure}

Using the above scaling hypothesis, we analyze the data from iPEPS simulations for $D=(5,6,7,8)$ and select $\chi$ from 17 up to 200 for $D=(5,6,7)$ and up to 147 for $D=8$ 
restricting to states with correlation lengths $1/\xi < 0.3$. This results in an estimate 
$m^2(1/\xi\rightarrow0) = 0.0949(2)$ which is very close to the best QMC estimate $m^2_{QMC} = 0.09451(2)$~\cite{sandvik1997}. 
Here, we improve upon this estimate by recognizing that the way magnetization scales with $\xi$ might possess a slight $D$-dependence
which vanishes for $D\gg1$. The extended scaling hypothesis reads
\begin{equation}
\label{eq:heisenberg_surface}
m^2 = m^2_0 + a/\xi + b/(D\xi) + O(1/\xi^2).
\end{equation}
Fitting this surface to the same data via non-linear least-squares leads to $m^2(1/D\rightarrow0,1/\xi\rightarrow0) = 0.0947(2)$
which is in better agreement with QMC. We show the fixed-$D$ cuts of the resulting surface (\ref{eq:heisenberg_surface}) and the comparison
of different thermodynamic estimates in Fig.~\ref{fig:surface_m2_vs_invxi_vs_invD}.

\paragraph{$J_1-J_3$ model at $(J_2,J_3) = (0, 0.35)$:}

\begin{figure}[tb]
\includegraphics[width=\columnwidth]{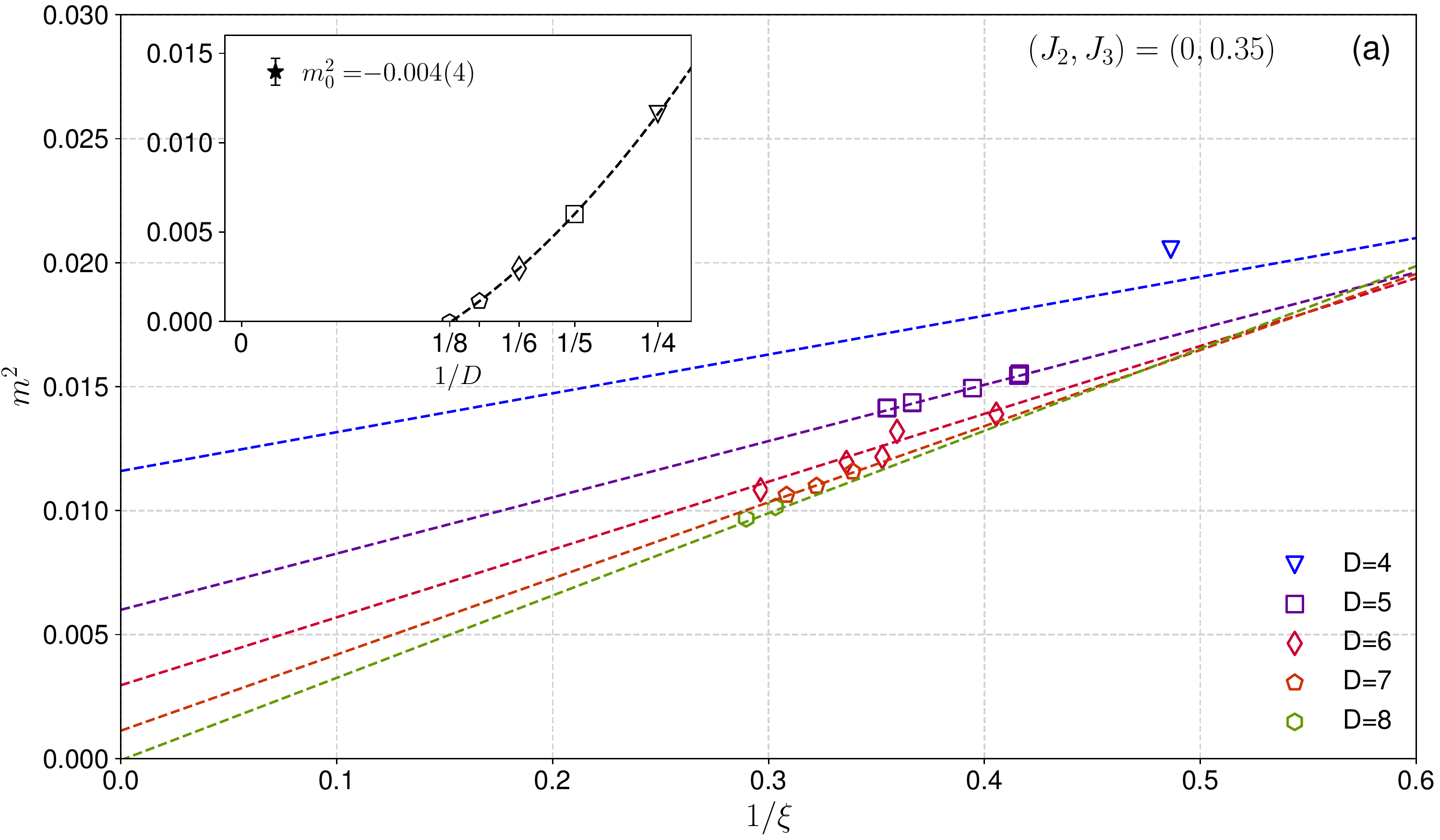}
\vspace{0.25cm}
\includegraphics[width=\columnwidth]{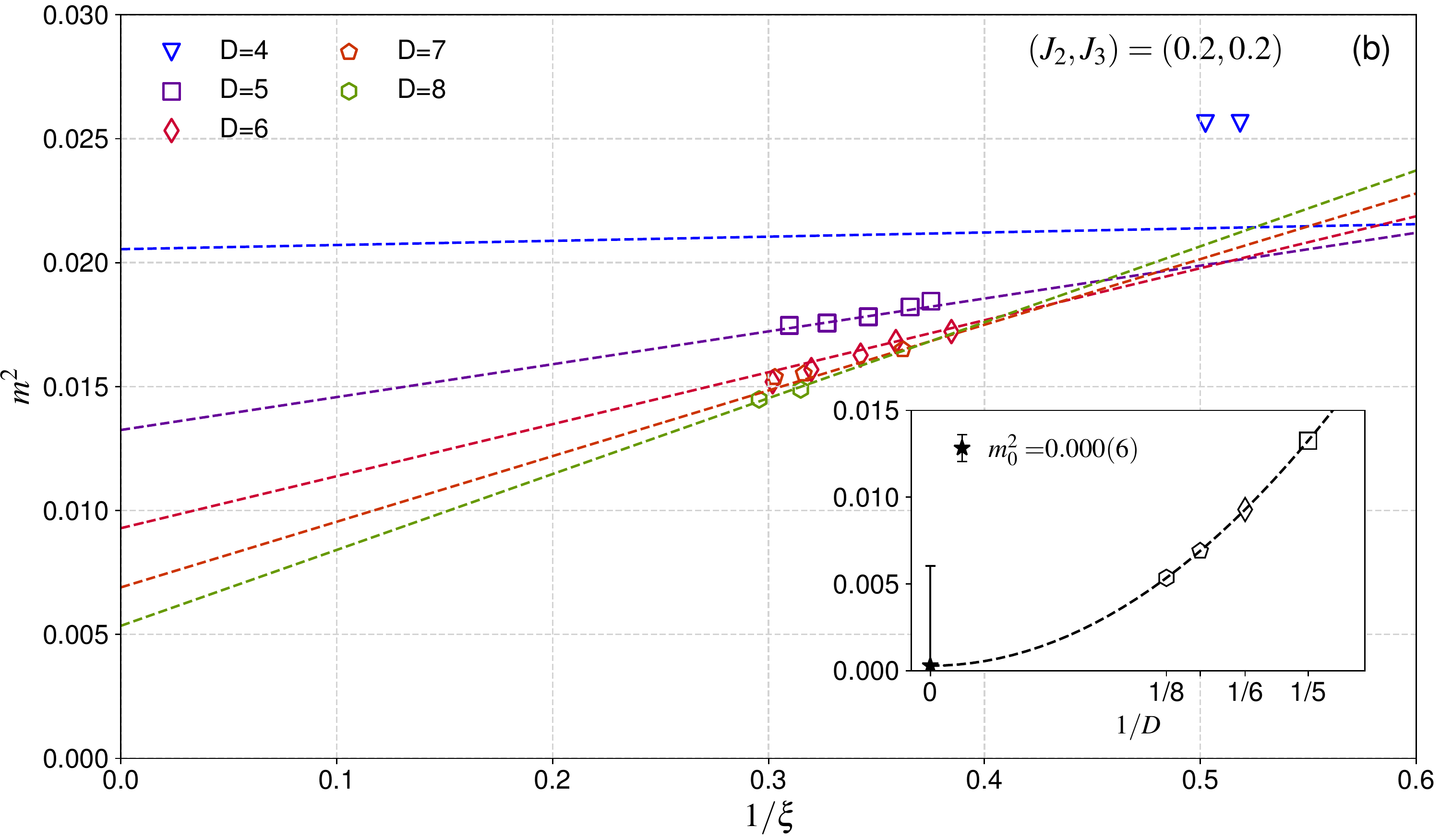}
\caption{\label{fig:surface_m2_vs_invxi_vs_invD_j20_j30} 
Scaling of magnetization at $(J_2,J_3)=(0,0.35)$ (a) and $(0.2,0.2)$ (b).
The coloured dashed lines are fixed-$D$ cuts (coloured dashed lines) of the surface $m^2(1/D,1/\xi)$ obtained by 
least-squares fit to Eq.~\ref{eq:surface_m2} ($D=4$ is excluded from the fit). 
The coloured symbols are iPEPS data for $D=(4,5,6,7,8)$ with environment dimensions $\chi$ up to 63, 151, 252, 196, and 160 respectively. In (b), additional data point $(D,\chi)=(5,300)$ is included.
The inset shows the dashed $m^2(1/D)$ curve obtained in the limit $1/\xi\rightarrow0$. The individual black points 
show \textit{hypothetical} value of $m^2$ reached in the limit of $1/\xi\rightarrow0$ for $D=(4,5,6,7,8)$ (see text). }
\end{figure}

\begin{figure}[htb]
\includegraphics[width=\columnwidth]{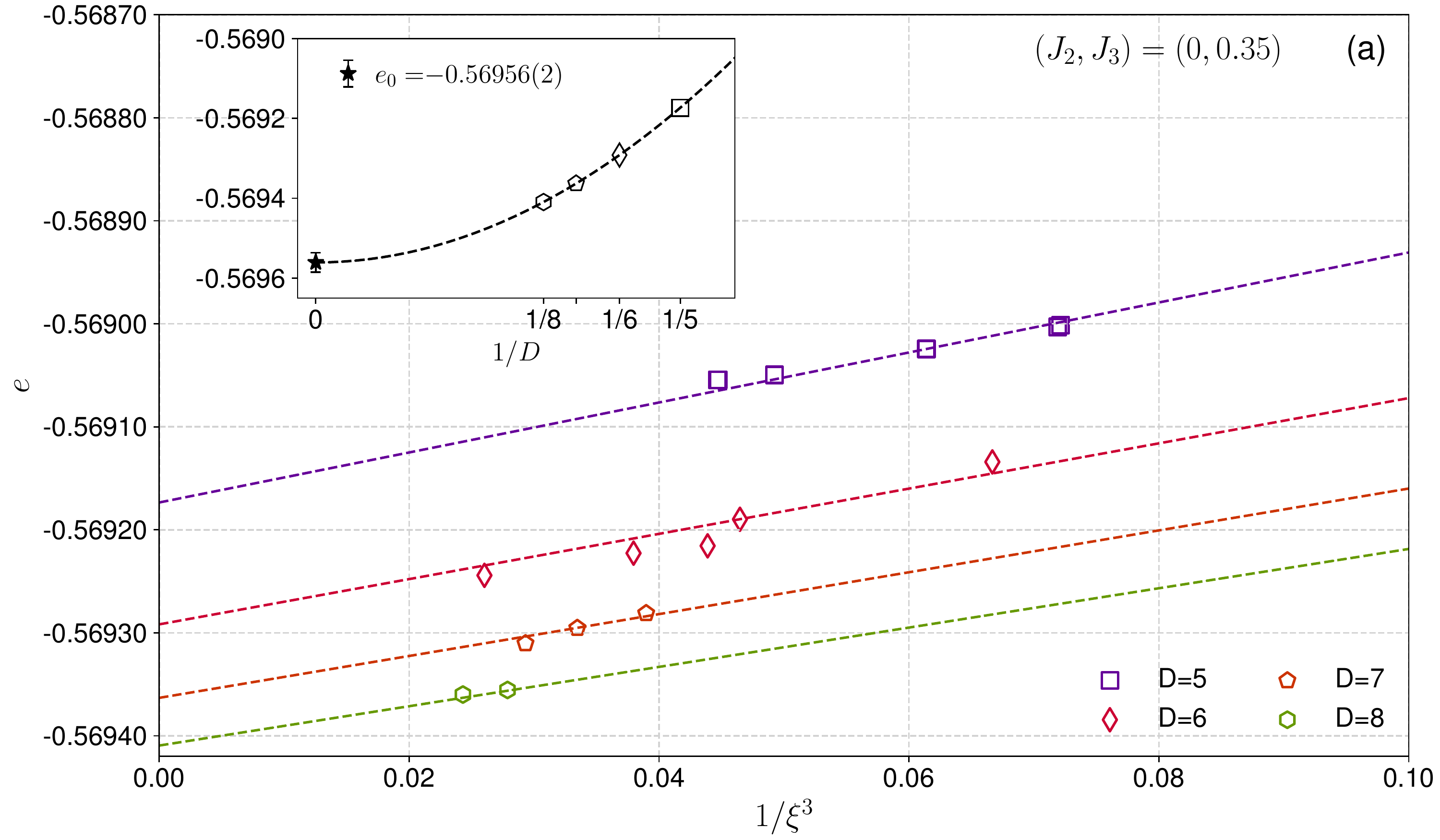}
\vspace{0.25cm}
\includegraphics[width=\columnwidth]{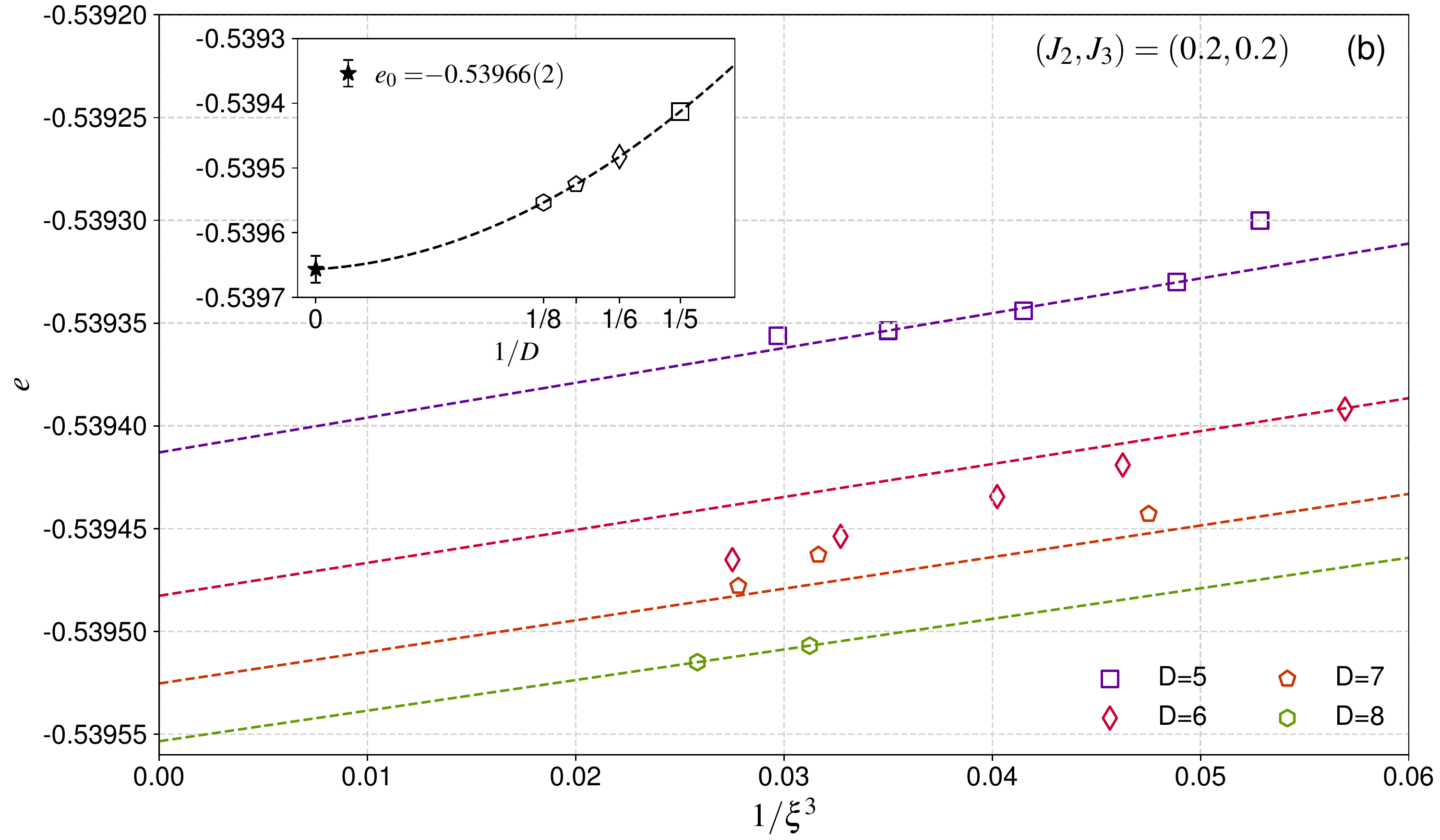}

\caption{\label{fig:surface_e_vs_invxi_vs_invD} 
Scaling of energy at $(J_2,J_3)=(0,0.35)$ (a) and $(0.2,0.2)$ (b).
The coloured dashed lines are fixed-$D$ cuts (coloured dashed lines) of the surface $e(1/D,1/\xi)$ obtained by least-squares fit to Eq.~\ref{eq:surface_e}. 
The coloured symbols are iPEPS data for $D=(5,6,7,8)$ with environment dimensions $\chi$ up to 151, 252, 196, and 160 respectively. In (b), additional data point $(D,\chi)=(5,300)$ is included.
The inset shows the dashed $e(1/D)$ curve obtained in the limit $1/\xi\rightarrow0$. The individual black points 
show \textit{hypothetical} value of $e$ reached in the limit of $1/\xi\rightarrow0$ for $D=(5,6,7,8)$ (see text).}
\end{figure}

The optimizations at this highly-frustrated point become considerably more demanding. In particular, for bond dimensions $D=(6,7,8)$ 
we observe that the necessary environment dimension for regular behaviour of optimizations is roughly $\chi/D^2 \gtrsim 2$. 
For bond dimensions $D=(6,7,8)$ we perform optimizations reaching environment dimensions $\chi$ up to 252, 196, and 160 respectively.
The technical limitations are set by the memory requirements of the intermediate steps in the construction of all RDMs needed to compute observables.

The resulting fixed-$(D,\chi)$ iPEPS data at $D=(5,6,7,8)$ for magnetization (energy) can no longer be described by simple scaling ansatz
$m^2=m^2_0 + a/\xi + O(1/\xi^2)$ ($e=e_0 + c/\xi^3 + O(1/\xi^4)$), at least in the regime accessible by our simulations. In contrast to AFM Heisenberg point,
where the optimized ansatz reached correlations lengths as large as $1/\xi \approx 0.125$, at $(J_2,J_3)=(0,0.35)$ the largest correlation lengths  
attained are only $1/\xi \approx 0.3$. We believe this is a manifestation of frustration, where equally good description of system on patches of characteristic size $\xi$
requires increasingly larger $D$ and $\chi$ compared to the unfrustrated case. In order to extract thermodynamic estimates for magnetization and energy from
our iPEPS data we thus adopt empirical approach following the idea behind improved fit of magnetization at AFM Heisenberg point. We postulate following scaling hypotheses
for magnetization and energy of optimal iPEPS ansatz as functions of both correlation length and bond dimension $D$
\begin{align} 
m^2 &= m^2_0 + \frac{m^2_1}{D} + \frac{m^2_2}{D^2} + \frac{a}{\xi} + \frac{b}{D\xi} + O(1/\xi^2), \label{eq:surface_m2} \\ 
e &= e_0 + \frac{e_1}{D} + \frac{e_2}{D^2} + \frac{c}{\xi^3} + \frac{d}{(D\xi^3)} + O(1/\xi^4). \label{eq:surface_e}
\end{align}
The functional form of these surfaces is motivated by the evidence that simple FCLS hypothesis for the Néel phase (\ref{eq:simple_FCLS_m2}) works 
appreciably well even close to the paramagnetic phase~\cite{hasik2021}. Moreover, for sufficiently large bond dimensions the finite-$D$ effects 
should become irrelevant and the simple scaling hypothesis is recovered. 
We fit these surfaces to the iPEPS data for magnetization and energy at $D=(5,6,7,8)$ and show the finite-$D$ cuts of the resulting surfaces in 
Fig.~\ref{fig:surface_m2_vs_invxi_vs_invD_j20_j30}(a) and Fig.~\ref{fig:surface_e_vs_invxi_vs_invD}(a) respectively. 

Our thermodynamic estimate for magnetization is $m^2(1/D\rightarrow0,1/\xi\rightarrow0)=-0.004(4)$ which is compatible with QSL phase at $(J_2,J_3)=(0,0.35)$. Similarly, the energy per site is estimated as $e(1/D\rightarrow0,1/\xi\rightarrow0)=-0.56956(2)$. The coefficients of the fitted surfaces are listed in Tab.~\ref{tab:surface_fits}.

Let us remark that while the limit $1/D,1/\xi\rightarrow0$ is unambiguous, the limit of surfaces Eq.~\ref{eq:surface_m2}-\ref{eq:surface_e} with only $1/\xi\rightarrow0$ 
for small finite $D$ might be unphysical since the optimal iPEPS instead realize finite correlation length
even for environment dimensions $\chi\rightarrow\infty$.

\paragraph{Point $(J_2,J_3) = (0.2, 0.2)$:}
The final point subjected to iPEPS analysis is highly-frustrated point $(J_2,J_3) = (0.2, 0.2)$ where both NNN and NNNN coupling 
play role. As in the case of $(J_2,J_3) = (0, 0.35)$, we find that for bond dimensions $D=(6,7,8)$ well-behaved optimizations require environment 
dimension of size at least $\chi/D^2 \gtrsim 2$. The surface fits, as shown in Fig.~\ref{fig:surface_m2_vs_invxi_vs_invD_j20_j30}(b) and Fig.~\ref{fig:surface_e_vs_invxi_vs_invD}(b), lead to thermodynamic estimates for magnetization 
$m^2(1/D\rightarrow0,1/\xi\rightarrow0)=0.000(6)$, compatible with SL phase at $(J_2,J_3)=(0.2,0.2)$, and the energy per site $e(1/D\rightarrow0,1/\xi\rightarrow0)=-0.53966(2)$. The coefficients of the fitted surfaces are listed in Tab.~\ref{tab:surface_fits}.
The error on the $m^2$ estimate remains large and is mainly due to the limited range of correlation lengths we could reach for the computationally accessible bond and environment dimensions $D,\chi$.

\begin{table}
\begin{tabular}{ c|c|c|c|c|c }
\hline
$(J_2,J_3)$ & $m^2_0$ & $m^2_1$ & $m^2_2$ & $a$ & $b$ \\
\hline
\hline
\multirow{2}{*}{$(0,0)$} & 0.09467 & - & - & 0.03642 & -0.01678 \\
                         & 0.00021 & - & - & 0.00192 &  0.00789 \\
\hline
\multirow{2}{*}{$(0,0.35)$} & -0.00396 & 0.00004 & 0.24876 & 0.05077 & -0.14044 \\
                            &  0.00391 & 0.03279 & 0.16913 & 0.02061 &  0.11525 \\
\hline
\multirow{2}{*}{$(0.2,0.2)$} & 0.00028 & 0       & 0.32416 & 0.05955 & -0.23147 \\
                             & 0.00576 & 0.04222 & 0.13837 & 0.02129 &  0.12325 \\
\hline
\multicolumn{6}{c}{} \\
\hline
& $e_0$ & $e_1$ & $e_2$ & $c$ & $d$ \\
\hline
\hline
\multirow{2}{*}{$(0,0.35)$} & -0.569561 &  0        &  0.00968 & 0.00104 & 0.00691 \\
                            &  0.000024 &  0.000307 &  0.00212 & 0.00198 & 0.01079 \\
\hline
\multirow{2}{*}{$(0.2,0.2)$} & -0.539656 & 0.000168 & 0.00525 & 0.00115 & 0.00273 \\
                             &  0.000021 & 0.000037 & 0.00083 & 0.00053 & 0.00325 \\
\hline

\end{tabular}
\caption{\label{tab:surface_fits} The least-squares fits of scaling hypotheses for magnetization and energy, 
defined in the Eqs.~\ref{eq:surface_m2}-~\ref{eq:surface_e} for studied $(J_2,J_3)$ points. 
For each coefficient, the top number is the estimated value and the bottom number is the error on the estimate based on the covariance matrix obtained from Jacobian of the least-squares cost function. The coefficient $m^2_1$ $(e_1)$ is bounded to be positive. }
\end{table}
%%%%% J1-J2-J3 iPEPS %%%%%

\begin{figure}[htbp]
	\centering
	\includegraphics[width=3.2in]{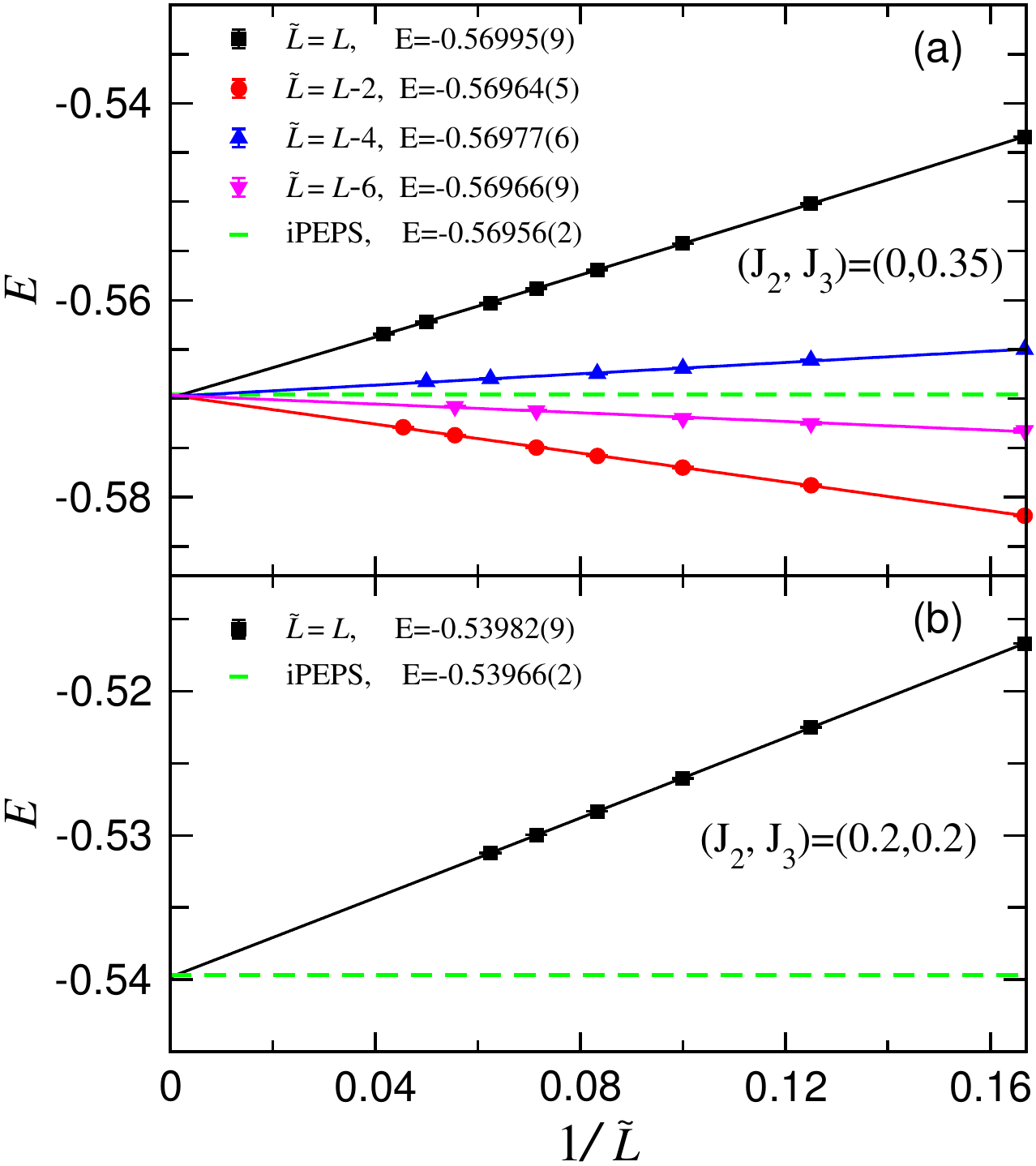}
	\caption{Energy comparison  between finite PEPS and iPEPS. Finite PEPS results on different system sizes are from $D=8$, and iPEPS results (dashed lines) are from finite correlation scaling. (a)  finite size scaling of different system size $L \times L$ up to $L=24$ at $(J_2, J_3)=(0, 0.35)$. Energies from different bulk choices $\tilde{L}\times \tilde{L}$ are shown for extrapolations.  A second order extrapolations for $\tilde{L}=L$ or linear fittings for other cases are shown. Extrapolated energies of second order fittings (not shown) for $\tilde{L}=L-2$ and $\tilde{L}=L-4$ cases are $-0.5695(1)$, $-0.5697(2)$, respectively.  (b) second order extrapolations of energies at $(J_2, J_3)=(0.2, 0.2)$. Corresponding iPEPS and extrapolated finite PEPS energies are given in the legend.  }
	\label{fig:fiPEPSvsiPEPS}
\end{figure}

\subsection{Comparison with finite PEPS results}

  We finish this Appendix by a detailed comparison of the iPEPS and finite PEPS results within the QSL phase. As shown in Fig.~\ref{fig:fiPEPSvsiPEPS} the PEPS energy (persite) computed on open $L\times L$ clusters shows finite-size effects  significantly larger than the finite-$\xi$ effects of  the iPEPS data -- as seen e.g. from a simple comparison of the energy scales used in  Figs.~\ref{fig:surface_e_vs_invxi_vs_invD} and \ref{fig:fiPEPSvsiPEPS}. 
  This is due to the fact that the leading correction of the iPEPS energy goes as $1/\xi^3$ while the leading finite PEPS correction goes as $1/L$. 
  However, the finite PEPS energy follows a very precise quadratic scaling in the inverse bulk-size $1/\tilde{L}$, for a given choice of central bulk $\tilde{L}\times \tilde{L}$, enabling very precise fits and an accurate extrapolation to the thermodynamic limit. We observe a very good agreement between the iPEPS and PEPS extrapolated energies with up to 4 significant digits. 
  
   \begin{figure}[htbp]
 \centering
 \includegraphics[width=3.4in]{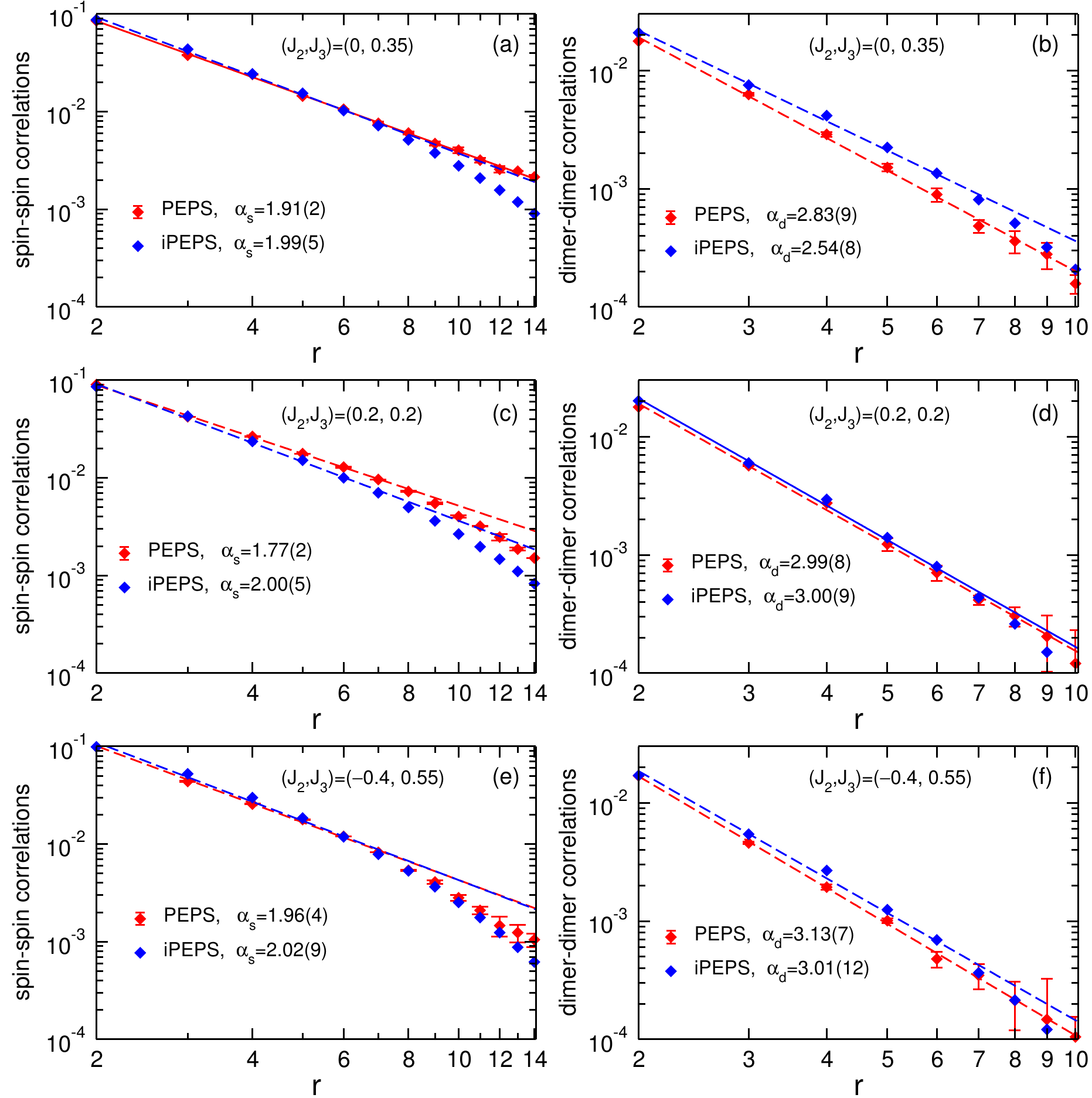}
 \caption{Comparison between finite PEPS and iPEPS correlations at $D=8$ (optimized as $\chi_{opt}=160$ and evaluated at $\chi$=400), plotted on a semi-log scale. In the case of  the finite PEPS correlations, a $20\times 28$ strip is used at $(J_2, J_3)=(0, 0.35)$, and a $12\times 28$ strip at $(J_2, J_3)=(0.2, 0.2)$ and $(J_2, J_3)=(-0.4, 0.55)$. Dashed lines are power law fits using data up to intermediate distances $r\leq 7$. }
\label{fig:iPEPScorrelation}
 \end{figure}      

We have also compared the iPEPS connected spin-spin and dimer-dimer correlations to the ones obtained on finite strips. 
The generic connected spin-spin correlation function is $(-1)^{\bf r}\big<{\bf S_i}\cdot{\bf S_{i+r}}\big>- m_0^2$ for magnetic and nonmagnetic phases where $m_0$ is the thermodynamic limit AFM order, which would give precise decay behaviour of spin correlations when the system size $L$ is sufficiently large.  In the QSL phase the thermodynamic limit AFM order is 0, and the connected spin-spin correlation is $(-1)^{\bf r}\big<{\bf S_i}\cdot{\bf S_{i+r}}\big>$ for finite size calculations. For iPEPS, the finite $D$ ($D=8$ here) shows a residual staggered magnetization $m_0$ and the connected spin-spin correlation function is $(-1)^{\bf r}\big<{\bf S_i}\cdot{\bf S_{i+r}}\big>- m_0^2$, which could provide a systematically improved description for the decay behaviour of the QSL phase and would be exact in the limit $D\rightarrow\infty$ with a vanishing magnetic order. The later finite PEPS and iPEPS spin-spin correlations are compared in Figs.~\ref{fig:iPEPScorrelation}(a), \ref{fig:iPEPScorrelation}(c) and \ref{fig:iPEPScorrelation}(e), showing a good agreement, and likely algebraic decays with similar exponents. We have also performed a similar comparison for the (staggered) dimer-dimer correlation function defined in Eq.~\ref{eq:dimer_cor}.
Figs.~\ref{fig:iPEPScorrelation}(b),\ref{fig:iPEPScorrelation}(d) and \ref{fig:iPEPScorrelation}(f) show the absolute value $|C_d(r)|$ vs $r$ and reveal again a good agreement between finite PEPS and iPEPS data, with similar algebraic decays. However, we note that the  correlations shown here are still subject to small finite $L$ / finite $D$ corrections. We observe that for increasing $L$ / $D$ the correlations decay less rapidly so that the exponents extracted here (see values on the plots) can be considered as upper bounds of the true exponents (see Fig~\ref{fig:QSLexponent} for more accurate values obtained from order parameter scaling).  

\clearpage
\bibliography{j1j2j3paper}

\begin{thebibliography}{72}%
\makeatletter
\providecommand \@ifxundefined [1]{%
 \@ifx{#1\undefined}
}%
\providecommand \@ifnum [1]{%
 \ifnum #1\expandafter \@firstoftwo
 \else \expandafter \@secondoftwo
 \fi
}%
\providecommand \@ifx [1]{%
 \ifx #1\expandafter \@firstoftwo
 \else \expandafter \@secondoftwo
 \fi
}%
\providecommand \natexlab [1]{#1}%
\providecommand \enquote  [1]{``#1''}%
\providecommand \bibnamefont  [1]{#1}%
\providecommand \bibfnamefont [1]{#1}%
\providecommand \citenamefont [1]{#1}%
\providecommand \href@noop [0]{\@secondoftwo}%
\providecommand \href [0]{\begingroup \@sanitize@url \@href}%
\providecommand \@href[1]{\@@startlink{#1}\@@href}%
\providecommand \@@href[1]{\endgroup#1\@@endlink}%
\providecommand \@sanitize@url [0]{\catcode `\\12\catcode `\$12\catcode
  `\&12\catcode `\#12\catcode `\^12\catcode `\_12\catcode `\%12\relax}%
\providecommand \@@startlink[1]{}%
\providecommand \@@endlink[0]{}%
\providecommand \url  [0]{\begingroup\@sanitize@url \@url }%
\providecommand \@url [1]{\endgroup\@href {#1}{\urlprefix }}%
\providecommand \urlprefix  [0]{URL }%
\providecommand \Eprint [0]{\href }%
\providecommand \doibase [0]{http://dx.doi.org/}%
\providecommand \selectlanguage [0]{\@gobble}%
\providecommand \bibinfo  [0]{\@secondoftwo}%
\providecommand \bibfield  [0]{\@secondoftwo}%
\providecommand \translation [1]{[#1]}%
\providecommand \BibitemOpen [0]{}%
\providecommand \bibitemStop [0]{}%
\providecommand \bibitemNoStop [0]{.\EOS\space}%
\providecommand \EOS [0]{\spacefactor3000\relax}%
\providecommand \BibitemShut  [1]{\csname bibitem#1\endcsname}%
\let\auto@bib@innerbib\@empty
%</preamble>
\bibitem [{\citenamefont {Balents}(2010)}]{balents2010}%
  \BibitemOpen
  \bibfield  {author} {\bibinfo {author} {\bibfnamefont {Leon}\ \bibnamefont
  {Balents}},\ }\bibfield  {title} {\enquote {\bibinfo {title} {Spin liquids in
  frustrated magnets},}\ }\href {\doibase https://doi.org/10.1038/nature08917}
  {\bibfield  {journal} {\bibinfo  {journal} {Nature}\ }\textbf {\bibinfo
  {volume} {464}},\ \bibinfo {pages} {199--208} (\bibinfo {year}
  {2010})}\BibitemShut {NoStop}%
\bibitem [{\citenamefont {Anderson}(1987)}]{Anderson1987}%
  \BibitemOpen
  \bibfield  {author} {\bibinfo {author} {\bibfnamefont {P.~W.}\ \bibnamefont
  {Anderson}},\ }\bibfield  {title} {\enquote {\bibinfo {title} {The resonating
  valence bond state in ${{\rm La_2CuO_4}}$ and superconductivity},}\ }\href
  {\doibase 10.1126/science.235.4793.1196} {\bibfield  {journal} {\bibinfo
  {journal} {Science}\ }\textbf {\bibinfo {volume} {235}},\ \bibinfo {pages}
  {1196--1198} (\bibinfo {year} {1987})}\BibitemShut {NoStop}%
\bibitem [{\citenamefont {Senthil}\ \emph
  {et~al.}(2004{\natexlab{a}})\citenamefont {Senthil}, \citenamefont
  {Vishwanath}, \citenamefont {Balents}, \citenamefont {Sachdev},\ and\
  \citenamefont {Fisher}}]{DQCP1}%
  \BibitemOpen
  \bibfield  {author} {\bibinfo {author} {\bibfnamefont {T.}~\bibnamefont
  {Senthil}}, \bibinfo {author} {\bibfnamefont {Ashvin}\ \bibnamefont
  {Vishwanath}}, \bibinfo {author} {\bibfnamefont {Leon}\ \bibnamefont
  {Balents}}, \bibinfo {author} {\bibfnamefont {Subir}\ \bibnamefont
  {Sachdev}}, \ and\ \bibinfo {author} {\bibfnamefont {Matthew P.~A.}\
  \bibnamefont {Fisher}},\ }\bibfield  {title} {\enquote {\bibinfo {title}
  {Deconfined quantum critical points},}\ }\href {\doibase
  10.1126/science.1091806} {\bibfield  {journal} {\bibinfo  {journal}
  {Science}\ }\textbf {\bibinfo {volume} {303}},\ \bibinfo {pages} {1490--1494}
  (\bibinfo {year} {2004}{\natexlab{a}})}\BibitemShut {NoStop}%
\bibitem [{\citenamefont {Senthil}\ \emph
  {et~al.}(2004{\natexlab{b}})\citenamefont {Senthil}, \citenamefont {Balents},
  \citenamefont {Sachdev}, \citenamefont {Vishwanath},\ and\ \citenamefont
  {Fisher}}]{DQCP2}%
  \BibitemOpen
  \bibfield  {author} {\bibinfo {author} {\bibfnamefont {T.}~\bibnamefont
  {Senthil}}, \bibinfo {author} {\bibfnamefont {Leon}\ \bibnamefont {Balents}},
  \bibinfo {author} {\bibfnamefont {Subir}\ \bibnamefont {Sachdev}}, \bibinfo
  {author} {\bibfnamefont {Ashvin}\ \bibnamefont {Vishwanath}}, \ and\ \bibinfo
  {author} {\bibfnamefont {Matthew P.~A.}\ \bibnamefont {Fisher}},\ }\bibfield
  {title} {\enquote {\bibinfo {title} {Quantum criticality beyond the
  landau-ginzburg-wilson paradigm},}\ }\href {\doibase
  10.1103/PhysRevB.70.144407} {\bibfield  {journal} {\bibinfo  {journal} {Phys.
  Rev. B}\ }\textbf {\bibinfo {volume} {70}},\ \bibinfo {pages} {144407}
  (\bibinfo {year} {2004}{\natexlab{b}})}\BibitemShut {NoStop}%
\bibitem [{\citenamefont {Sandvik}(2007)}]{JQ2007}%
  \BibitemOpen
  \bibfield  {author} {\bibinfo {author} {\bibfnamefont {Anders~W.}\
  \bibnamefont {Sandvik}},\ }\bibfield  {title} {\enquote {\bibinfo {title}
  {Evidence for deconfined quantum criticality in a two-dimensional
  \text{Heisenberg} model with four-spin interactions},}\ }\href {\doibase
  10.1103/PhysRevLett.98.227202} {\bibfield  {journal} {\bibinfo  {journal}
  {Phys. Rev. Lett.}\ }\textbf {\bibinfo {volume} {98}},\ \bibinfo {pages}
  {227202} (\bibinfo {year} {2007})}\BibitemShut {NoStop}%
\bibitem [{\citenamefont {Nahum}\ \emph
  {et~al.}(2015{\natexlab{a}})\citenamefont {Nahum}, \citenamefont {Serna},
  \citenamefont {Chalker}, \citenamefont {Ortu\~no},\ and\ \citenamefont
  {Somoza}}]{loopmodel2}%
  \BibitemOpen
  \bibfield  {author} {\bibinfo {author} {\bibfnamefont {Adam}\ \bibnamefont
  {Nahum}}, \bibinfo {author} {\bibfnamefont {P.}~\bibnamefont {Serna}},
  \bibinfo {author} {\bibfnamefont {J.~T.}\ \bibnamefont {Chalker}}, \bibinfo
  {author} {\bibfnamefont {M.}~\bibnamefont {Ortu\~no}}, \ and\ \bibinfo
  {author} {\bibfnamefont {A.~M.}\ \bibnamefont {Somoza}},\ }\bibfield  {title}
  {\enquote {\bibinfo {title} {Emergent so(5) symmetry at the n\'eel to
  valence-bond-solid transition},}\ }\href {\doibase
  10.1103/PhysRevLett.115.267203} {\bibfield  {journal} {\bibinfo  {journal}
  {Phys. Rev. Lett.}\ }\textbf {\bibinfo {volume} {115}},\ \bibinfo {pages}
  {267203} (\bibinfo {year} {2015}{\natexlab{a}})}\BibitemShut {NoStop}%
\bibitem [{\citenamefont {Sreejith}\ \emph {et~al.}(2019)\citenamefont
  {Sreejith}, \citenamefont {Powell},\ and\ \citenamefont
  {Nahum}}]{sreejith2019}%
  \BibitemOpen
  \bibfield  {author} {\bibinfo {author} {\bibfnamefont {G.~J.}\ \bibnamefont
  {Sreejith}}, \bibinfo {author} {\bibfnamefont {Stephen}\ \bibnamefont
  {Powell}}, \ and\ \bibinfo {author} {\bibfnamefont {Adam}\ \bibnamefont
  {Nahum}},\ }\bibfield  {title} {\enquote {\bibinfo {title} {Emergent so(5)
  symmetry at the columnar ordering transition in the classical cubic dimer
  model},}\ }\href {\doibase 10.1103/PhysRevLett.122.080601} {\bibfield
  {journal} {\bibinfo  {journal} {Phys. Rev. Lett.}\ }\textbf {\bibinfo
  {volume} {122}},\ \bibinfo {pages} {080601} (\bibinfo {year}
  {2019})}\BibitemShut {NoStop}%
\bibitem [{\citenamefont {Melko}\ and\ \citenamefont {Kaul}(2008)}]{JQ2008}%
  \BibitemOpen
  \bibfield  {author} {\bibinfo {author} {\bibfnamefont {Roger~G.}\
  \bibnamefont {Melko}}\ and\ \bibinfo {author} {\bibfnamefont {Ribhu~K.}\
  \bibnamefont {Kaul}},\ }\bibfield  {title} {\enquote {\bibinfo {title}
  {Scaling in the fan of an unconventional quantum critical point},}\ }\href
  {\doibase 10.1103/PhysRevLett.100.017203} {\bibfield  {journal} {\bibinfo
  {journal} {Phys. Rev. Lett.}\ }\textbf {\bibinfo {volume} {100}},\ \bibinfo
  {pages} {017203} (\bibinfo {year} {2008})}\BibitemShut {NoStop}%
\bibitem [{\citenamefont {Jiang}\ \emph
  {et~al.}(2008{\natexlab{a}})\citenamefont {Jiang}, \citenamefont {Nyfeler},
  \citenamefont {Chandrasekharan},\ and\ \citenamefont {Wiese}}]{JQ2008_2}%
  \BibitemOpen
  \bibfield  {author} {\bibinfo {author} {\bibfnamefont {F-J}\ \bibnamefont
  {Jiang}}, \bibinfo {author} {\bibfnamefont {M}~\bibnamefont {Nyfeler}},
  \bibinfo {author} {\bibfnamefont {S}~\bibnamefont {Chandrasekharan}}, \ and\
  \bibinfo {author} {\bibfnamefont {U-J}\ \bibnamefont {Wiese}},\ }\bibfield
  {title} {\enquote {\bibinfo {title} {From an antiferromagnet to a valence
  bond solid: evidence for a first-order phase transition},}\ }\href {\doibase
  10.1088/1742-5468/2008/02/p02009} {\bibfield  {journal} {\bibinfo  {journal}
  {Journal of Statistical Mechanics: Theory and Experiment}\ }\textbf {\bibinfo
  {volume} {2008}},\ \bibinfo {pages} {P02009} (\bibinfo {year}
  {2008}{\natexlab{a}})}\BibitemShut {NoStop}%
\bibitem [{\citenamefont {Lou}\ \emph {et~al.}(2009)\citenamefont {Lou},
  \citenamefont {Sandvik},\ and\ \citenamefont {Kawashima}}]{JQ2009}%
  \BibitemOpen
  \bibfield  {author} {\bibinfo {author} {\bibfnamefont {Jie}\ \bibnamefont
  {Lou}}, \bibinfo {author} {\bibfnamefont {Anders~W.}\ \bibnamefont
  {Sandvik}}, \ and\ \bibinfo {author} {\bibfnamefont {Naoki}\ \bibnamefont
  {Kawashima}},\ }\bibfield  {title} {\enquote {\bibinfo {title}
  {Antiferromagnetic to valence-bond-solid transitions in two-dimensional
  $\text{SU}(n)$ \text{Heisenberg} models with multispin interactions},}\
  }\href {\doibase 10.1103/PhysRevB.80.180414} {\bibfield  {journal} {\bibinfo
  {journal} {Phys. Rev. B}\ }\textbf {\bibinfo {volume} {80}},\ \bibinfo
  {pages} {180414} (\bibinfo {year} {2009})}\BibitemShut {NoStop}%
\bibitem [{\citenamefont {Sandvik}(2010{\natexlab{a}})}]{JQ2010}%
  \BibitemOpen
  \bibfield  {author} {\bibinfo {author} {\bibfnamefont {Anders~W.}\
  \bibnamefont {Sandvik}},\ }\bibfield  {title} {\enquote {\bibinfo {title}
  {Continuous quantum phase transition between an antiferromagnet and a
  valence-bond solid in two dimensions: Evidence for logarithmic corrections to
  scaling},}\ }\href {\doibase 10.1103/PhysRevLett.104.177201} {\bibfield
  {journal} {\bibinfo  {journal} {Phys. Rev. Lett.}\ }\textbf {\bibinfo
  {volume} {104}},\ \bibinfo {pages} {177201} (\bibinfo {year}
  {2010}{\natexlab{a}})}\BibitemShut {NoStop}%
\bibitem [{\citenamefont {Kaul}(2011)}]{JQ2011}%
  \BibitemOpen
  \bibfield  {author} {\bibinfo {author} {\bibfnamefont {Ribhu~K.}\
  \bibnamefont {Kaul}},\ }\bibfield  {title} {\enquote {\bibinfo {title}
  {Quantum criticality in su(3) and su(4) antiferromagnets},}\ }\href {\doibase
  10.1103/PhysRevB.84.054407} {\bibfield  {journal} {\bibinfo  {journal} {Phys.
  Rev. B}\ }\textbf {\bibinfo {volume} {84}},\ \bibinfo {pages} {054407}
  (\bibinfo {year} {2011})}\BibitemShut {NoStop}%
\bibitem [{\citenamefont {Block}\ \emph {et~al.}(2013)\citenamefont {Block},
  \citenamefont {Melko},\ and\ \citenamefont {Kaul}}]{JQ2013}%
  \BibitemOpen
  \bibfield  {author} {\bibinfo {author} {\bibfnamefont {Matthew~S.}\
  \bibnamefont {Block}}, \bibinfo {author} {\bibfnamefont {Roger~G.}\
  \bibnamefont {Melko}}, \ and\ \bibinfo {author} {\bibfnamefont {Ribhu~K.}\
  \bibnamefont {Kaul}},\ }\bibfield  {title} {\enquote {\bibinfo {title} {Fate
  of $\mathbb{CP}^{N\ensuremath{-}1}$ fixed points with $q$ monopoles},}\
  }\href {\doibase 10.1103/PhysRevLett.111.137202} {\bibfield  {journal}
  {\bibinfo  {journal} {Phys. Rev. Lett.}\ }\textbf {\bibinfo {volume} {111}},\
  \bibinfo {pages} {137202} (\bibinfo {year} {2013})}\BibitemShut {NoStop}%
\bibitem [{\citenamefont {Harada}\ \emph {et~al.}(2013)\citenamefont {Harada},
  \citenamefont {Suzuki}, \citenamefont {Okubo}, \citenamefont {Matsuo},
  \citenamefont {Lou}, \citenamefont {Watanabe}, \citenamefont {Todo},\ and\
  \citenamefont {Kawashima}}]{JQ2013_2}%
  \BibitemOpen
  \bibfield  {author} {\bibinfo {author} {\bibfnamefont {Kenji}\ \bibnamefont
  {Harada}}, \bibinfo {author} {\bibfnamefont {Takafumi}\ \bibnamefont
  {Suzuki}}, \bibinfo {author} {\bibfnamefont {Tsuyoshi}\ \bibnamefont
  {Okubo}}, \bibinfo {author} {\bibfnamefont {Haruhiko}\ \bibnamefont
  {Matsuo}}, \bibinfo {author} {\bibfnamefont {Jie}\ \bibnamefont {Lou}},
  \bibinfo {author} {\bibfnamefont {Hiroshi}\ \bibnamefont {Watanabe}},
  \bibinfo {author} {\bibfnamefont {Synge}\ \bibnamefont {Todo}}, \ and\
  \bibinfo {author} {\bibfnamefont {Naoki}\ \bibnamefont {Kawashima}},\
  }\bibfield  {title} {\enquote {\bibinfo {title} {Possibility of deconfined
  criticality in su($n$) heisenberg models at small $n$},}\ }\href {\doibase
  10.1103/PhysRevB.88.220408} {\bibfield  {journal} {\bibinfo  {journal} {Phys.
  Rev. B}\ }\textbf {\bibinfo {volume} {88}},\ \bibinfo {pages} {220408}
  (\bibinfo {year} {2013})}\BibitemShut {NoStop}%
\bibitem [{\citenamefont {Chen}\ \emph {et~al.}(2013)\citenamefont {Chen},
  \citenamefont {Huang}, \citenamefont {Deng}, \citenamefont {Kuklov},
  \citenamefont {Prokof'ev},\ and\ \citenamefont {Svistunov}}]{JQ2013_3}%
  \BibitemOpen
  \bibfield  {author} {\bibinfo {author} {\bibfnamefont {Kun}\ \bibnamefont
  {Chen}}, \bibinfo {author} {\bibfnamefont {Yuan}\ \bibnamefont {Huang}},
  \bibinfo {author} {\bibfnamefont {Youjin}\ \bibnamefont {Deng}}, \bibinfo
  {author} {\bibfnamefont {A.~B.}\ \bibnamefont {Kuklov}}, \bibinfo {author}
  {\bibfnamefont {N.~V.}\ \bibnamefont {Prokof'ev}}, \ and\ \bibinfo {author}
  {\bibfnamefont {B.~V.}\ \bibnamefont {Svistunov}},\ }\bibfield  {title}
  {\enquote {\bibinfo {title} {Deconfined criticality flow in the heisenberg
  model with ring-exchange interactions},}\ }\href {\doibase
  10.1103/PhysRevLett.110.185701} {\bibfield  {journal} {\bibinfo  {journal}
  {Phys. Rev. Lett.}\ }\textbf {\bibinfo {volume} {110}},\ \bibinfo {pages}
  {185701} (\bibinfo {year} {2013})}\BibitemShut {NoStop}%
\bibitem [{\citenamefont {Pujari}\ \emph {et~al.}(2015)\citenamefont {Pujari},
  \citenamefont {Alet},\ and\ \citenamefont {Damle}}]{JQ2015}%
  \BibitemOpen
  \bibfield  {author} {\bibinfo {author} {\bibfnamefont {Sumiran}\ \bibnamefont
  {Pujari}}, \bibinfo {author} {\bibfnamefont {Fabien}\ \bibnamefont {Alet}}, \
  and\ \bibinfo {author} {\bibfnamefont {Kedar}\ \bibnamefont {Damle}},\
  }\bibfield  {title} {\enquote {\bibinfo {title} {Transitions to valence-bond
  solid order in a honeycomb lattice antiferromagnet},}\ }\href {\doibase
  10.1103/PhysRevB.91.104411} {\bibfield  {journal} {\bibinfo  {journal} {Phys.
  Rev. B}\ }\textbf {\bibinfo {volume} {91}},\ \bibinfo {pages} {104411}
  (\bibinfo {year} {2015})}\BibitemShut {NoStop}%
\bibitem [{\citenamefont {Sandvik}\ and\ \citenamefont {Zhao}(2020)}]{JQ2020}%
  \BibitemOpen
  \bibfield  {author} {\bibinfo {author} {\bibfnamefont {Anders~W.}\
  \bibnamefont {Sandvik}}\ and\ \bibinfo {author} {\bibfnamefont {Bowen}\
  \bibnamefont {Zhao}},\ }\bibfield  {title} {\enquote {\bibinfo {title}
  {Consistent scaling exponents at the deconfined quantum-critical point},}\
  }\href {\doibase 10.1088/0256-307x/37/5/057502} {\bibfield  {journal}
  {\bibinfo  {journal} {Chinese Physics Letters}\ }\textbf {\bibinfo {volume}
  {37}},\ \bibinfo {pages} {057502} (\bibinfo {year} {2020})}\BibitemShut
  {NoStop}%
\bibitem [{\citenamefont {Nahum}\ \emph
  {et~al.}(2015{\natexlab{b}})\citenamefont {Nahum}, \citenamefont {Chalker},
  \citenamefont {Serna}, \citenamefont {Ortu\~no},\ and\ \citenamefont
  {Somoza}}]{loopmodel1}%
  \BibitemOpen
  \bibfield  {author} {\bibinfo {author} {\bibfnamefont {Adam}\ \bibnamefont
  {Nahum}}, \bibinfo {author} {\bibfnamefont {J.~T.}\ \bibnamefont {Chalker}},
  \bibinfo {author} {\bibfnamefont {P.}~\bibnamefont {Serna}}, \bibinfo
  {author} {\bibfnamefont {M.}~\bibnamefont {Ortu\~no}}, \ and\ \bibinfo
  {author} {\bibfnamefont {A.~M.}\ \bibnamefont {Somoza}},\ }\bibfield  {title}
  {\enquote {\bibinfo {title} {Deconfined quantum criticality, scaling
  violations, and classical loop models},}\ }\href {\doibase
  10.1103/PhysRevX.5.041048} {\bibfield  {journal} {\bibinfo  {journal} {Phys.
  Rev. X}\ }\textbf {\bibinfo {volume} {5}},\ \bibinfo {pages} {041048}
  (\bibinfo {year} {2015}{\natexlab{b}})}\BibitemShut {NoStop}%
\bibitem [{\citenamefont {Charrier}\ and\ \citenamefont
  {Alet}(2010)}]{charrier2010}%
  \BibitemOpen
  \bibfield  {author} {\bibinfo {author} {\bibfnamefont {D.}~\bibnamefont
  {Charrier}}\ and\ \bibinfo {author} {\bibfnamefont {F.}~\bibnamefont
  {Alet}},\ }\bibfield  {title} {\enquote {\bibinfo {title} {Phase diagram of
  an extended classical dimer model},}\ }\href {\doibase
  10.1103/PhysRevB.82.014429} {\bibfield  {journal} {\bibinfo  {journal} {Phys.
  Rev. B}\ }\textbf {\bibinfo {volume} {82}},\ \bibinfo {pages} {014429}
  (\bibinfo {year} {2010})}\BibitemShut {NoStop}%
\bibitem [{\citenamefont {Liu}\ \emph {et~al.}(2019)\citenamefont {Liu},
  \citenamefont {Wang}, \citenamefont {Sato}, \citenamefont {Hohenadler},
  \citenamefont {Wang}, \citenamefont {Guo},\ and\ \citenamefont
  {Assaad}}]{liuyuhai2019}%
  \BibitemOpen
  \bibfield  {author} {\bibinfo {author} {\bibfnamefont {Y.}~\bibnamefont
  {Liu}}, \bibinfo {author} {\bibfnamefont {Z.}~\bibnamefont {Wang}}, \bibinfo
  {author} {\bibfnamefont {T.}~\bibnamefont {Sato}}, \bibinfo {author}
  {\bibfnamefont {M.}~\bibnamefont {Hohenadler}}, \bibinfo {author}
  {\bibfnamefont {C.}~\bibnamefont {Wang}}, \bibinfo {author} {\bibfnamefont
  {W.}~\bibnamefont {Guo}}, \ and\ \bibinfo {author} {\bibfnamefont {F.F.}\
  \bibnamefont {Assaad}},\ }\bibfield  {title} {\enquote {\bibinfo {title}
  {Superconductivity from the condensation of topological defects in a quantum
  spin-hall insulator},}\ }\href {https://doi.org/10.1038/s41467-019-10372-0}
  {\bibfield  {journal} {\bibinfo  {journal} {Nat Commun}\ }\textbf {\bibinfo
  {volume} {10}},\ \bibinfo {pages} {2658} (\bibinfo {year}
  {2019})}\BibitemShut {NoStop}%
\bibitem [{\citenamefont {Shao}\ \emph {et~al.}(2016)\citenamefont {Shao},
  \citenamefont {Guo},\ and\ \citenamefont {Sandvik}}]{JQ2016}%
  \BibitemOpen
  \bibfield  {author} {\bibinfo {author} {\bibfnamefont {Hui}\ \bibnamefont
  {Shao}}, \bibinfo {author} {\bibfnamefont {Wenan}\ \bibnamefont {Guo}}, \
  and\ \bibinfo {author} {\bibfnamefont {Anders~W.}\ \bibnamefont {Sandvik}},\
  }\bibfield  {title} {\enquote {\bibinfo {title} {Quantum criticality with two
  length scales},}\ }\href {\doibase 10.1126/science.aad5007} {\bibfield
  {journal} {\bibinfo  {journal} {Science}\ }\textbf {\bibinfo {volume}
  {352}},\ \bibinfo {pages} {213--216} (\bibinfo {year} {2016})}\BibitemShut
  {NoStop}%
\bibitem [{\citenamefont {Wang}\ \emph {et~al.}(2017)\citenamefont {Wang},
  \citenamefont {Nahum}, \citenamefont {Metlitski}, \citenamefont {Xu},\ and\
  \citenamefont {Senthil}}]{wang2017}%
  \BibitemOpen
  \bibfield  {author} {\bibinfo {author} {\bibfnamefont {Chong}\ \bibnamefont
  {Wang}}, \bibinfo {author} {\bibfnamefont {Adam}\ \bibnamefont {Nahum}},
  \bibinfo {author} {\bibfnamefont {Max~A.}\ \bibnamefont {Metlitski}},
  \bibinfo {author} {\bibfnamefont {Cenke}\ \bibnamefont {Xu}}, \ and\ \bibinfo
  {author} {\bibfnamefont {T.}~\bibnamefont {Senthil}},\ }\bibfield  {title}
  {\enquote {\bibinfo {title} {Deconfined quantum critical points: Symmetries
  and dualities},}\ }\href {\doibase 10.1103/PhysRevX.7.031051} {\bibfield
  {journal} {\bibinfo  {journal} {Phys. Rev. X}\ }\textbf {\bibinfo {volume}
  {7}},\ \bibinfo {pages} {031051} (\bibinfo {year} {2017})}\BibitemShut
  {NoStop}%
\bibitem [{\citenamefont {Gorbenko}\ \emph
  {et~al.}(2018{\natexlab{a}})\citenamefont {Gorbenko}, \citenamefont
  {Rychkov},\ and\ \citenamefont {Zan}}]{gorbenko2018_1}%
  \BibitemOpen
  \bibfield  {author} {\bibinfo {author} {\bibfnamefont {V.}~\bibnamefont
  {Gorbenko}}, \bibinfo {author} {\bibfnamefont {S.}~\bibnamefont {Rychkov}}, \
  and\ \bibinfo {author} {\bibfnamefont {B.}~\bibnamefont {Zan}},\ }\bibfield
  {title} {\enquote {\bibinfo {title} {Walking, weak first-order transitions,
  and complex cfts},}\ }\href {\doibase 10.1007/JHEP10(2018)108} {\bibfield
  {journal} {\bibinfo  {journal} {J. High Energ. Phys.}\ }\textbf {\bibinfo
  {volume} {2018}} (\bibinfo {year} {2018}{\natexlab{a}}),\
  10.1007/JHEP10(2018)108}\BibitemShut {NoStop}%
\bibitem [{\citenamefont {Gorbenko}\ \emph
  {et~al.}(2018{\natexlab{b}})\citenamefont {Gorbenko}, \citenamefont
  {Rychkov},\ and\ \citenamefont {Zan}}]{gorbenko2018_2}%
  \BibitemOpen
  \bibfield  {author} {\bibinfo {author} {\bibfnamefont {Victor}\ \bibnamefont
  {Gorbenko}}, \bibinfo {author} {\bibfnamefont {Slava}\ \bibnamefont
  {Rychkov}}, \ and\ \bibinfo {author} {\bibfnamefont {Bernardo}\ \bibnamefont
  {Zan}},\ }\bibfield  {title} {\enquote {\bibinfo {title} {{Walking, Weak
  first-order transitions, and Complex CFTs II. Two-dimensional Potts model at
  $Q>4$}},}\ }\href {\doibase 10.21468/SciPostPhys.5.5.050} {\bibfield
  {journal} {\bibinfo  {journal} {SciPost Phys.}\ }\textbf {\bibinfo {volume}
  {5}},\ \bibinfo {pages} {50} (\bibinfo {year}
  {2018}{\natexlab{b}})}\BibitemShut {NoStop}%
\bibitem [{\citenamefont {Zhijin}(2018)}]{lizhijin2018}%
  \BibitemOpen
  \bibfield  {author} {\bibinfo {author} {\bibfnamefont {Li}~\bibnamefont
  {Zhijin}},\ }\bibfield  {title} {\enquote {\bibinfo {title} {Solving $qed_3$
  with conformal bootstrap},}\ }\href@noop {} {\  (\bibinfo {year} {2018})},\
  \Eprint {http://arxiv.org/abs/2018.09281} {arXiv:2018.09281
  [cond-mat.str-el]} \BibitemShut {NoStop}%
\bibitem [{\citenamefont {Lee}\ \emph {et~al.}(2019)\citenamefont {Lee},
  \citenamefont {You}, \citenamefont {Sachdev},\ and\ \citenamefont
  {Vishwanath}}]{Ashvin2019}%
  \BibitemOpen
  \bibfield  {author} {\bibinfo {author} {\bibfnamefont {Jong~Yeon}\
  \bibnamefont {Lee}}, \bibinfo {author} {\bibfnamefont {Yi-Zhuang}\
  \bibnamefont {You}}, \bibinfo {author} {\bibfnamefont {Subir}\ \bibnamefont
  {Sachdev}}, \ and\ \bibinfo {author} {\bibfnamefont {Ashvin}\ \bibnamefont
  {Vishwanath}},\ }\bibfield  {title} {\enquote {\bibinfo {title} {Signatures
  of a deconfined phase transition on the shastry-sutherland lattice:
  Applications to quantum critical
  ${\mathrm{srcu}}_{2}({\mathrm{bo}}_{3}{)}_{2}$},}\ }\href {\doibase
  10.1103/PhysRevX.9.041037} {\bibfield  {journal} {\bibinfo  {journal} {Phys.
  Rev. X}\ }\textbf {\bibinfo {volume} {9}},\ \bibinfo {pages} {041037}
  (\bibinfo {year} {2019})}\BibitemShut {NoStop}%
\bibitem [{\citenamefont {Ma}\ and\ \citenamefont {Wang}(2020)}]{Wang2020}%
  \BibitemOpen
  \bibfield  {author} {\bibinfo {author} {\bibfnamefont {Ruochen}\ \bibnamefont
  {Ma}}\ and\ \bibinfo {author} {\bibfnamefont {Chong}\ \bibnamefont {Wang}},\
  }\bibfield  {title} {\enquote {\bibinfo {title} {Theory of deconfined
  pseudocriticality},}\ }\href {\doibase 10.1103/PhysRevB.102.020407}
  {\bibfield  {journal} {\bibinfo  {journal} {Phys. Rev. B}\ }\textbf {\bibinfo
  {volume} {102}},\ \bibinfo {pages} {020407} (\bibinfo {year}
  {2020})}\BibitemShut {NoStop}%
\bibitem [{\citenamefont {Nahum}(2020)}]{Nahum2020}%
  \BibitemOpen
  \bibfield  {author} {\bibinfo {author} {\bibfnamefont {Adam}\ \bibnamefont
  {Nahum}},\ }\bibfield  {title} {\enquote {\bibinfo {title} {Note on
  wess-zumino-witten models and quasiuniversality in $2+1$ dimensions},}\
  }\href {\doibase 10.1103/PhysRevB.102.201116} {\bibfield  {journal} {\bibinfo
   {journal} {Phys. Rev. B}\ }\textbf {\bibinfo {volume} {102}},\ \bibinfo
  {pages} {201116} (\bibinfo {year} {2020})}\BibitemShut {NoStop}%
\bibitem [{\citenamefont {Zhao}\ \emph
  {et~al.}(2020{\natexlab{a}})\citenamefont {Zhao}, \citenamefont {Takahashi},\
  and\ \citenamefont {Sandvik}}]{sandvik2020}%
  \BibitemOpen
  \bibfield  {author} {\bibinfo {author} {\bibfnamefont {Bowen}\ \bibnamefont
  {Zhao}}, \bibinfo {author} {\bibfnamefont {Jun}\ \bibnamefont {Takahashi}}, \
  and\ \bibinfo {author} {\bibfnamefont {Anders~W.}\ \bibnamefont {Sandvik}},\
  }\bibfield  {title} {\enquote {\bibinfo {title} {Multicritical deconfined
  quantum criticality and lifshitz point of a helical valence-bond phase},}\
  }\href {\doibase 10.1103/PhysRevLett.125.257204} {\bibfield  {journal}
  {\bibinfo  {journal} {Phys. Rev. Lett.}\ }\textbf {\bibinfo {volume} {125}},\
  \bibinfo {pages} {257204} (\bibinfo {year} {2020}{\natexlab{a}})}\BibitemShut
  {NoStop}%
\bibitem [{\citenamefont {He}\ \emph {et~al.}(2021)\citenamefont {He},
  \citenamefont {Rong},\ and\ \citenamefont {Su}}]{He2021}%
  \BibitemOpen
  \bibfield  {author} {\bibinfo {author} {\bibfnamefont {Yin-Chen}\
  \bibnamefont {He}}, \bibinfo {author} {\bibfnamefont {Junchen}\ \bibnamefont
  {Rong}}, \ and\ \bibinfo {author} {\bibfnamefont {Ning}\ \bibnamefont {Su}},\
  }\bibfield  {title} {\enquote {\bibinfo {title} {{Non-Wilson-Fisher kinks of
  $O(N)$ numerical bootstrap: from the deconfined phase transition to a
  putative new family of CFTs}},}\ }\href {\doibase
  10.21468/SciPostPhys.10.5.115} {\bibfield  {journal} {\bibinfo  {journal}
  {SciPost Phys.}\ }\textbf {\bibinfo {volume} {10}},\ \bibinfo {pages} {115}
  (\bibinfo {year} {2021})}\BibitemShut {NoStop}%
\bibitem [{\citenamefont {Wang}\ \emph {et~al.}(2021)\citenamefont {Wang},
  \citenamefont {Zaletel}, \citenamefont {Mong},\ and\ \citenamefont
  {Assaad}}]{Fakher2021}%
  \BibitemOpen
  \bibfield  {author} {\bibinfo {author} {\bibfnamefont {Zhenjiu}\ \bibnamefont
  {Wang}}, \bibinfo {author} {\bibfnamefont {Michael~P.}\ \bibnamefont
  {Zaletel}}, \bibinfo {author} {\bibfnamefont {Roger S.~K.}\ \bibnamefont
  {Mong}}, \ and\ \bibinfo {author} {\bibfnamefont {Fakher~F.}\ \bibnamefont
  {Assaad}},\ }\bibfield  {title} {\enquote {\bibinfo {title} {Phases of the
  ($2+1$) dimensional so(5) nonlinear sigma model with topological term},}\
  }\href {\doibase 10.1103/PhysRevLett.126.045701} {\bibfield  {journal}
  {\bibinfo  {journal} {Phys. Rev. Lett.}\ }\textbf {\bibinfo {volume} {126}},\
  \bibinfo {pages} {045701} (\bibinfo {year} {2021})}\BibitemShut {NoStop}%
\bibitem [{\citenamefont {Yang}\ \emph {et~al.}(2021)\citenamefont {Yang},
  \citenamefont {Sandvik},\ and\ \citenamefont {Wang}}]{wangling2021}%
  \BibitemOpen
  \bibfield  {author} {\bibinfo {author} {\bibfnamefont {Jianwei}\ \bibnamefont
  {Yang}}, \bibinfo {author} {\bibfnamefont {Anders~W.}\ \bibnamefont
  {Sandvik}}, \ and\ \bibinfo {author} {\bibfnamefont {Ling}\ \bibnamefont
  {Wang}},\ }\bibfield  {title} {\enquote {\bibinfo {title} {Quantum
  criticality and spin liquid phase in the shastry-sutherland model},}\
  }\href@noop {} {\  (\bibinfo {year} {2021})},\ \Eprint
  {http://arxiv.org/abs/2104.08887v2} {arXiv:2104.08887v2 [cond-mat.str-el]}
  \BibitemShut {NoStop}%
\bibitem [{\citenamefont {Anderson}(1973)}]{Anderson1973}%
  \BibitemOpen
  \bibfield  {author} {\bibinfo {author} {\bibfnamefont {P.W.}\ \bibnamefont
  {Anderson}},\ }\bibfield  {title} {\enquote {\bibinfo {title} {Resonating
  valence bonds: A new kind of insulator?}}\ }\href {\doibase
  https://doi.org/10.1016/0025-5408(73)90167-0} {\bibfield  {journal} {\bibinfo
   {journal} {Materials Research Bulletin}\ }\textbf {\bibinfo {volume} {8}},\
  \bibinfo {pages} {153--160} (\bibinfo {year} {1973})}\BibitemShut {NoStop}%
\bibitem [{\citenamefont {Liu}\ \emph {et~al.}(2020)\citenamefont {Liu},
  \citenamefont {Gong}, \citenamefont {Li}, \citenamefont {Poilblanc},
  \citenamefont {Chen},\ and\ \citenamefont {Gu}}]{liuQSL}%
  \BibitemOpen
  \bibfield  {author} {\bibinfo {author} {\bibfnamefont {Wen-Yuan}\
  \bibnamefont {Liu}}, \bibinfo {author} {\bibfnamefont {Shou-Shu}\
  \bibnamefont {Gong}}, \bibinfo {author} {\bibfnamefont {Yu-Bin}\ \bibnamefont
  {Li}}, \bibinfo {author} {\bibfnamefont {Didier}\ \bibnamefont {Poilblanc}},
  \bibinfo {author} {\bibfnamefont {Wei-Qiang}\ \bibnamefont {Chen}}, \ and\
  \bibinfo {author} {\bibfnamefont {Zheng-Cheng}\ \bibnamefont {Gu}},\
  }\href@noop {} {\enquote {\bibinfo {title} {Gapless quantum spin liquid and
  global phase diagram of the spin-1/2 $j_1$-$j_2$ square antiferromagnetic
  heisenberg model},}\ } (\bibinfo {year} {2020}),\ \Eprint
  {http://arxiv.org/abs/2009.01821} {arXiv:2009.01821 [cond-mat.str-el]}
  \BibitemShut {NoStop}%
\bibitem [{\citenamefont {Rastelli}\ \emph {et~al.}(1986)\citenamefont
  {Rastelli}, \citenamefont {L.},\ and\ \citenamefont {Tassi}}]{rastelli1986}%
  \BibitemOpen
  \bibfield  {author} {\bibinfo {author} {\bibfnamefont {E.}~\bibnamefont
  {Rastelli}}, \bibinfo {author} {\bibfnamefont {Reatto}\ \bibnamefont {L.}}, \
  and\ \bibinfo {author} {\bibfnamefont {A.}~\bibnamefont {Tassi}},\ }\bibfield
   {title} {\enquote {\bibinfo {title} {Quantum fluctuations and phase diagram
  of heisenberg models with competing interactions},}\ }\href {\doibase
  10.1088/0022-3719/19/33/011} {\bibfield  {journal} {\bibinfo  {journal}
  {Journal of Physics C: Solid State Physics}\ }\textbf {\bibinfo {volume}
  {19}},\ \bibinfo {pages} {6623--6633} (\bibinfo {year} {1986})}\BibitemShut
  {NoStop}%
\bibitem [{\citenamefont {Chandra}\ and\ \citenamefont
  {Doucot}(1988)}]{spinwave1988}%
  \BibitemOpen
  \bibfield  {author} {\bibinfo {author} {\bibfnamefont {P.}~\bibnamefont
  {Chandra}}\ and\ \bibinfo {author} {\bibfnamefont {B.}~\bibnamefont
  {Doucot}},\ }\bibfield  {title} {\enquote {\bibinfo {title} {Possible
  spin-liquid state at large $s$ for the frustrated square heisenberg
  lattice},}\ }\href {\doibase 10.1103/PhysRevB.38.9335} {\bibfield  {journal}
  {\bibinfo  {journal} {Phys. Rev. B}\ }\textbf {\bibinfo {volume} {38}},\
  \bibinfo {pages} {9335--9338} (\bibinfo {year} {1988})}\BibitemShut {NoStop}%
\bibitem [{\citenamefont {Ioffe}\ and\ \citenamefont {Larkin}(1988)}]{RG1988}%
  \BibitemOpen
  \bibfield  {author} {\bibinfo {author} {\bibfnamefont {L.~B.}\ \bibnamefont
  {Ioffe}}\ and\ \bibinfo {author} {\bibfnamefont {A.~I.}\ \bibnamefont
  {Larkin}},\ }\bibfield  {title} {\enquote {\bibinfo {title} {Effective action
  of a two-dimensional antiferromagnet},}\ }\href {\doibase
  https://doi.org/10.1142/S0217979288000160} {\bibfield  {journal} {\bibinfo
  {journal} {Int. J. Mod. Phys. B}\ }\textbf {\bibinfo {volume} {2}},\ \bibinfo
  {pages} {203--219} (\bibinfo {year} {1988})}\BibitemShut {NoStop}%
\bibitem [{\citenamefont {Figueirido}\ \emph {et~al.}(1990)\citenamefont
  {Figueirido}, \citenamefont {Karlhede}, \citenamefont {Kivelson},
  \citenamefont {Sondhi}, \citenamefont {Rocek},\ and\ \citenamefont
  {Rokhsar}}]{ed1990}%
  \BibitemOpen
  \bibfield  {author} {\bibinfo {author} {\bibfnamefont {F.}~\bibnamefont
  {Figueirido}}, \bibinfo {author} {\bibfnamefont {A.}~\bibnamefont
  {Karlhede}}, \bibinfo {author} {\bibfnamefont {S.}~\bibnamefont {Kivelson}},
  \bibinfo {author} {\bibfnamefont {S.}~\bibnamefont {Sondhi}}, \bibinfo
  {author} {\bibfnamefont {M.}~\bibnamefont {Rocek}}, \ and\ \bibinfo {author}
  {\bibfnamefont {D.~S.}\ \bibnamefont {Rokhsar}},\ }\bibfield  {title}
  {\enquote {\bibinfo {title} {Exact diagonalization of finite frustrated
  spin-(1/2 heisenberg models},}\ }\href {\doibase 10.1103/PhysRevB.41.4619}
  {\bibfield  {journal} {\bibinfo  {journal} {Phys. Rev. B}\ }\textbf {\bibinfo
  {volume} {41}},\ \bibinfo {pages} {4619--4632} (\bibinfo {year}
  {1990})}\BibitemShut {NoStop}%
\bibitem [{\citenamefont {Read}\ and\ \citenamefont
  {Sachdev}(1991)}]{largeN1991}%
  \BibitemOpen
  \bibfield  {author} {\bibinfo {author} {\bibfnamefont {N.}~\bibnamefont
  {Read}}\ and\ \bibinfo {author} {\bibfnamefont {Subir}\ \bibnamefont
  {Sachdev}},\ }\bibfield  {title} {\enquote {\bibinfo {title} {Large-n
  expansion for frustrated quantum antiferromagnets},}\ }\href {\doibase
  10.1103/PhysRevLett.66.1773} {\bibfield  {journal} {\bibinfo  {journal}
  {Phys. Rev. Lett.}\ }\textbf {\bibinfo {volume} {66}},\ \bibinfo {pages}
  {1773--1776} (\bibinfo {year} {1991})}\BibitemShut {NoStop}%
\bibitem [{\citenamefont {Rastelli}\ and\ \citenamefont
  {Tassi}(1992)}]{nonlineareffect1992}%
  \BibitemOpen
  \bibfield  {author} {\bibinfo {author} {\bibfnamefont {E.}~\bibnamefont
  {Rastelli}}\ and\ \bibinfo {author} {\bibfnamefont {A.}~\bibnamefont
  {Tassi}},\ }\bibfield  {title} {\enquote {\bibinfo {title} {Nonlinear effects
  in the spin-liquid phase},}\ }\href {\doibase 10.1103/PhysRevB.46.10793}
  {\bibfield  {journal} {\bibinfo  {journal} {Phys. Rev. B}\ }\textbf {\bibinfo
  {volume} {46}},\ \bibinfo {pages} {10793--10799} (\bibinfo {year}
  {1992})}\BibitemShut {NoStop}%
\bibitem [{\citenamefont {Ferrer}(1993)}]{ferrer1993}%
  \BibitemOpen
  \bibfield  {author} {\bibinfo {author} {\bibfnamefont {Jaime}\ \bibnamefont
  {Ferrer}},\ }\bibfield  {title} {\enquote {\bibinfo {title} {Spin-liquid
  phase for the frustrated quantum heisenberg antiferromagnet on a square
  lattice},}\ }\href {\doibase 10.1103/PhysRevB.47.8769} {\bibfield  {journal}
  {\bibinfo  {journal} {Phys. Rev. B}\ }\textbf {\bibinfo {volume} {47}},\
  \bibinfo {pages} {8769--8782} (\bibinfo {year} {1993})}\BibitemShut {NoStop}%
\bibitem [{\citenamefont {Leung}\ and\ \citenamefont {Lam}(1996)}]{ed1996}%
  \BibitemOpen
  \bibfield  {author} {\bibinfo {author} {\bibfnamefont {P.~W.}\ \bibnamefont
  {Leung}}\ and\ \bibinfo {author} {\bibfnamefont {Ngar-wing}\ \bibnamefont
  {Lam}},\ }\bibfield  {title} {\enquote {\bibinfo {title} {Numerical evidence
  for the spin-peierls state in the frustrated quantum antiferromagnet},}\
  }\href {\doibase 10.1103/PhysRevB.53.2213} {\bibfield  {journal} {\bibinfo
  {journal} {Phys. Rev. B}\ }\textbf {\bibinfo {volume} {53}},\ \bibinfo
  {pages} {2213--2216} (\bibinfo {year} {1996})}\BibitemShut {NoStop}%
\bibitem [{\citenamefont {Capriotti}\ \emph {et~al.}(2004)\citenamefont
  {Capriotti}, \citenamefont {Scalapino},\ and\ \citenamefont
  {White}}]{DMRG2004}%
  \BibitemOpen
  \bibfield  {author} {\bibinfo {author} {\bibfnamefont {Luca}\ \bibnamefont
  {Capriotti}}, \bibinfo {author} {\bibfnamefont {Douglas~J.}\ \bibnamefont
  {Scalapino}}, \ and\ \bibinfo {author} {\bibfnamefont {Steven~R.}\
  \bibnamefont {White}},\ }\bibfield  {title} {\enquote {\bibinfo {title}
  {Spin-liquid versus dimerized ground states in a frustrated heisenberg
  antiferromagnet},}\ }\href {\doibase 10.1103/PhysRevLett.93.177004}
  {\bibfield  {journal} {\bibinfo  {journal} {Phys. Rev. Lett.}\ }\textbf
  {\bibinfo {volume} {93}},\ \bibinfo {pages} {177004} (\bibinfo {year}
  {2004})}\BibitemShut {NoStop}%
\bibitem [{\citenamefont {Capriotti}\ and\ \citenamefont
  {Sachdev}(2004)}]{classical2004}%
  \BibitemOpen
  \bibfield  {author} {\bibinfo {author} {\bibfnamefont {Luca}\ \bibnamefont
  {Capriotti}}\ and\ \bibinfo {author} {\bibfnamefont {Subir}\ \bibnamefont
  {Sachdev}},\ }\bibfield  {title} {\enquote {\bibinfo {title} {Low-temperature
  broken-symmetry phases of spiral antiferromagnets},}\ }\href {\doibase
  10.1103/PhysRevLett.93.257206} {\bibfield  {journal} {\bibinfo  {journal}
  {Phys. Rev. Lett.}\ }\textbf {\bibinfo {volume} {93}},\ \bibinfo {pages}
  {257206} (\bibinfo {year} {2004})}\BibitemShut {NoStop}%
\bibitem [{\citenamefont {Mambrini}\ \emph {et~al.}(2006)\citenamefont
  {Mambrini}, \citenamefont {L\"auchli}, \citenamefont {Poilblanc},\ and\
  \citenamefont {Mila}}]{PVB4}%
  \BibitemOpen
  \bibfield  {author} {\bibinfo {author} {\bibfnamefont {Matthieu}\
  \bibnamefont {Mambrini}}, \bibinfo {author} {\bibfnamefont {Andreas}\
  \bibnamefont {L\"auchli}}, \bibinfo {author} {\bibfnamefont {Didier}\
  \bibnamefont {Poilblanc}}, \ and\ \bibinfo {author} {\bibfnamefont
  {Fr\'ed\'eric}\ \bibnamefont {Mila}},\ }\bibfield  {title} {\enquote
  {\bibinfo {title} {Plaquette valence-bond crystal in the frustrated
  \text{Heisenberg} quantum antiferromagnet on the square lattice},}\ }\href
  {\doibase 10.1103/PhysRevB.74.144422} {\bibfield  {journal} {\bibinfo
  {journal} {Phys. Rev. B}\ }\textbf {\bibinfo {volume} {74}},\ \bibinfo
  {pages} {144422} (\bibinfo {year} {2006})}\BibitemShut {NoStop}%
\bibitem [{\citenamefont {Sindzingre}\ \emph {et~al.}(2010)\citenamefont
  {Sindzingre}, \citenamefont {Shannon},\ and\ \citenamefont {Momoi}}]{ed2010}%
  \BibitemOpen
  \bibfield  {author} {\bibinfo {author} {\bibfnamefont {Philippe}\
  \bibnamefont {Sindzingre}}, \bibinfo {author} {\bibfnamefont {Nic}\
  \bibnamefont {Shannon}}, \ and\ \bibinfo {author} {\bibfnamefont {Tsutomu}\
  \bibnamefont {Momoi}},\ }\bibfield  {title} {\enquote {\bibinfo {title}
  {Phase diagram of the spin-1/2 ${J}_{1}$-${J}_{2}$-${J}_{3}$ heisenberg model
  on the square lattice},}\ }\href {\doibase 10.1088/1742-6596/200/2/022058}
  {\bibfield  {journal} {\bibinfo  {journal} {Journal of Physics: Conference
  Series}\ }\textbf {\bibinfo {volume} {200}},\ \bibinfo {pages} {022058}
  (\bibinfo {year} {2010})}\BibitemShut {NoStop}%
\bibitem [{\citenamefont {Zhao}\ \emph
  {et~al.}(2020{\natexlab{b}})\citenamefont {Zhao}, \citenamefont {Takahashi},\
  and\ \citenamefont {Sandvik}}]{zhao2020}%
  \BibitemOpen
  \bibfield  {author} {\bibinfo {author} {\bibfnamefont {Bowen}\ \bibnamefont
  {Zhao}}, \bibinfo {author} {\bibfnamefont {Jun}\ \bibnamefont {Takahashi}}, \
  and\ \bibinfo {author} {\bibfnamefont {Anders~W.}\ \bibnamefont {Sandvik}},\
  }\bibfield  {title} {\enquote {\bibinfo {title} {Comment on ``gapless spin
  liquid ground state of the spin-$1/2$ ${J}_{1}$-${J}_{2}$ \text{Heisenberg}
  model on square lattices''},}\ }\href {\doibase 10.1103/PhysRevB.101.157101}
  {\bibfield  {journal} {\bibinfo  {journal} {Phys. Rev. B}\ }\textbf {\bibinfo
  {volume} {101}},\ \bibinfo {pages} {157101} (\bibinfo {year}
  {2020}{\natexlab{b}})}\BibitemShut {NoStop}%
\bibitem [{\citenamefont {Sandvik}(2010{\natexlab{b}})}]{FSS1}%
  \BibitemOpen
  \bibfield  {author} {\bibinfo {author} {\bibfnamefont {Anders~W.}\
  \bibnamefont {Sandvik}},\ }\bibfield  {title} {\enquote {\bibinfo {title}
  {Computational studies of quantum spin systems},}\ }\href {\doibase
  10.1063/1.3518900} {\bibfield  {journal} {\bibinfo  {journal} {AIP Conference
  Proceedings}\ }\textbf {\bibinfo {volume} {1297}},\ \bibinfo {pages}
  {135--338} (\bibinfo {year} {2010}{\natexlab{b}})}\BibitemShut {NoStop}%
\bibitem [{\citenamefont {Xu}\ \emph {et~al.}(2019)\citenamefont {Xu},
  \citenamefont {Qi}, \citenamefont {Zhang}, \citenamefont {Assaad},
  \citenamefont {Xu},\ and\ \citenamefont {Meng}}]{u1squarelattice2019}%
  \BibitemOpen
  \bibfield  {author} {\bibinfo {author} {\bibfnamefont {Xiao~Yan}\
  \bibnamefont {Xu}}, \bibinfo {author} {\bibfnamefont {Yang}\ \bibnamefont
  {Qi}}, \bibinfo {author} {\bibfnamefont {Long}\ \bibnamefont {Zhang}},
  \bibinfo {author} {\bibfnamefont {Fakher~F.}\ \bibnamefont {Assaad}},
  \bibinfo {author} {\bibfnamefont {Cenke}\ \bibnamefont {Xu}}, \ and\ \bibinfo
  {author} {\bibfnamefont {Zi~Yang}\ \bibnamefont {Meng}},\ }\bibfield  {title}
  {\enquote {\bibinfo {title} {Monte carlo study of lattice compact quantum
  electrodynamics with fermionic matter: The parent state of quantum phases},}\
  }\href {\doibase 10.1103/PhysRevX.9.021022} {\bibfield  {journal} {\bibinfo
  {journal} {Phys. Rev. X}\ }\textbf {\bibinfo {volume} {9}},\ \bibinfo {pages}
  {021022} (\bibinfo {year} {2019})}\BibitemShut {NoStop}%
\bibitem [{\citenamefont {Janssen}\ \emph {et~al.}(2020)\citenamefont
  {Janssen}, \citenamefont {Wang}, \citenamefont {Scherer}, \citenamefont
  {Meng},\ and\ \citenamefont {Xu}}]{u1squarelattice2020}%
  \BibitemOpen
  \bibfield  {author} {\bibinfo {author} {\bibfnamefont {Lukas}\ \bibnamefont
  {Janssen}}, \bibinfo {author} {\bibfnamefont {Wei}\ \bibnamefont {Wang}},
  \bibinfo {author} {\bibfnamefont {Michael~M.}\ \bibnamefont {Scherer}},
  \bibinfo {author} {\bibfnamefont {Zi~Yang}\ \bibnamefont {Meng}}, \ and\
  \bibinfo {author} {\bibfnamefont {Xiao~Yan}\ \bibnamefont {Xu}},\ }\bibfield
  {title} {\enquote {\bibinfo {title} {Confinement transition in the
  ${\mathrm{qed}}_{3}$-gross-neveu-xy universality class},}\ }\href {\doibase
  10.1103/PhysRevB.101.235118} {\bibfield  {journal} {\bibinfo  {journal}
  {Phys. Rev. B}\ }\textbf {\bibinfo {volume} {101}},\ \bibinfo {pages}
  {235118} (\bibinfo {year} {2020})}\BibitemShut {NoStop}%
\bibitem [{\citenamefont {Shackleton}\ \emph {et~al.}(2021)\citenamefont
  {Shackleton}, \citenamefont {Thomson},\ and\ \citenamefont
  {Sachdev}}]{shackleton2021}%
  \BibitemOpen
  \bibfield  {author} {\bibinfo {author} {\bibfnamefont {Henry}\ \bibnamefont
  {Shackleton}}, \bibinfo {author} {\bibfnamefont {Alex}\ \bibnamefont
  {Thomson}}, \ and\ \bibinfo {author} {\bibfnamefont {Subir}\ \bibnamefont
  {Sachdev}},\ }\bibfield  {title} {\enquote {\bibinfo {title} {Deconfined
  criticality and a gapless \textbf{$Z_2$} spin liquid in the square-lattice
  antiferromagnet},}\ }\href {\doibase 10.1103/PhysRevB.104.045110} {\bibfield
  {journal} {\bibinfo  {journal} {Phys. Rev. B}\ }\textbf {\bibinfo {volume}
  {104}},\ \bibinfo {pages} {045110} (\bibinfo {year} {2021})}\BibitemShut
  {NoStop}%
\bibitem [{\citenamefont {Wen}(2002)}]{wen2002}%
  \BibitemOpen
  \bibfield  {author} {\bibinfo {author} {\bibfnamefont {Xiao-Gang}\
  \bibnamefont {Wen}},\ }\bibfield  {title} {\enquote {\bibinfo {title}
  {Quantum orders and symmetric spin liquids},}\ }\href {\doibase
  10.1103/PhysRevB.65.165113} {\bibfield  {journal} {\bibinfo  {journal} {Phys.
  Rev. B}\ }\textbf {\bibinfo {volume} {65}},\ \bibinfo {pages} {165113}
  (\bibinfo {year} {2002})}\BibitemShut {NoStop}%
\bibitem [{\citenamefont {Ferrari}\ and\ \citenamefont
  {Becca}(2020)}]{ferrari2020}%
  \BibitemOpen
  \bibfield  {author} {\bibinfo {author} {\bibfnamefont {Francesco}\
  \bibnamefont {Ferrari}}\ and\ \bibinfo {author} {\bibfnamefont {Federico}\
  \bibnamefont {Becca}},\ }\bibfield  {title} {\enquote {\bibinfo {title}
  {Gapless spin liquid and valence-bond solid in the ${J}_{1}$-${J}_{2}$
  heisenberg model on the square lattice: Insights from singlet and triplet
  excitations},}\ }\href {\doibase 10.1103/PhysRevB.102.014417} {\bibfield
  {journal} {\bibinfo  {journal} {Phys. Rev. B}\ }\textbf {\bibinfo {volume}
  {102}},\ \bibinfo {pages} {014417} (\bibinfo {year} {2020})}\BibitemShut
  {NoStop}%
\bibitem [{\citenamefont {Demler}\ \emph {et~al.}(2004)\citenamefont {Demler},
  \citenamefont {Hanke},\ and\ \citenamefont {Zhang}}]{SO5theory}%
  \BibitemOpen
  \bibfield  {author} {\bibinfo {author} {\bibfnamefont {Eugene}\ \bibnamefont
  {Demler}}, \bibinfo {author} {\bibfnamefont {Werner}\ \bibnamefont {Hanke}},
  \ and\ \bibinfo {author} {\bibfnamefont {Shou-Cheng}\ \bibnamefont {Zhang}},\
  }\bibfield  {title} {\enquote {\bibinfo {title} {$\mathit{SO}(5)$ theory of
  antiferromagnetism and superconductivity},}\ }\href {\doibase
  10.1103/RevModPhys.76.909} {\bibfield  {journal} {\bibinfo  {journal} {Rev.
  Mod. Phys.}\ }\textbf {\bibinfo {volume} {76}},\ \bibinfo {pages} {909--974}
  (\bibinfo {year} {2004})}\BibitemShut {NoStop}%
\bibitem [{\citenamefont {Ebadi}\ \emph {et~al.}(2021)\citenamefont {Ebadi},
  \citenamefont {Wang}, \citenamefont {Levine},\ and\ \citenamefont
  {et~al.}}]{lukin2021}%
  \BibitemOpen
  \bibfield  {author} {\bibinfo {author} {\bibfnamefont {S.}~\bibnamefont
  {Ebadi}}, \bibinfo {author} {\bibfnamefont {T.T.}\ \bibnamefont {Wang}},
  \bibinfo {author} {\bibfnamefont {H.}~\bibnamefont {Levine}}, \ and\ \bibinfo
  {author} {\bibnamefont {et~al.}},\ }\bibfield  {title} {\enquote {\bibinfo
  {title} {Quantum phases of matter on a 256-atom programmable quantum
  simulator},}\ }\href {\doibase https://doi.org/10.1038/s41586-021-03582-4}
  {\bibfield  {journal} {\bibinfo  {journal} {Nature}\ }\textbf {\bibinfo
  {volume} {595}},\ \bibinfo {pages} {227--232} (\bibinfo {year}
  {2021})}\BibitemShut {NoStop}%
\bibitem [{\citenamefont {Scholl}\ \emph {et~al.}(2021)\citenamefont {Scholl},
  \citenamefont {Schuler}, \citenamefont {Williams},\ and\ \citenamefont
  {et~al.}}]{browaeys2021}%
  \BibitemOpen
  \bibfield  {author} {\bibinfo {author} {\bibfnamefont {P.}~\bibnamefont
  {Scholl}}, \bibinfo {author} {\bibfnamefont {M.}~\bibnamefont {Schuler}},
  \bibinfo {author} {\bibfnamefont {H.J.}\ \bibnamefont {Williams}}, \ and\
  \bibinfo {author} {\bibnamefont {et~al.}},\ }\bibfield  {title} {\enquote
  {\bibinfo {title} {Quantum simulation of 2d antiferromagnets with hundreds of
  rydberg atoms.}}\ }\href {\doibase
  https://doi.org/10.1038/s41586-021-03585-1} {\bibfield  {journal} {\bibinfo
  {journal} {Nature}\ }\textbf {\bibinfo {volume} {595}},\ \bibinfo {pages}
  {233--238} (\bibinfo {year} {2021})}\BibitemShut {NoStop}%
\bibitem [{\citenamefont {Semeghini}\ \emph {et~al.}(2021)\citenamefont
  {Semeghini}, \citenamefont {Levine}, \citenamefont {Keesling}, \citenamefont
  {Ebadi},\ and\ \citenamefont {et~al.}}]{topologicalExp1}%
  \BibitemOpen
  \bibfield  {author} {\bibinfo {author} {\bibfnamefont {G.}~\bibnamefont
  {Semeghini}}, \bibinfo {author} {\bibfnamefont {H.}~\bibnamefont {Levine}},
  \bibinfo {author} {\bibfnamefont {A.}~\bibnamefont {Keesling}}, \bibinfo
  {author} {\bibfnamefont {S.}~\bibnamefont {Ebadi}}, \ and\ \bibinfo {author}
  {\bibnamefont {et~al.}},\ }\bibfield  {title} {\enquote {\bibinfo {title}
  {Probing topological spin liquids on a programmable quantum simulator},}\
  }\href {\doibase 10.1126/science.abi8794} {\bibfield  {journal} {\bibinfo
  {journal} {Science}\ }\textbf {\bibinfo {volume} {374}},\ \bibinfo {pages}
  {1242--1247} (\bibinfo {year} {2021})}\BibitemShut {NoStop}%
\bibitem [{\citenamefont {White}(1992)}]{white1992}%
  \BibitemOpen
  \bibfield  {author} {\bibinfo {author} {\bibfnamefont {Steven~R.}\
  \bibnamefont {White}},\ }\bibfield  {title} {\enquote {\bibinfo {title}
  {Density matrix formulation for quantum renormalization groups},}\ }\href
  {\doibase 10.1103/PhysRevLett.69.2863} {\bibfield  {journal} {\bibinfo
  {journal} {Phys. Rev. Lett.}\ }\textbf {\bibinfo {volume} {69}},\ \bibinfo
  {pages} {2863--2866} (\bibinfo {year} {1992})}\BibitemShut {NoStop}%
\bibitem [{\citenamefont {Verstraete}\ \emph {et~al.}(2008)\citenamefont
  {Verstraete}, \citenamefont {Murg},\ and\ \citenamefont
  {Cirac}}]{verstraete2008}%
  \BibitemOpen
  \bibfield  {author} {\bibinfo {author} {\bibfnamefont {F.}~\bibnamefont
  {Verstraete}}, \bibinfo {author} {\bibfnamefont {V.}~\bibnamefont {Murg}}, \
  and\ \bibinfo {author} {\bibfnamefont {J.I.}\ \bibnamefont {Cirac}},\
  }\bibfield  {title} {\enquote {\bibinfo {title} {Matrix product states,
  projected entangled pair states, and variational renormalization group
  methods for quantum spin systems},}\ }\href {\doibase
  10.1080/14789940801912366} {\bibfield  {journal} {\bibinfo  {journal}
  {Advances in Physics}\ }\textbf {\bibinfo {volume} {57}},\ \bibinfo {pages}
  {143--224} (\bibinfo {year} {2008})}\BibitemShut {NoStop}%
\bibitem [{\citenamefont {McCulloch}\ and\ \citenamefont
  {Gul{\'a}csi}(2002)}]{su2}%
  \BibitemOpen
  \bibfield  {author} {\bibinfo {author} {\bibfnamefont {I.~P.}\ \bibnamefont
  {McCulloch}}\ and\ \bibinfo {author} {\bibfnamefont {M.}~\bibnamefont
  {Gul{\'a}csi}},\ }\bibfield  {title} {\enquote {\bibinfo {title} {The
  non-abelian density matrix renormalization group algorithm},}\ }\href
  {http://iopscience.iop.org/0295-5075/57/6/852} {\bibfield  {journal}
  {\bibinfo  {journal} {Europhysics Letters}\ }\textbf {\bibinfo {volume}
  {57}},\ \bibinfo {pages} {852--858} (\bibinfo {year} {2002})}\BibitemShut
  {NoStop}%
\bibitem [{\citenamefont {Sandvik}\ and\ \citenamefont
  {Vidal}(2007)}]{sandvik2007}%
  \BibitemOpen
  \bibfield  {author} {\bibinfo {author} {\bibfnamefont {A.~W.}\ \bibnamefont
  {Sandvik}}\ and\ \bibinfo {author} {\bibfnamefont {G.}~\bibnamefont
  {Vidal}},\ }\bibfield  {title} {\enquote {\bibinfo {title} {Variational
  quantum monte carlo simulations with tensor-network states},}\ }\href
  {\doibase 10.1103/PhysRevLett.99.220602} {\bibfield  {journal} {\bibinfo
  {journal} {Phys. Rev. Lett.}\ }\textbf {\bibinfo {volume} {99}},\ \bibinfo
  {pages} {220602} (\bibinfo {year} {2007})}\BibitemShut {NoStop}%
\bibitem [{\citenamefont {Schuch}\ \emph {et~al.}(2008)\citenamefont {Schuch},
  \citenamefont {Wolf}, \citenamefont {Verstraete},\ and\ \citenamefont
  {Cirac}}]{schuch2008}%
  \BibitemOpen
  \bibfield  {author} {\bibinfo {author} {\bibfnamefont {Norbert}\ \bibnamefont
  {Schuch}}, \bibinfo {author} {\bibfnamefont {Michael~M.}\ \bibnamefont
  {Wolf}}, \bibinfo {author} {\bibfnamefont {Frank}\ \bibnamefont
  {Verstraete}}, \ and\ \bibinfo {author} {\bibfnamefont {J.~Ignacio}\
  \bibnamefont {Cirac}},\ }\bibfield  {title} {\enquote {\bibinfo {title}
  {Simulation of quantum many-body systems with strings of operators and monte
  carlo tensor contractions},}\ }\href {\doibase
  10.1103/PhysRevLett.100.040501} {\bibfield  {journal} {\bibinfo  {journal}
  {Phys. Rev. Lett.}\ }\textbf {\bibinfo {volume} {100}},\ \bibinfo {pages}
  {040501} (\bibinfo {year} {2008})}\BibitemShut {NoStop}%
\bibitem [{\citenamefont {Liu}\ \emph {et~al.}(2017)\citenamefont {Liu},
  \citenamefont {Dong}, \citenamefont {Han}, \citenamefont {Guo},\ and\
  \citenamefont {He}}]{liu2017}%
  \BibitemOpen
  \bibfield  {author} {\bibinfo {author} {\bibfnamefont {Wen-Yuan}\
  \bibnamefont {Liu}}, \bibinfo {author} {\bibfnamefont {Shao-Jun}\
  \bibnamefont {Dong}}, \bibinfo {author} {\bibfnamefont {Yong-Jian}\
  \bibnamefont {Han}}, \bibinfo {author} {\bibfnamefont {Guang-Can}\
  \bibnamefont {Guo}}, \ and\ \bibinfo {author} {\bibfnamefont {Lixin}\
  \bibnamefont {He}},\ }\bibfield  {title} {\enquote {\bibinfo {title}
  {Gradient optimization of finite projected entangled pair states},}\ }\href
  {\doibase 10.1103/PhysRevB.95.195154} {\bibfield  {journal} {\bibinfo
  {journal} {Phys. Rev. B}\ }\textbf {\bibinfo {volume} {95}},\ \bibinfo
  {pages} {195154} (\bibinfo {year} {2017})}\BibitemShut {NoStop}%
\bibitem [{\citenamefont {Liu}\ \emph {et~al.}(2021)\citenamefont {Liu},
  \citenamefont {Huang}, \citenamefont {Gong},\ and\ \citenamefont
  {Gu}}]{obcPEPS}%
  \BibitemOpen
  \bibfield  {author} {\bibinfo {author} {\bibfnamefont {Wen-Yuan}\
  \bibnamefont {Liu}}, \bibinfo {author} {\bibfnamefont {Yi-Zhen}\ \bibnamefont
  {Huang}}, \bibinfo {author} {\bibfnamefont {Shou-Shu}\ \bibnamefont {Gong}},
  \ and\ \bibinfo {author} {\bibfnamefont {Zheng-Cheng}\ \bibnamefont {Gu}},\
  }\bibfield  {title} {\enquote {\bibinfo {title} {Accurate simulation for
  finite projected entangled pair states in two dimensions},}\ }\href {\doibase
  10.1103/PhysRevB.103.235155} {\bibfield  {journal} {\bibinfo  {journal}
  {Phys. Rev. B}\ }\textbf {\bibinfo {volume} {103}},\ \bibinfo {pages}
  {235155} (\bibinfo {year} {2021})}\BibitemShut {NoStop}%
\bibitem [{\citenamefont {Jiang}\ \emph
  {et~al.}(2008{\natexlab{b}})\citenamefont {Jiang}, \citenamefont {Weng},\
  and\ \citenamefont {Xiang}}]{jiang2008}%
  \BibitemOpen
  \bibfield  {author} {\bibinfo {author} {\bibfnamefont {H.~C.}\ \bibnamefont
  {Jiang}}, \bibinfo {author} {\bibfnamefont {Z.~Y.}\ \bibnamefont {Weng}}, \
  and\ \bibinfo {author} {\bibfnamefont {T.}~\bibnamefont {Xiang}},\ }\bibfield
   {title} {\enquote {\bibinfo {title} {Accurate determination of tensor
  network state of quantum lattice models in two dimensions},}\ }\href
  {\doibase 10.1103/PhysRevLett.101.090603} {\bibfield  {journal} {\bibinfo
  {journal} {Phys. Rev. Lett.}\ }\textbf {\bibinfo {volume} {101}},\ \bibinfo
  {pages} {090603} (\bibinfo {year} {2008}{\natexlab{b}})}\BibitemShut
  {NoStop}%
\bibitem [{\citenamefont {Hasik}\ \emph {et~al.}(2021)\citenamefont {Hasik},
  \citenamefont {Poilblanc},\ and\ \citenamefont {Becca}}]{hasik2021}%
  \BibitemOpen
  \bibfield  {author} {\bibinfo {author} {\bibfnamefont {Juraj}\ \bibnamefont
  {Hasik}}, \bibinfo {author} {\bibfnamefont {Didier}\ \bibnamefont
  {Poilblanc}}, \ and\ \bibinfo {author} {\bibfnamefont {Federico}\
  \bibnamefont {Becca}},\ }\bibfield  {title} {\enquote {\bibinfo {title}
  {Investigation of the néel phase of the frustrated heisenberg
  antiferromagnet by differentiable symmetric tensor networks},}\ }\href
  {\doibase 10.21468/SciPostPhys.10.1.012} {\bibfield  {journal} {\bibinfo
  {journal} {SciPost Phys.}\ }\textbf {\bibinfo {volume} {10}},\ \bibinfo
  {pages} {12} (\bibinfo {year} {2021})}\BibitemShut {NoStop}%
\bibitem [{Note1()}]{Note1}%
  \BibitemOpen
  \bibinfo {note} {The antiferromagnetic order is incorporated into this
  translationally invariant wavefunction by rotation of the physical space on
  each sublattice-B site: $\protect \mathbf {S}\cdot \protect \mathbf {S}\
  \rightarrow \ \protect \mathbf {S}\cdot \protect \mathbf {\protect
  \mathaccentV {tilde}07E{S}}$ with $\protect \mathaccentV {tilde}07E{S}^\alpha
  = -\sigma _y S^\alpha (\sigma _y)^T$}\BibitemShut {NoStop}%
\bibitem [{\citenamefont {Hasik}\ and\ \citenamefont
  {Mbeng}(2020)}]{Hasik2021pepstorch}%
  \BibitemOpen
  \bibfield  {author} {\bibinfo {author} {\bibfnamefont {J.}~\bibnamefont
  {Hasik}}\ and\ \bibinfo {author} {\bibfnamefont {G.~B.}\ \bibnamefont
  {Mbeng}},\ }\href {https://github.com/jurajHasik/peps-torch} {\enquote
  {\bibinfo {title} {peps-torch: A differentiable tensor network library for
  two-dimensional lattice models},}\ } (\bibinfo {year} {2020})\BibitemShut
  {NoStop}%
\bibitem [{\citenamefont {Corboz}\ \emph {et~al.}(2018)\citenamefont {Corboz},
  \citenamefont {Czarnik}, \citenamefont {Kapteijns},\ and\ \citenamefont
  {Tagliacozzo}}]{Corboz2018}%
  \BibitemOpen
  \bibfield  {author} {\bibinfo {author} {\bibfnamefont {Philippe}\
  \bibnamefont {Corboz}}, \bibinfo {author} {\bibfnamefont {Piotr}\
  \bibnamefont {Czarnik}}, \bibinfo {author} {\bibfnamefont {Geert}\
  \bibnamefont {Kapteijns}}, \ and\ \bibinfo {author} {\bibfnamefont {Luca}\
  \bibnamefont {Tagliacozzo}},\ }\bibfield  {title} {\enquote {\bibinfo {title}
  {Finite correlation length scaling with infinite projected entangled-pair
  states},}\ }\href {\doibase 10.1103/PhysRevX.8.031031} {\bibfield  {journal}
  {\bibinfo  {journal} {Phys. Rev. X}\ }\textbf {\bibinfo {volume} {8}},\
  \bibinfo {pages} {031031} (\bibinfo {year} {2018})}\BibitemShut {NoStop}%
\bibitem [{\citenamefont {Rader}\ and\ \citenamefont
  {L\"auchli}(2018)}]{rader2018}%
  \BibitemOpen
  \bibfield  {author} {\bibinfo {author} {\bibfnamefont {Michael}\ \bibnamefont
  {Rader}}\ and\ \bibinfo {author} {\bibfnamefont {Andreas~M.}\ \bibnamefont
  {L\"auchli}},\ }\bibfield  {title} {\enquote {\bibinfo {title} {Finite
  correlation length scaling in lorentz-invariant gapless \text{iPEPS} wave
  functions},}\ }\href {\doibase 10.1103/PhysRevX.8.031030} {\bibfield
  {journal} {\bibinfo  {journal} {Phys. Rev. X}\ }\textbf {\bibinfo {volume}
  {8}},\ \bibinfo {pages} {031030} (\bibinfo {year} {2018})}\BibitemShut
  {NoStop}%
\bibitem [{\citenamefont {Vanhecke}\ \emph {et~al.}(2021)\citenamefont
  {Vanhecke}, \citenamefont {Hasik}, \citenamefont {Verstraete},\ and\
  \citenamefont {Vanderstraeten}}]{vanhecke2021scaling}%
  \BibitemOpen
  \bibfield  {author} {\bibinfo {author} {\bibfnamefont {Bram}\ \bibnamefont
  {Vanhecke}}, \bibinfo {author} {\bibfnamefont {Juraj}\ \bibnamefont {Hasik}},
  \bibinfo {author} {\bibfnamefont {Frank}\ \bibnamefont {Verstraete}}, \ and\
  \bibinfo {author} {\bibfnamefont {Laurens}\ \bibnamefont {Vanderstraeten}},\
  }\href@noop {} {\enquote {\bibinfo {title} {A scaling hypothesis for
  projected entangled-pair states},}\ } (\bibinfo {year} {2021}),\ \Eprint
  {http://arxiv.org/abs/2102.03143} {arXiv:2102.03143 [quant-ph]} \BibitemShut
  {NoStop}%
\bibitem [{\citenamefont {Sandvik}(1997)}]{sandvik1997}%
  \BibitemOpen
  \bibfield  {author} {\bibinfo {author} {\bibfnamefont {Anders~W.}\
  \bibnamefont {Sandvik}},\ }\bibfield  {title} {\enquote {\bibinfo {title}
  {Finite-size scaling of the ground-state parameters of the two-dimensional
  heisenberg model},}\ }\href {\doibase 10.1103/PhysRevB.56.11678} {\bibfield
  {journal} {\bibinfo  {journal} {Phys. Rev. B}\ }\textbf {\bibinfo {volume}
  {56}},\ \bibinfo {pages} {11678--11690} (\bibinfo {year} {1997})}\BibitemShut
  {NoStop}%
\end{thebibliography}%
\end{document}